\documentclass[journal ]{new-aiaa}
\usepackage[utf8]{inputenc}
\usepackage{textcomp}

\usepackage{graphicx}
\usepackage{amsmath}
\usepackage{bm}
\usepackage[version=4]{mhchem}
\usepackage{siunitx}
\usepackage{longtable,tabularx}
\usepackage{textgreek}
\usepackage{caption}
\usepackage{subcaption}
\title{Density-based in-orbit collision risk model extension to any impact geometry}

\author{Lorenzo Giudici \footnote{Ph.D. Student, Department of Aerospace Science and Technology,  lorenzo1.giudici@polimi.it.}, 
Juan Luis Gonzalo \footnote{Ph.D. Assistant Professor, Department of Aerospace Science and Technology,  juanluis.gonzalo@polimi.it.}, 
Camilla Colombo \footnote{Ph.D. Associate Professor, Department of Aerospace Science and Technology,  camilla.colombo@polimi.it.}}
\affil{Politecnico di Milano, Via la Masa 34, 20156 Milano, Italy}

\begin{document}

\maketitle

\begin{abstract}
Neglecting small fragments in space debris evolutionary models can lead to a significant underestimation of the collision risk for operational satellites. However, when scaling down to the millimeter range, the debris population grows to over one million objects, making deterministic approaches too computationally expensive. On the contrary, probabilistic models provide a more efficient alternative, which however typically work under some simplifying assumptions on the dynamics, limiting their field of applicability. This work proposes an extension of the density-based collision risk models to any orbital dynamics and impact geometry. The impact rate with a target satellite is derived from a multi-dimensional phase space density function in Keplerian elements, which discretely varies over both phase space and time. The assumption of a bin-wise constant cloud density allows for the analytical transformation of the six-dimensional distribution in orbital elements into the three-dimensional spatial density function, guaranteeing an efficient and accurate evaluation of the fragments flux. The proposed method is applied to the assessment of the collision risk posed by real fragmentation clouds in different orbital regions on a high-risk target object. The effect on the impact rate of the additional model features, compared to previous probabilistic formulations, is discussed.

\end{abstract}

\section*{Nomenclature}

{\renewcommand\arraystretch{1.0}
\noindent\begin{longtable*}{@{}l @{\quad=\quad} l@{}}
$A_c$  & Cross-sectional area, m$^2$ \\
$A/M$ & Area-to-mass ratio, m$^2$/kg\\
$a$  & Semi-major axis, km \\
$\mathcal{D}_{(\cdot)}$ & Domain in the given variable \\
$e$ & Eccentricity \\
$\mathrm{F}$ & Incomplete elliptic integral of the first kind \\
$\bm{F}$ & Orbital dynamics function\\
$f$ & True anomaly, rad \\
$\mathrm{H}$ & Hessian tensor \\
$i$& Inclination, rad \\
$\mathrm{J}$ & Jacobian matrix \\
$\mathrm{K}$ & Complete elliptic integral of the first kind \\
$M$ & Mean anomaly, rad \\
$N$ & Number of fragments \\
$n_{(\cdot)}$  & Fragments density in the given variables \\
$P$ & Cumulative density function \\
$p$ & Probability density function\\
$r$ & Orbital radius, km \\
$r_p$ & Perigee radius, km \\
$r_a$ & Apogee radius, km \\
$\bm{r}$ & Position vector, km \\
$S$ & Explosion factor \\
$u$ & Argument of latitude, rad \\
$V$ & Bin volume \\
$v_\mathrm{rel}$ & Relative velocity, km/s \\
$\bm{v}$ & Velocity vector, km/s \\
$\bm{\alpha}$ & Subset of Keplerian elements $(a,e,i)$ \\
$\bm{\beta}$ & Subset of Keplerian elements $(\Omega,\omega,f)$ \\
$\gamma$ & In-plane ejection velocity angle, rad\\
$\bm{\gamma}$ & Slow-varying Keplerian elements $(a,e,i,\Omega,\omega)$\\
$\Delta(\cdot)$ & Variation in the given variable \\
$\delta$ & Angle between velocity vectors of impacting objects, rad\\
$\delta(\cdot)$ & Step-size in the given variable \\
$\zeta$ & Fraction of fragments captured by the model \\
$\eta$ & Number of impacts \\
$\dot{\eta}$ & Impact rate, 1/s\\
$\tau$ & Semi-major axis, eccentricity to perigee, apogee radii coordinates transformation \\
$\phi$ & Latitude, rad \\
$\lambda$ & Logarithm to base 10 of characteristic length, m\\
$\nu$ & Logarithm to base 10 of ejection velocity magnitude, km/s\\
$\varphi$ & Out-of plane ejection velocity angle, rad\\
$\chi$ & Logarithm to base 10 of area-to-mass ratio, m$^2$/kg\\
$\psi$ & Cartesian to Keplerian coordinates transformation\\
$\Omega$ & Right ascension of the ascending node, rad \\
$\omega$ & Argument of periapsis, rad \\
\multicolumn{2}{@{}l}{Subscripts}\\
P & Parent object\\
T & Target object\\
\end{longtable*}}

\section{Introduction}
\label{Introduction}

\lettrine{T}{he} number of services provided by in-orbit satellites is massively increasing and, together with that, our exploitation of the space environment. As of today, more than thirty thousand objects are tracked by the space surveillance network, of which only one-third are operational satellites~\cite{esaReport}. As motivated in~\cite{klinkrad2006}, the trackability of space objects strictly depends on the orbital altitude. Due to the current limitations in the sensitivity of radars and telescopes, the lower size threshold for an object to be detected is 1-10 cm in Low-Earth Orbit (LEO), and on the order of 1 m in the Geostationary ring (GEO)~\cite{klinkrad2006}. However, excluding smaller fragments when studying the effects of remediation measures for the space debris problem, results in an underestimation of the collision risk for the active satellites population~\cite{White2014}. Experimental data proves that impacts from millimeter-sized particles may compromise the functionality of certain equipment, while centimeter-size projectiles can potentially destroy a satellite in case of a collision~\cite{Drolshagen2008,Krag2017}. Therefore, to have a representative picture of the health and future trend of the space environment, the debris models must have the capability to estimate the actual threat posed to space operations by such small objects. However, when scaling down to the millimeter-size, the population of objects grows to over 100 million~\cite{Johnson2010}, representing a barrier even to the computational power of modern computers.

The assessment of the collision risk posed by a population of fragments was historically performed through the \"Opik's theory~\cite{Opik1951}. It provides equations to derive the probability of collision between two objects, based on their orbital elements, under the assumption of zero eccentricity and inclination for one of the two. This assumption was eliminated in a later extension by Wetherill~\cite{Wetherill1967}. Kessler extended the model to compute the impact risk from the spatial density of a debris population, assuming the fragments orbit as uniformly distributed over the Euler angles, and characterized by fixed semi-major axis, eccentricity and inclination~\cite{Kessler1978, Kessler1981}. Because of the uniform distribution of the particles in longitude, the spatial density was evaluated according to a two-dimensional discretization of the physical space in altitude and latitude. However, as motivated in~\cite{Rossi2009}, such approximation does not allow the applicability of the model to any orbital regime. Indeed, under some initial conditions, randomization may take long to occur and the fragments orbit shape and orientation may be considerably distorted by orbital perturbations.

Semi-deterministic debris evolutionary models were developed to answer this necessity. Firstly introduced by Rossi et al.~\cite{Rossi1996,Rossi1998}, these methods accomplished the propagation of debris populations under any force model, by gathering the fragments in some representative samples, whose orbit is propagated in time~\cite{Lewis2001,Liou2004,Virgili2016}. The relative simplicity of the method makes it very flexible to include additional complexities, as predictive launch models or active debris removal. The CUBE algorithm was developed for the accurate evaluation of the collision hazard caused by the evolving fragments population~\cite{Liou2006}. The model divides the physical space into sufficiently small volumes and estimates the spatial density of objects through a uniform sampling of the system in time. Whenever two objects share the same cube, the collision probability is estimated according to their spatial density, the relative velocity, the collision cross-section, and the volume of the cube. 

The bottleneck of the semi-deterministic approaches remains the computational cost. For this reason, efficient analytical and semi-analytical probabilistic debris models were also developed. Through an analogy with fluid dynamics, these methods treat the fragments no longer as individual pieces but as a fluid, which continuously deforms under the effect of external disturbances, as the orbital perturbations. The evolution of the cloud density is retrieved through the time integration of the continuity equation~\cite{McInnes1993,Letizia2015a,Letizia2015b}. In the work by Letizia et al.~\cite{Letizia2015b}, the collision probability was evaluated according to the formulation by Kessler~\cite{Kessler1978}, but from a one-dimensional time-varying spatial density function dependent on orbital radius. The method assumes the debris cloud as randomized over the Euler angles and the target object moving on a circular orbit, which constrains its applicability to LEO. Instead, the dependency on the fragments distribution over longitude was included in a dedicated GEO model proposed in~\cite{McKnight2013}. 

The STARLING suite was developed at Politecnico di Milano with the aim of extending the continuum approach to any orbital regime~\cite{Frey2019}. The suite numerically integrates the continuity equation through the method of characteristics in the phase space of Keplerian elements and area-to-mass ratio, and retrieves the density distribution by fitting a Gaussian Mixture Model (GMM) to the propagated bulk of samples. However, when the third-body perturbation and solar radiation pressure have a predominant effect on the cloud evolution, they may induce bifurcations on a small subset of the phase space, branching out part of the distribution from the main bulk of characteristics. When such condition occurs, the interpolation through the GMM fails~\cite{Frey2019}, limiting the applicability of the model to simplified dynamical regimes. In~\cite{Frey2021}, Frey et al. derived an equation for the estimation of the impact rate between a fragmentation cloud and a target satellite, directly from the six-dimensional phase space density in Keplerian elements. Nevertheless, due to the current limitation of the model, its use was limited to LEO, still assuming fixed fragments inclination and cloud randomization over the Euler angles.

This work aims to extend the density-based collision probability models to any orbital regime. It adopts the continuum propagation method proposed in~\cite{Giudici2023} for the estimation of the debris cloud evolution in time, under the effect of any force model. The resulting evolving density function discretely varies over the phase space of Keplerian elements, and time. Under this condition, the impact rate with a target satellite is conviniently evaluated through the semi-analytical integration of the fragments flux over the target area. This allows an efficient and very accurate estimation of the effect of a fragmentation event on the space environment, applicable to any orbital region.

The paper is organized as follows:
\begin{itemize}
    \item[-] Section II gives an overview of the cloud propagator method presented in~\cite{Giudici2023}, and adopted as input for the collision risk model within this article.
    \item[-] Section III is the core of this work. Firstly, the collision risk model for LEO fragmentation proposed in~\cite{Letizia2015b} is explained, to stress the similarity with the developed approach. In addition, we take the chance to correct a conceptual error in the derivation of the theory. Secondly, the novel and more advanced collision risk assessment method, valid under any orbital regime, is presented.
    \item[-] Section IV is devoted to the application of the theory to real fragmentation events. The additional features, which the proposed method is able to characterize, because of the multidimensional description of the cloud dynamics, are included once at a time, to evaluate their effect on the estimated collision probability.
    \item[-] Section V recaps the main results and achievements of the work.
\end{itemize}

\section{Fragmentation modeling and propagation}
\label{Fragmentation modeling and propagation}

\begin{figure}[!ht]
\centering
\includegraphics[width=0.9\textwidth]{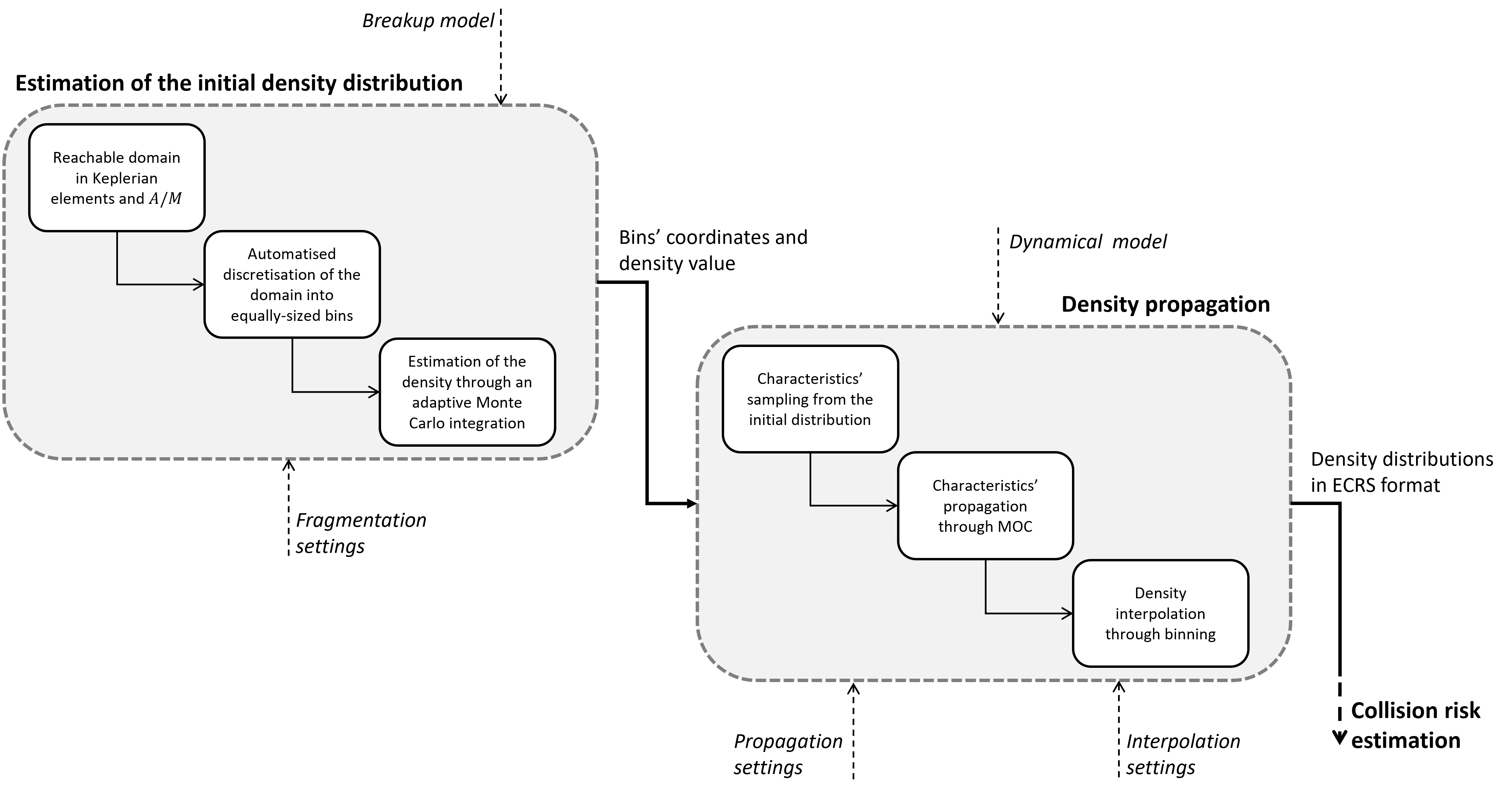}
\caption{Block diagram of the cloud propagation model.}
\label{fig: scheme_cloud_prop}
\end{figure}

This section recaps the method for the modeling and propagation of a potential cloud of fragments generated by either an in-orbit collision or explosion, which was presented in a past work by the authors~\cite{Giudici2023}. The same model is here adopted as input for the collision risk assessment method proposed in Section~\ref{Debris density-based collision risk assessment}. The cloud propagator is divided into two main parts, as shown in Figure~\ref{fig: scheme_cloud_prop}. Each of them is explained in the following sections.

\subsection{Estimation of the initial density distribution}
\label{Estimation of the initial density distribution}

The debris density distribution at fragmentation epoch is approximated through a binning approach. The multidimensional phase space is partitioned into equally-sized rectangular-shaped volumes, here referred to as bins, and a density value is assigned to each of them. If the grid is fine enough, despite of its discontinuous nature, the discretized density distribution resembles the actual fragments distribution. The bottleneck of the approach is the computational load; indeed, it must be understood that the number of bins grows as an exponential function of the number of phase space dimensions. Therefore, it is crucial to precisely define the domain occupied by the fragmentation event, a priori, and to estimate the phase space density in such a region only.

\subsubsection{Breakup model}
\label{Brekup model}

The model adopts the NASA Standard Breakup Model (NASA SBM) for characterizing the fragments ejected by the fragmentation event~\cite{Johnson2001, Krisko2011}. In~\cite{Frey2021}, the method was reformulated in a probabilistic fashion; the ejected debris are described according to three Probability Density Functions (PDFs) in logarithm to base 10 of:
\begin{itemize}
    \item Characteristic length, $\lambda$.
    \item Area-to-mass ratio, $\chi$, given $\lambda$.
    \item Ejection velocity, $\nu$, given $\chi$.
\end{itemize}
The three PDFs will be referred to as $p_\lambda$, $p_{\chi|\lambda}$, $p_{\nu|\chi}$, respectively. As detailed in~\cite{Giudici2023}, the fragments distribution over characteristic length can be filtered out through marginalization, i.e., integrating the conditional density distribution $p_{\chi|\lambda}$ over $\lambda$, to obtain $p_\chi$. Indeed, the dynamical model considered in this paper only depends on area-to-mass ratio as physical property. 

It is worth noting that the method for the estimation of the initial distribution presented in this paper does not strictly depend on the breakup model adopted. The only constraint is that the fragments must be described through PDFs in area-to-mass ratio $A/M$ and ejection velocity $\Delta v$ (or functions of them). 

\subsubsection{Phase space domain of the breakup event}
\label{Phase space domain of the breakup event}

The domain probabilistically reachable by the ejected fragments is set according to the Cumulative Density Functions (CDFs) of the PDFs presented in Section~\ref{Brekup model}. As it will be detailed in Section~\ref{Continuity-equation based fragments density propagation}, the model adopts a variation of parameters approach~\cite{Vallado2007} for the propagation of the fragments density. Therefore, the density distribution at fragmentation epoch must be function of Keplerian elements and area-to-mass ratio. Consequently, the domain must be defined in such a phase space, according to the following steps:
\begin{enumerate}
    \item Definition of the boundaries in logarithm to base 10 of area-to-mass ratio and ejection velocity, $\mathcal{D_{\chi,\nu}}$.
    \item Transformation into a domain in Keplerian elements and area-to-mass ratio, $\mathcal{D}_{\bm{\alpha},A/M}$.
\end{enumerate}

As demonstrated in~\cite{Giudici2023}, if the domain in $\chi$ is split into $N_\chi$ bins, the 2D domain $\mathcal{D}_{\chi,\nu}$ hosting a fraction $\zeta$ of the total number of fragments can be bounded by two limit values in $\chi$, $\chi_0$ and $\chi_{N_\chi}$, and $N_\chi$ values in $\nu$, $\nu_j$. Indeed, note that the velocity distribution $p_{\nu|\chi}$ is conditional and, thus, it varies depending on the value of area-to-mass ratio. As a result, each area-to-mass ratio bin is characterized by a different limit in ejection velocity. The $N_\chi+2$ limit values can be found by solving the following system of $N_\chi+2$ non-linear equations~\cite{Giudici2023}:
\begin{equation}
\begin{cases}
    \sum_{j=1}^{N_\chi}\Phi^{(j)}\left(\nu_j\right)P_\chi\left(\chi_{j-1}<\chi\leq\chi_j\right)=\zeta\\
    p_{\nu|\chi}\left(\nu_1,\overline{\chi}_1\right)\overline{p}_\chi^{(1)}=p_{\nu|\chi}\left(\nu_{2},\overline{\chi}_{2}\right)\overline{p}_\chi^{(2)}\\
    \qquad\qquad\vdots\\
    p_{\nu|\chi}\left(\nu_{N_\chi-1},\overline{\chi}_{N_\chi-1}\right)\overline{p}_\chi^{(N_\chi-1)}=p_{\nu|\chi}\left(\nu_{N_\chi},\overline{\chi}_{N_\chi}\right)\overline{p}_\chi^{(N_\chi)}\\
    P_\chi\left(\chi_0<\chi\leq\chi_{N_\chi}\right)=\frac{\sqrt{1+8\zeta}}{2}-\frac{1}{2}\\
    p_\chi\left(\chi_0\right)=p_\chi\left(\chi_{N_\chi}\right)
\end{cases}
\label{eq: nu,chi bounds}
\end{equation}
where $P_\chi$ and $P_{\nu|\chi}$ are the CDFs of $p_\chi$ and $p_{\nu|\chi}$, and $\overline{\chi}_j$ is the middle point of the bin in logarithm to base 10 of area-to-mass ratio, with edges $\chi_{j-1}$ and $\chi_j$. Note that, in order for the system to be solvable, the PDF in $\chi$, $p_\chi$, is assumed to be piece-wise constant; this is the reason why in Eq.~(\ref{eq: nu,chi bounds}) $p_\chi$ is replaced by $\overline{p}^{(j)}_\chi$, which represents the average value of $p_\chi$ over the j$^{\textrm{th}}$ bin in $\chi$ with step-size $\delta \chi_j$, i.e.:
\begin{equation}
    \overline{p}^{(j)}_\chi=\frac{1}{\delta \chi_j}\int_{\chi_{j-1}}^{\chi_j}p_\chi\, \mathrm{d}\chi=\frac{P_\chi\left(\chi_{j-1}<\chi\leq\chi_j\right)}{\delta \chi_j}
\end{equation}
Eq.~(\ref{eq: nu,chi bounds}) is solved combing root finding and non-linear programming algorithms. Full detail of the theory can be found in~\cite{Giudici2023}. 

The ejected fragments are assumed to share the same initial position $\bm{r}_P$, which corresponds to the parent object position vector at fragmentation epoch. This implies that the orbits on which they are injected depend only on the variation of velocity vector caused by the fragmentation event. As a result, three out of the six Keplerian elements can freely vary in order for the new orbits to intersect the parent orbit in the fragmentation point~\cite{Frey2021}. The remaining three dependent elements are function of the position vector $\bm{r}_P$ and of the independent ones. In this paper, semi-major axis $a$, eccentricity $e$ and inclination $i$ are chosen as independent Keplerian elements $\bm{\alpha}$.

The NASA SBM provides information on the ejection velocity magnitude; however, it does not define how the velocity distribution is characterized in terms of direction of the impulse. As commonly done in the literature~\cite{Letizia2015c}, the ejection velocity vector is assumed to be isotropically distributed in direction. As explained in~\cite{Frey2021}, this approximation translates in the following PDFs for the in-plane $\gamma$ and out-of-plane $\varphi$ ejection velocity angles:
\begin{subequations}
\begin{align}
    &p_\gamma=\frac{1}{2\pi}\\
    &p_\varphi=\frac{\cos \varphi}{2}
\end{align}
\end{subequations}
A one-to-one correspondence, that maps each limit in ejection velocity, $\Delta v_j=10^{\nu_j}$, into a domain in Keplerian elements, $\mathcal{D}_{\bm{\alpha}}$, is needed. As motivated in~\cite{Giudici2023}, even though the most obvious approach would be to compute the maximum variation of the Keplerian elements caused by an ejection velocity vector with magnitude as high as $\Delta v_j$, this would cause the domain to be unnecessarily vast. Indeed, it must be understood that the fragments orbit resulting from the maximization would be achievable with a single combination of $\gamma$ and $\varphi$, and an ejection velocity magnitude as high as $\Delta v_j$. As a result, the minimum PDF value at the edge of the domain $\mathcal{D}_{\nu,\chi}$ would be scaled by a factor $p_\gamma p_\varphi=\cos\varphi/4\pi\leq1/4\pi$, which means that it would add a negligible contribution to the fragments distribution.
Instead, the domain in the subset of Keplerian elements $\bm{\alpha}$ is more conveniently computed as follows.
\begin{equation}
    \mathcal{D}_{\bm{\alpha}}=\left[\bm{\alpha}_P+f_s\,\overline{\Delta \bm{\alpha}^-},\;\bm{\alpha}_P+f_s\,\overline{\Delta \bm{\alpha}^+}\right]
\end{equation}
where $f_s$ is a safety factor, and the negative $\overline{\Delta \bm{\alpha}^-}$ and positive $\overline{\Delta \bm{\alpha}^+}$ variations of the Keplerian elements are obtained through: 
\begin{itemize}
    \item[-] Averaging of the orbital elements variation $\Delta\bm{\alpha}$ over the in-plane $\gamma$ and out-of-plane $\varphi$ ejection velocity angles.
    \item[-] Maximization of the variation $\Delta\bm{\alpha}$ over the ejection velocity magnitude $\Delta v_j$.
\end{itemize}
The procedure translates in the following set of equations:
\begin{subequations}
\begin{align}
    & \overline{\Delta\alpha^+}_i^{(j)}\left(\Delta v_j\right)=\max_{\Delta v\leq\Delta v_j}\int_{\gamma_{1_i}^+}^{\gamma_{2_i}^+}\int_{\varphi_{1_i}^+}^{\varphi_{2_i}^+}\Delta \alpha_i(\Delta v,\gamma,\varphi)\, p_\gamma p_\varphi\;\mathrm{d}\varphi\mathrm{d}\gamma\\
    & \overline{\Delta\alpha^-}_i^{(j)}\left(\Delta v_j\right)=\min_{\Delta v\leq\Delta v_j}\int_{\gamma_{1_i}^-}^{\gamma_{2_i}^-}\int_{\varphi_{1_i}^-}^{\varphi_{2_i}^-}\Delta \alpha_i(\Delta v,\gamma,\varphi)\, p_\gamma p_\varphi\;\mathrm{d}\varphi\mathrm{d}\gamma
\end{align}
\label{eq:dkep_av}%
\end{subequations}
where the plus and minus signs indicate the angles $\gamma$ and $\varphi$ that vary positively and negatively the Keplerian element $\alpha_i$. In other words, Eqs.~(\ref{eq:dkep_av}) define the maximum average variation (positive and negative) in the independent Keplerian elements $\bm{\alpha}$ for fragments ejected with a maximum ejection velocity magnitude as high as $\Delta v_j$. The non-linear relations that allow the computation of the variation of the independent Keplerian elements $\Delta \bm{\alpha}$, given $\Delta v$, $\gamma$, $\varphi$, can be found in~\cite{Giudici2023}.

\subsubsection{Domain discretization}
\label{Domain discretization}

The model presented in Section~\ref{Phase space domain of the breakup event} allows the accurate identification of the domain in Keplerian elements and area-to-mass ratio probabilistically reachable by the fragments ejected by an explosion or collision in orbit. This section explains the way the computed domain is partitioned into equally-sized rectangular-shaped volumes, through an automatized computation of the step-size in each variable, according to fragmentation properties and location.

The accuracy of a binning approach for the approximation of a general function over a multi-dimensional phase space is inherently linked to its rate of change over a bin volume. The gradient of the probability density function $p_{\nu,\chi}$ with respect to the independent Keplerian elements $\bm{\alpha}$, $\nabla_{\bm{\alpha}}p_{\nu,\chi}$, indicates how fast the density is changing locally. Hence, the gradient is used for setting autonomously the step-sizes $\bm{\delta \alpha}$ in the independent Keplerian elements. Since an equally-sized binning approach is implemented, the step-sizes are set on the basis of the average absolute value of the gradient of the density over the domain, according to the following relation~\cite{Giudici2023}:
\begin{equation}
    \bm{\delta \alpha}=\frac{1}{f_a}\max_{\chi,\nu\;\in\;\mathcal{D}_{\chi,\nu}}\left(p_{\nu,\chi}\right)\,\overline{\left|\nabla_{\bm{\alpha}}p_{\nu,\chi} \right|}^{-1}
\label{eq: step-sizes}
\end{equation}
where $f_a$ is a factor that can be tuned to control the model accuracy. Eq.~(\ref{eq: step-sizes}) constrains the average variation of the PDF $p_{\nu,\chi}$ across a bin to a fraction of the maximum density value over the considered domain $\mathcal{D}_{\bm{\alpha}}$.

\subsubsection{Density estimation through an adaptive Monte Carlo integration}
\label{Density estimation through an adaptive Monte Carlo integration}

After having partitioned the domain $\mathcal{D}_{\bm{\alpha}}$ according to the step-sizes of Eq.~(\ref{eq: step-sizes}), a density value is assigned to each bin so that the discretized density distribution best resembles the fragments distribution. The PDF $p_{\nu,\chi}$ must be translated into a probability density function in the adopted phase space, i.e., in the independent Keplerian elements $\bm{\alpha}$ and area-to-mass ratio $A/M$. The transformation is obtained through change of variables~\cite{Frey_thesis}, as follows.
\begin{equation}
    p_{\bm{x}}=\frac{p_{\nu,\chi}\left(\psi_{\bm{v}\rightarrow \bm{\alpha}}^{-1}(\bm{\alpha})\right)}{\left|\det \mathrm{J}_{\bm{v}\rightarrow \bm{\alpha}}\right|}
\label{eq:p_x}
\end{equation}
where the subscript $\bm{x}$ refers to the phase space variables $(a,e,i,A/M)$, $\psi_{\bm{v}\rightarrow \bm{\alpha}}$ indicates the transformation from the velocity vector $\bm{v}=\bm{v}_P+\bm{\Delta v}$ to the subset of Keplerian elements $\bm{\alpha}$, and $\mathrm{J}_{\bm{v}\rightarrow \bm{\alpha}}$ is the Jacobian of the transformation, which can be found in~\cite{Gonzalo2021}.

The density value assigned to each bin is the integral mean of $p_{\bm{x}}$ over the bin volume $V_{\bm{x}}$, i.e.:
\begin{equation}
    \overline{p_{\bm{x}}}=\frac{1}{V_{\bm{x}}}\iiiint_{V_{\bm{x}}}p_{\bm{x}}\;\textrm{d}\bm{x}
\label{eq:n_av}
\end{equation}
No closed form solution has been found for the integral of Eq.~(\ref{eq:n_av}); therefore, it is solved through Monte Carlo integration. Since the number of bins is typically of the order of $10^5-10^6$, such a procedure is considerably demanding from a computational cost point of view. To address this problem, the number of samples for the integration is varied according to the local value of the density gradient~\cite{Giudici2023}; as a result, the flatter the PDF $p_{\bm{x}}$ over a given bin, the fewer samples are extracted to compute the mean value, dramatically increasing the computational efficiency of the code.

\subsection{Continuity-equation based fragments density propagation}
\label{Continuity-equation based fragments density propagation}

Once the computation of the density distribution at fragmentation epoch, presented in Section~\ref{Estimation of the initial density distribution}, is concluded, each bin of the domain has an associated density value $p_{\bm{x}}$. Following the same approach pursued by other authors~\cite{McInnes1993,Letizia2015a,Frey2019}, the fragments density is propagated by applying the Method Of Characteristics (MOC)~\cite{Jhon1981} to the continuity equation. For a first-order Partial Differential Equation (PDE), like the continuity equation, the MOC discovers curves in the phase space, the characteristic curves, along which a PDE transforms into a system of Ordinary Differential Equations (ODE). The system of ODE can be integrated numerically, providing the solution along the characteristic curves. As a result, to know the evolution of the fragments density in the whole phase space, a bunch of characteristics has to be propagated and, eventually, interpolated.

\subsubsection{Samples propagation through the method of characteristics}
\label{Characteristics propagation through the method of characteristics}

The proposed model aims to represent with the same accuracy the high- and low-density regions. For this reason, the characteristics to be propagated are randomly sampled from the phase space, extracting one sample from each bin of the domain $\mathcal{D}_{\bm{\alpha}, A/M}$. For each extracted sample, the system of ODE to be integrated reads as:
\begin{equation}
\begin{cases}
    \frac{\mathrm{d}\bm{y}}{\mathrm{d}t}=\bm{F}\\
    \frac{\mathrm{d}n_{\bm{x}}}{\mathrm{d}t}=-n_{\bm{x}}\nabla_{\bm{y}}\cdot\bm{F}
\end{cases}
\end{equation}
where $\bm{y}$ are the phase space variables, $t$ is time, $\bm{F}$ is the force model, and $n_{\bm{x}}$ the phase space density, which is computed from the PDF $p_{\bm{x}}$ of Eq.~(\ref{eq:p_x}) according to the following relation:
\begin{equation}
    n_{\bm{x}}=Np_{\bm{x}}
\end{equation}
with $N$ total number of fragments generated by the fragmentation event. It is worth noticing that the variables $\bm{y}$ differ from the subset $\bm{x}$ upon which the density depends. They represent the least number of variables, whose dynamical evolution is either of interest or is needed for evaluating the force model $\bm{F}$. The characteristics are propagated semi-analytically, as a consequence of the averaging of the dynamics equations over the fast angular variable, through the software PlanODyn~\cite{Colombo2016}. It provides the trace of the Jacobian of the averaged dynamics with respect to the mean elements for atmospheric drag, J$_2$ perturbation, solar radiation pressure and luni-solar perturbations. Note that, to improve the computational efficiency, the software allows for a second averaging procedure, over the disturber orbital period. Full detail of the implemented dynamical model can be found in~\cite{Giudici2022}. Note that, because of the averaged dynamics considered, the set of variables $\bm{y}$ includes the five slow-varying orbital elements (i.e., semi-major axis $a$, eccentricity $e$, inclination $i$, right ascension of the ascending node $\Omega$, and argument of periapsis $\omega$) and area-to-mass ratio $A/M$. For a generic characteristic, the phase space variables $\bm{y}$ are obtained from the independent variables $\bm{x}$, by imposing intersection with the parent object orbit in the fragmentation point $\bm{r}_P$~\cite{Giudici2023}.

\subsubsection{Density interpolation through binning}
\label{Density interpolation through binning}

If propagated to the same epoch, the characteristics form a scattered point cloud in the phase space. Therefore, they must be interpolated to retrieve the density distribution in the whole domain. The interpolation is carried out through binning in an (up to) 6D phase space of slow-varying Keplerian elements and area-to-mass ratio, at specified time epochs, summing up the contribution of all the characteristics that share the same volume~\cite{Giudici2023}. As a result, the computed density distribution is bin-wise constant in space and evolves discretely in time. Note that the collision risk assessment model, which will be discussed in Section~\ref{Debris density-based collision risk assessment}, takes as input a phase space density defined in the whole set of Keplerian elements. Called $\bm{\beta}$ the subset $(\Omega,\omega,M)$, the density $n_{\bm{\alpha},\bm{\beta}}$ in the j$^\mathrm{th}$ bin is obtained from the density $n_{\bm{x}}$, according to the following relation:
\begin{equation}
    n_{\bm{\alpha},\bm{\beta}}^{(j)}=\frac{1}{2\pi}\frac{n^{(j)}_{\bm{x}}}{\delta\Omega\,\delta\omega}\,\delta A/M
\end{equation}
with $\delta\Omega$, $\delta\omega$ and $\delta A/M$ step-sizes in right ascension of the ascending node, argument of periapsis and area-to-mass ratio. The division by $2\pi$ is due to the fact that the mean anomaly $M$ is not included as interpolation variables, i.e., the fragments are always assumed to be randomized over $M$. 

As motivated in~\cite{Themis}, the use of a binning approach in such a multidimensional phase space is a tremendous challenge from a memory usage point of view. However, in most of the fragmentation scenarios, the cloud evolves occupying just a portion of the square domain, thus leaving part of the bins empty. In~\cite{Themis} it was demonstrated how this can be exploited to reduce the memory consumption, by applying well-known techniques for the storage of sparse arrays. The density distributions are stored in the extended Karnaugh map representation-Compressed Row Storage (CRS) format~\cite{Lin2002,Lin2005}, which is the extension of the CRS technique to a multidimensional array.

\section{Debris density-based collision risk assessment}
\label{Debris density-based collision risk assessment}

This section is devoted to explaining the physical and mathematical model developed for the estimation of the probability of collision between a potential fragments cloud, which is described through a phase space density function, and a selected target object. 

Firstly, a simplified approach, where the fragments are described through a spatial density function dependent on orbital radius only, is introduced. This method, which was firstly presented in~\cite{Letizia2015b}, lacking of information on the relative velocity between the target and the fragments, assumes the latter moving on circular orbits. On the contrary, the cloud propagation model explained in Section~\ref{Fragmentation modeling and propagation} allows to estimate the evolution in time of a 6D phase space density function in slow-varying Keplerian elements and area-to-mass ratio, which ensures the accurate evaluation of the relative velocity of collision. Nevertheless, the simplified collision probability model is presented to stress the similarity between the two methods. Furthermore, we take the chance to correct an error in the derivation of the method proposed in~\cite{Letizia2015b}, as well as to introduce the non-linear dependency of the spatial density function on latitude in the computation of the impact rate.

The model is then extended to consider the 6D phase space density function. The discrete nature of the density distribution in space and time, resulting from the modeling of the debris cloud through the continuum formulation proposed in Section~\ref{Fragmentation modeling and propagation}, is exploited to derive a semi-analytical approach to compute the integral of the fragments flux over the target area, as a discrete function of time. The knowledge on the impact rate is eventually translated into the cumulative probability of collision with the target, according to a Poisson distribution~\cite{McKnight1990}:
\begin{equation}
    P_c(t)=1-\exp{\left(-\eta(t)\right)}
\label{eq:Pc}
\end{equation}
with $\eta$ cumulative number of collision over time $t$. Eq.~(\ref{eq:Pc}) comes from the common analogy with the gas kinetic theory.

\subsection{Impact rate from a 1D spatial density distribution in orbital radius}
\label{Impact rate from a 1D spatial density distribution in orbital radius}

\begin{figure}[!ht]
\centering
\includegraphics[width=0.7\textwidth]{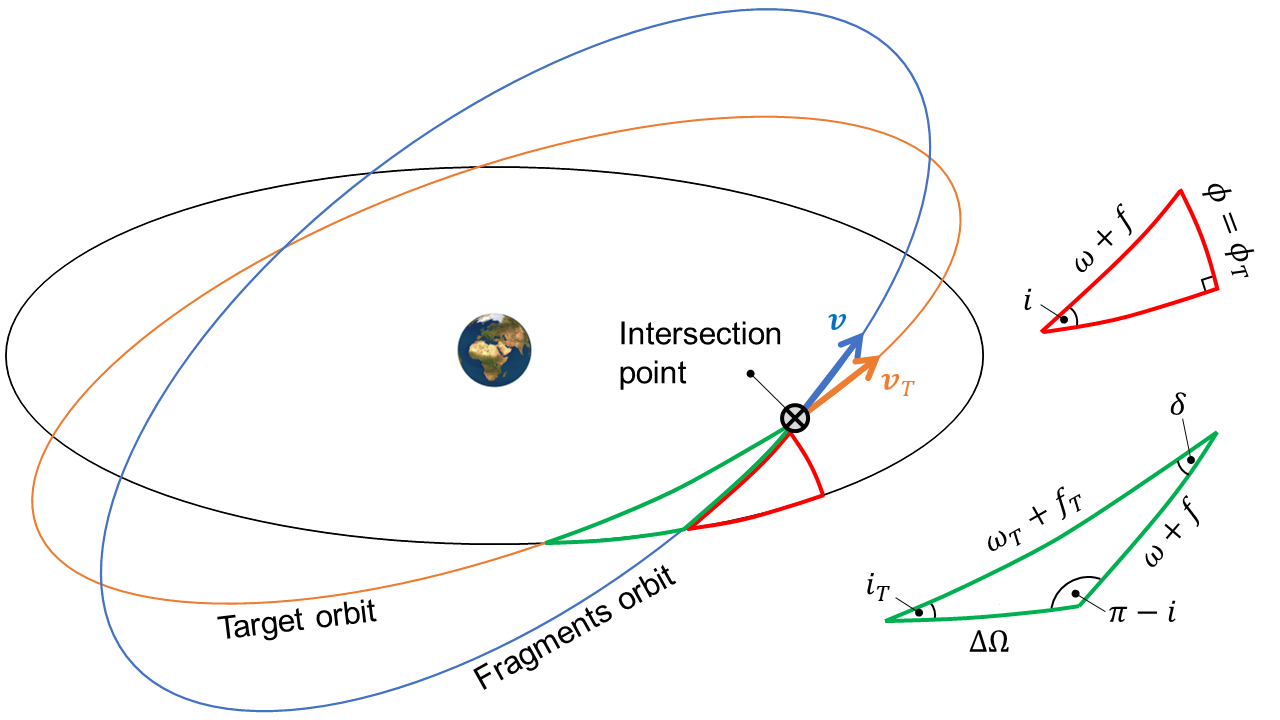}
\caption{Spherical triangle of the intersection between the target and fragments orbit.}
\label{fig: intersection}
\end{figure}

The average impact rate $\overline{\dot{\eta}}$ between a fragments cloud, described through a 1D spatial density function $n_{\bm{r}}(r)$, and a target moving on a Keplerian orbit can be approximated as~\cite{Letizia2015b}:
\begin{equation}
    \overline{\dot{\eta}}=A_c n_{\bm{r}}(r)\overline{v}_\mathrm{rel}
\label{eq: eta_dot_av_c}
\end{equation}
with $A_c$ cross-sectional area of the target, and $\overline{v}_\mathrm{rel}$ average impact velocity. Letizia et al.~\cite{Letizia2015b} pointed out that, for a fragmentation in LEO, the debris cloud forms a band around the Earth in a relatively short period of time. As a result, if the objective is to evaluate the long-term effect of such a cloud, it can be considered randomized over right ascension of the ascending node $\Omega$ and argument of periapsis $\omega$. The relative velocity between two objects having a conjunction can be written as function of the velocity moduli and the angle $\delta$ between the velocity vectors, as follows.
\begin{equation}
    v_\mathrm{rel}=\sqrt{v_T^2+v^2-2v_Tv\cos{\delta}}
\end{equation}
where $v_T$ and $v$ refer to the velocity moduli of the target and the fragments, respectively. The angle $\delta$ depends on the target, $i_T$, and fragments, $i$, inclination, and the difference in right ascension of the ascending node $\Delta \Omega$ between the two orbits, according to the following relation:
\begin{equation}
    \cos\delta=\sin{i_T}\sin{i}\cos{\Delta\Omega}+\cos{i_T}\cos{i}
\label{eq:cos_delta}
\end{equation}
Eq.~(\ref{eq:cos_delta}) derives from the application of the cosine rule to the green spherical triangle depicted in Figure~\ref{fig: intersection}.

\begin{figure}[!ht]
     \centering
     \includegraphics[width=0.5\textwidth]{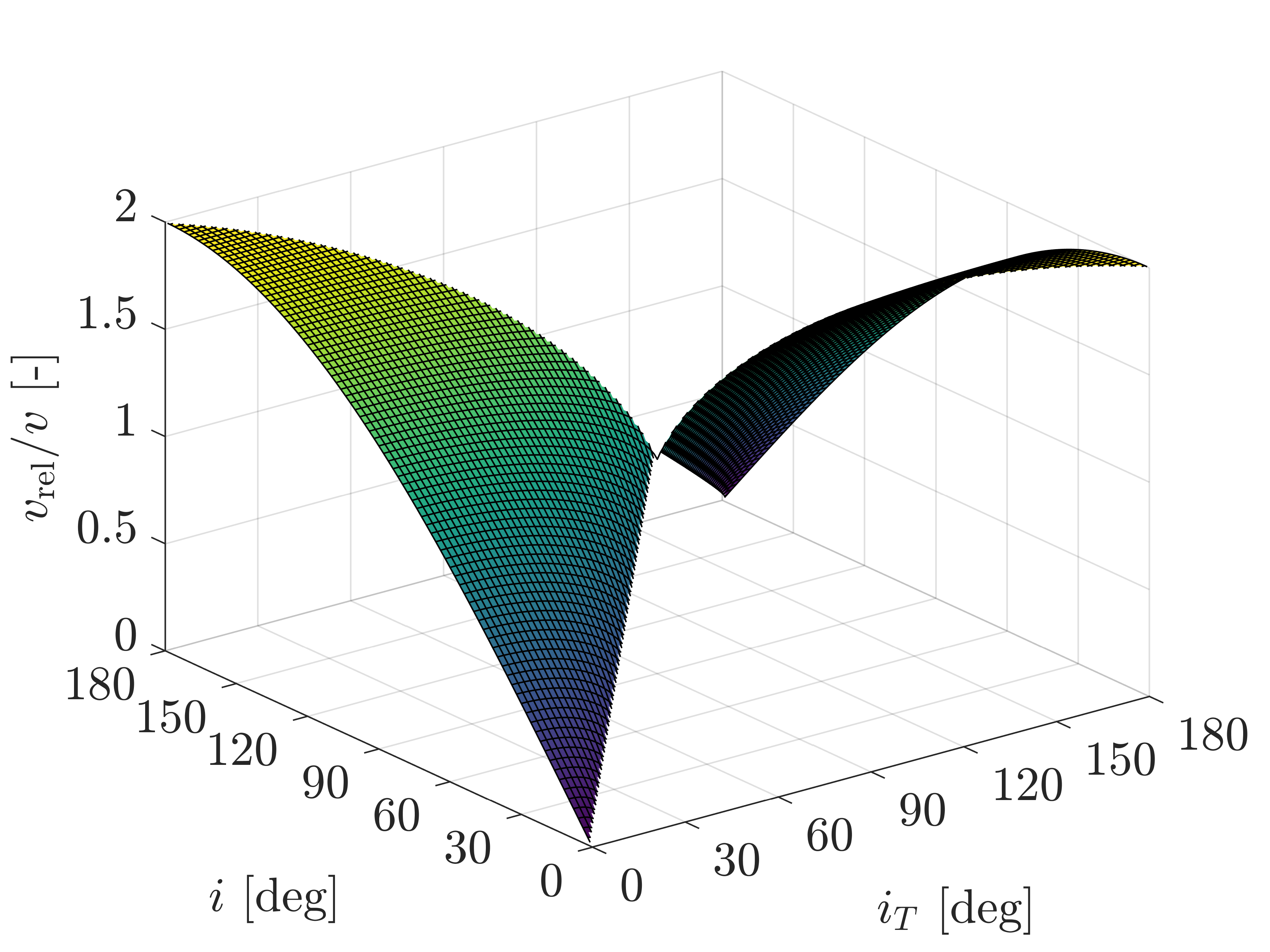}
     \caption{Impact velocity as function of fragments and target inclination.}
     \label{fig:dv_i_iT}
\end{figure}

In~\cite{Letizia2015b}, the randomization over right ascension of the ascending node $\Omega$ was misinterpreted as if each $\Delta \Omega$ is equally probable. As a result, the average impact velocity was computed as~\cite{Letizia2015b}:
\begin{equation}
    \overline{v}_\mathrm{rel}=\frac{1}{2\pi}\int_0^{2\pi}v_\mathrm{rel}\left(\Delta\Omega\right)\,\mathrm{d}\Delta\Omega
\label{eq:vrel_Letizia}
\end{equation}
However, it must be understood that the average impact velocity $\overline{v}_\mathrm{rel}$ corresponds to the mean relative velocity between the fragments and the target, as the latter moves on a fixed Keplerian orbit. Hence, the averaging procedure must be carried out over the mean anomaly $M_T$ of the target, to consider every possible conjunction geometry. For each target mean anomaly $M_T$, there exist two possible intersecting circular fragments orbits, shifted in right ascension of the ascending node with respect to the target orbit of $\Delta \Omega$, satisfying the cotangent law applied to the  green spherical triangle of Figure~\ref{fig: intersection}, i.e.:
\begin{equation}
    \cos{\Delta\Omega}\cos{i_T}=\cot{u_T}\sin{\Delta\Omega}-\cot{i}\sin{i_T}
\label{eq:Delta_Om}
\end{equation}
where $u_T=\omega_T+f_T$ is the argument of latitude of the target. Thus, the average relative velocity between the target and the fragments can be computed as:
\begin{equation}
    \overline{v}_\mathrm{rel}=\frac{1}{2\pi}\int_0^{2\pi}\frac{1}{2}\sum_{j=1}^2v_\mathrm{rel}\Bigl(\Delta\Omega_j(f_T)\Bigl)\,\frac{\mathrm{d}M_T}{\mathrm{d}f_T}\,\mathrm{d}f_T
\label{eq:Delta_v_av}
\end{equation}
with:
\begin{equation}
    \frac{\mathrm{d}M_T}{\mathrm{d}f_T}=\frac{\left(1-e_T^2\right)^{3/2}}{\left(1+e_T\cos{f_T}\right)^2}
\end{equation}
Note that $\Delta \Omega_j$ of Eq.~(\ref{eq:Delta_v_av}) indicates one of the two possible solutions of Eq.~(\ref{eq:Delta_Om}). Figure~\ref{fig:dv_i_iT} shows the average velocity of impact $\overline{v}_\mathrm{rel}$, normalized by the velocity on a circular orbit at the altitude of the target, as function of fragments, $i$, and target, $i_T$, inclination, assuming they move on circular orbits.

As it can be observed, the surface of solutions is symmetric about the point $(i,i_T)=(90,90)\;\mathrm{deg}$. As expected, the impact velocity is larger when the fragments move on a prograde orbit with the target on a retrograde one, or viceversa. In this case, the normalized average impact velocity is higher than the unity, approaching two when fragments and target lie on the equatorial plane, but are characterized by opposite direction of rotation. On the contrary, when the objects share the same direction of rotation, the normalized relative velocity is always smaller than one.

Finally, it is worth stressing that Eq.~(\ref{eq: eta_dot_av_c}) applies only if the target is also moving on a circular orbit. Indeed, the spatial density $n_{\bm{r}}$ is assumed to be constant for each target position along its orbit. If, however, the target orbit is elliptical, then the orbital radius varies as function of the target mean anomaly $M_T$. As a result, the spatial density must be included in the integration over $M_T$, as follows.
\begin{equation}
    \overline{\dot{\eta}}=\frac{A_c}{2\pi}\int_0^{2\pi}n_{\bm{r}}\Bigl(r_T(f_T)\Bigr)\,v_\mathrm{rel}^*\Bigl(\Delta \Omega(f_T)\Bigr)\,\frac{\mathrm{d}M_T}{\mathrm{d}f_T}\,\mathrm{d}f_T
\label{eq:eta_dot_c}
\end{equation}
where $v_\mathrm{rel}^*$ indicates the mean between the two possible impact velocities, given the target mean anomaly $M_T$.

\begin{figure*}[!ht]
     \centering
     \begin{subfigure}[b]{0.33\textwidth}
         \centering
         \includegraphics[width=\textwidth]{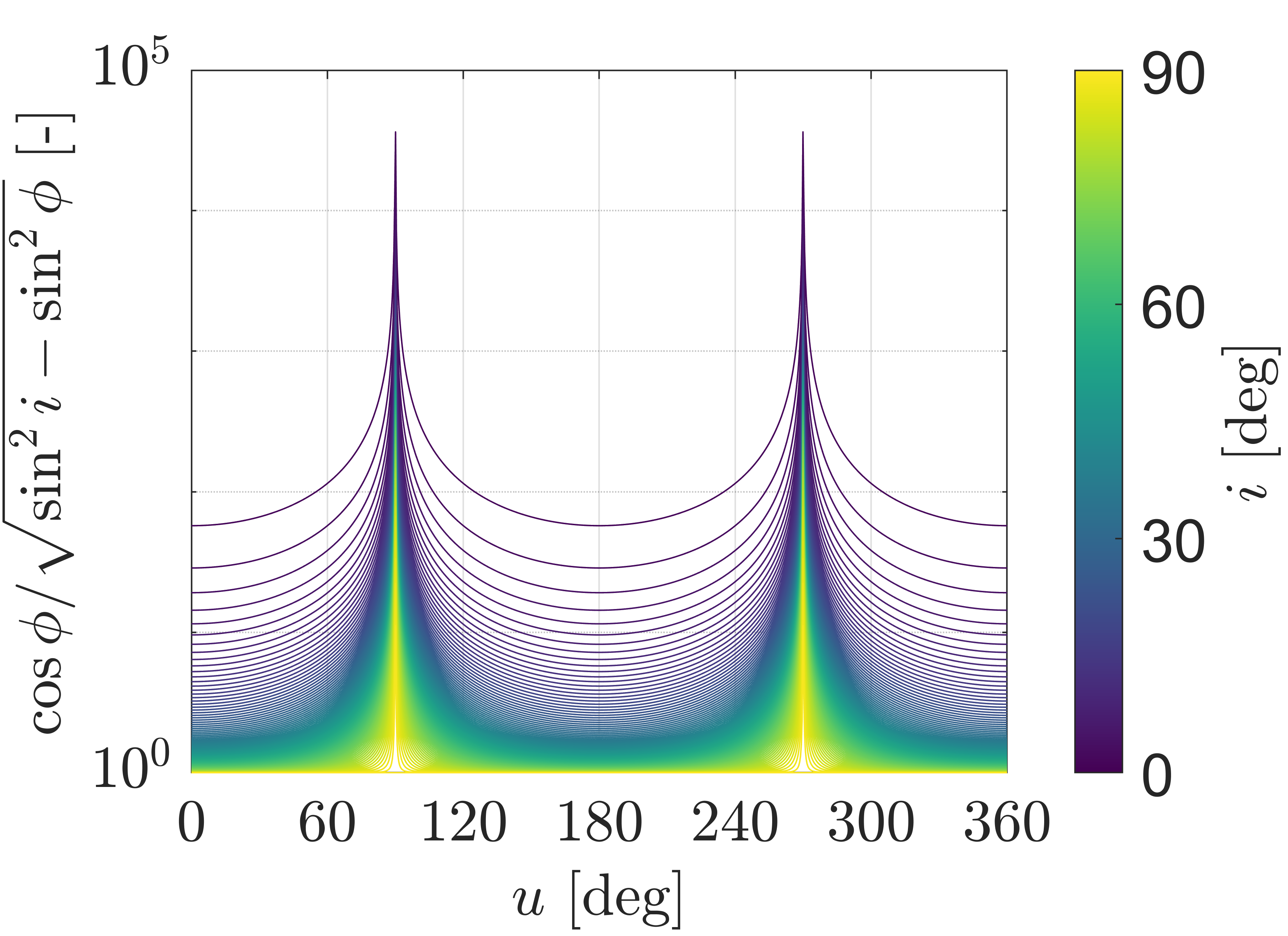}
         \caption{True anomaly-latitude relation}
         \label{fig:beta1}
     \end{subfigure}
     \begin{subfigure}[b]{0.33\textwidth}
         \centering
         \includegraphics[width=\textwidth]{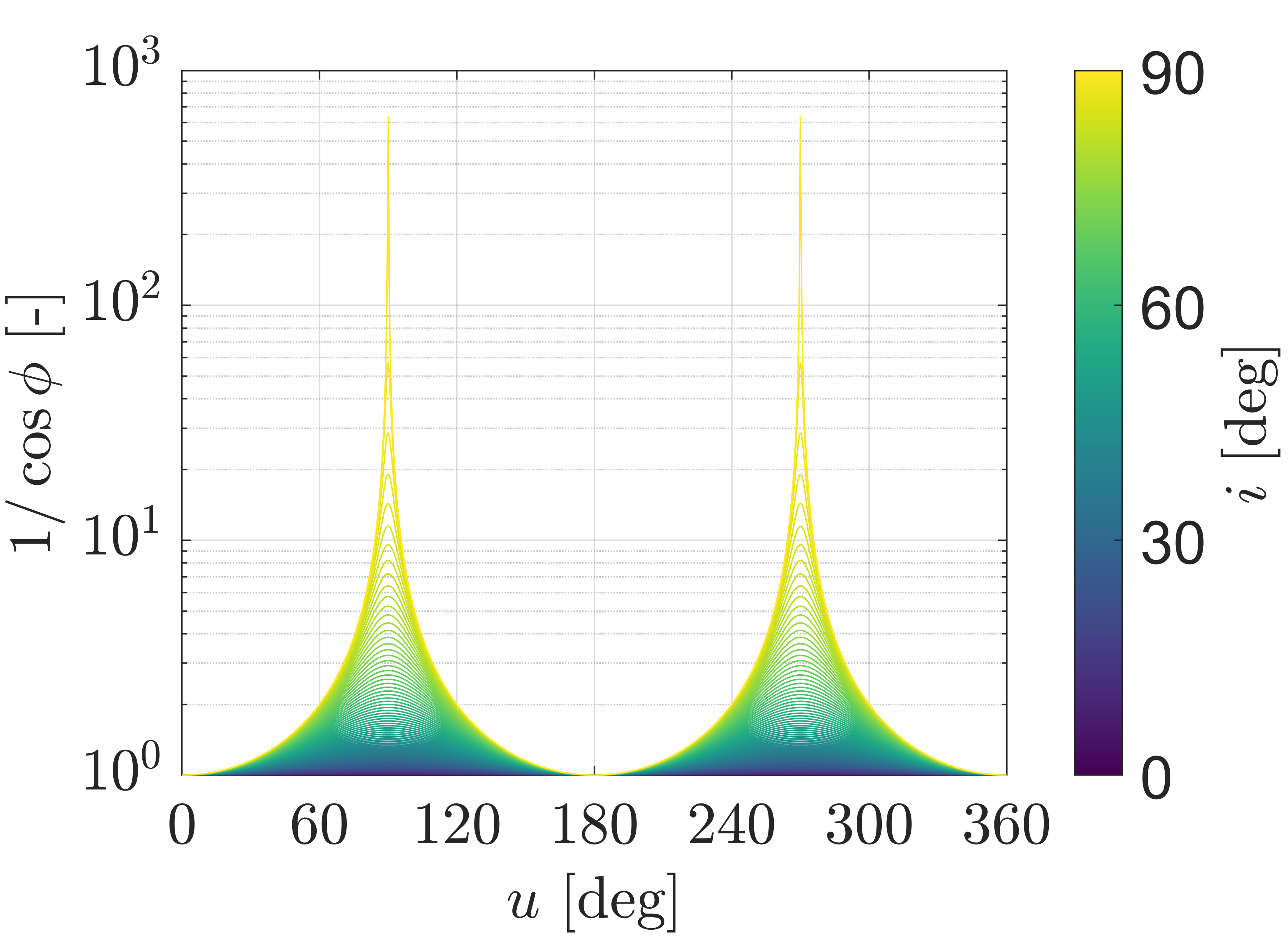}
         \caption{Circular band area-latitude relation}
         \label{fig:beta2}
     \end{subfigure}
     \begin{subfigure}[b]{0.33\textwidth}
         \centering
         \includegraphics[width=\textwidth]{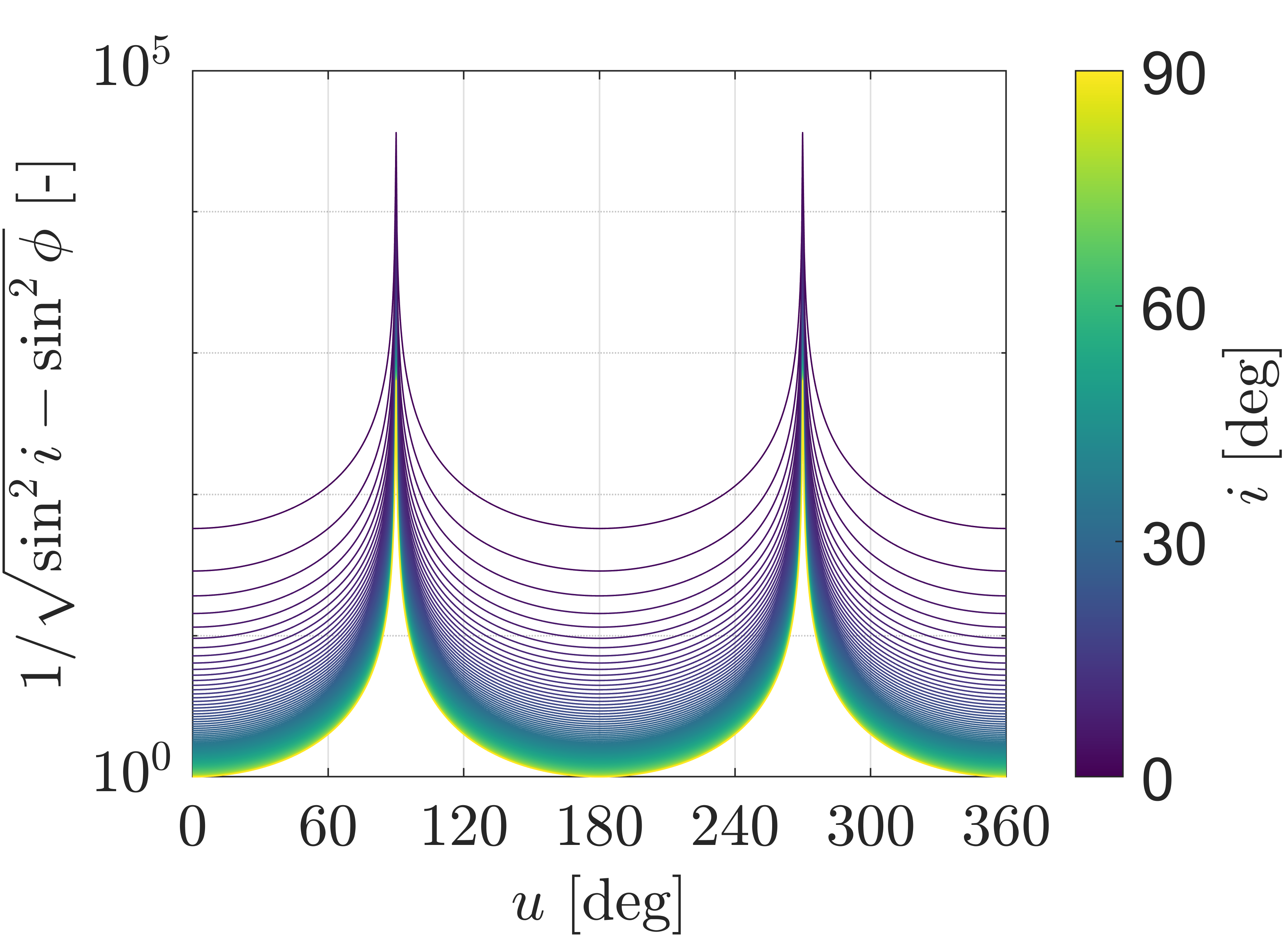}
         \caption{Combined effect}
         \label{fig:beta}
     \end{subfigure}
     \caption{Dependency of the spatial density function on latitude.}
     \label{fig:beta1,beta2,beta}
\end{figure*}

A further limitation of the model proposed in~\cite{Letizia2015b} is that the spatial density function is assumed to be randomized over both longitude $\lambda$ and latitude $\phi$. However, as motivated in~\cite{Kessler1978}, even if the fragments are assumed to share the same orbital inclination, and to be uniformly distributed in a band around the Earth, the spatial density function has a non-linear dependency on latitude $\phi$. Indeed, the spatial density in a band with infinitesimal thickness in radial distance $\mathrm{d}r$ and angular amplitude $\mathrm{d}\phi$, as function of orbital radius $r$ and latitude $\phi$, can be computed as:
\begin{equation}
    n_{\bm{r}}(r,\phi)=\frac{\mathrm{d}N(r,\phi)}{\mathrm{d}A(r,\phi)\mathrm{d}r}
\label{eq:n_r,phi}
\end{equation}
where $N$ is the number of fragments and $A$ is the surface area of the band around the Earth. The infinitesimal number of fragments can be computed according to the following equation:
\begin{equation}
    \mathrm{d}N(r,\phi)=N(r)\frac{\mathrm{d}M}{\pi}=\frac{N(r)}{\pi}\frac{\mathrm{d}M}{\mathrm{d}f}\frac{\mathrm{d}f}{\mathrm{d}\phi}\mathrm{d}\phi
\label{eq:dN}
\end{equation}
Assuming the fragments orbit as circular for simplicity, Eq.~(\ref{eq:dN}) simplifies as follows.
\begin{equation}
    dN(r,\phi)=\frac{N(r)}{\pi}\frac{\mathrm{d}f}{\mathrm{d}\phi}\mathrm{d}\phi=\frac{N(r)}{\pi}\frac{\cos\phi}{\sqrt{\sin^2i-\sin^2\phi}}\,\mathrm{d}\phi
\label{eq:dN1}
\end{equation}
Instead, the infinitesimal area $\mathrm{d}A$ takes the following form:
\begin{equation}
    dA(r,\phi)=2\pi r^2\cos\phi\,\mathrm{d}\phi
\label{eq:dA}
\end{equation}
Therefore, the 2D spatial density function of Eq.~(\ref{eq:n_r,phi}) modifies as follows.
\begin{equation}
    n_{\bm{r}}(r,\phi)=\frac{N(r)}{2\pi r^2}\frac{1}{\pi\sqrt{\sin^2i-\sin^2\phi}}=n_{\bm{r}}(r)\frac{2}{\pi\sqrt{\sin^2i-\sin^2\phi}}=n_{\bm{r}}(r)\beta(\phi)
\label{eq:n_r,phi1}
\end{equation}
which coincides with the expression found by Kessler~\cite{Kessler1978}. As it can be noted, the non-linear dependency of the spatial density function on latitude is introduced by both the true anomaly-latitude relation of Eq.~(\ref{eq:dN1}) and the dependency of the area of the circular band on latitude, expressed by Eq.~(\ref{eq:dA}). In Figure~\ref{fig:beta1,beta2,beta} the latitude-dependent parts of Eq.~(\ref{eq:dN1}), Eq.~(\ref{eq:dA}) and Eq.~(\ref{eq:n_r,phi1}) are shown as function of the argument of latitude $u$, for different values of inclination $i$. Note that latitude $\phi$ and argument of latitude $u$ are linked by the sine rule applied to the red spherical triangle of Figure~\ref{fig: intersection}, which provides the following expression:
\begin{equation}
    \sin\phi=\sin u\sin i
\label{eq:relation_phi_i_u}
\end{equation}

Replacing the 1D spatial density function in Eq.~(\ref{eq: eta_dot_av_c}) with the 2D density function of Eq.~(\ref{eq:n_r,phi1}), evaluated at the target orbital radius $r_T$ and latitude $\phi_T$, allows a more accurate estimation of the impact rate, which accounts for the fragments distribution over latitude. The resulting expression is here reported for completeness:
\begin{equation}
    \overline{\dot{\eta}}=\frac{A_c}{2\pi}\int_0^{2\pi}n_{\bm{r}}\Bigl(r_T(f_T)\Bigr)\,\beta\Bigl(\phi_T(f_T)\Bigr)\,v_\mathrm{rel}^*\Bigl(\Delta \Omega(f_T)\Bigr)\,\frac{\mathrm{d}M_T}{\mathrm{d}f_T}\,\mathrm{d}f_T
\label{eq:eta_dot_c1}
\end{equation}

\subsection{Impact rate from a 6D density distribution in Keplerian elements}
\label{Impact rate from a 6D density distribution in Keplerian elements}

In general, the average impact rate $\overline{\dot{\eta}}$ with a target moving on an orbit with slow-varying Keplerian elements $\bm{\gamma}_T$ can be estimated by integrating the impact rate over target mean anomaly $M_T$, as follows.
\begin{equation}
    \overline{\dot{\eta}}=\frac{1}{2\pi}\int_0^{2\pi}\dot{\eta}\left(\bm{\gamma}_T,M_T\right)\,\mathrm{d}M_T
\label{eq:eta_dot_av}
\end{equation}
If the fragments distribution is represented through a 6D density function in Cartesian coordinates $(\bm{r},\bm{v})$, $n_{\bm{r},\bm{v}}$, the impact rate associated to a fixed value of mean anomaly $M_T$ can be approximated as follows~\cite{Frey2021}.
\begin{equation}
    \dot{\eta}\left(\bm{\gamma}_T,M_T\right)=A_c\iiint_{	\mathbb{R}^3}n_{\bm{r},\bm{v}}(\bm{r}_T,\bm{v})\,v_\mathrm{rel}\,\mathrm{d}\bm{v}
\label{eq:eta_dot}
\end{equation}
where $\bm{r}$ and $\bm{v}$ indicate the position and velocity vectors of the fragments. The main assumption behind Eq.~(\ref{eq:eta_dot}) is that the fragments cross section $A_c$ are assumed to be negligible if compared to the target one.
As demonstrated in~\cite{Frey2021}, the 6D density function in Cartesian coordinates, evaluated at the target position $\bm{r}_T$, can be related to the phase space density in Keplerian elements, $n_{\bm{\alpha},\bm{\beta}}$, according to the following equation:
\begin{equation}
    n_{\bm{r},\bm{v}}(\bm{r}_T,\bm{v})\,\mathrm{d}\bm{v}=\sum_{k=1}^4\frac{n_{\bm{\alpha}, \bm{\beta}}\left(\bm{\alpha},\bm{\beta}^{(k)}\right)}{\left|\det{\mathrm{J}_{\bm{r}\rightarrow\bm{\beta}}^{(k)}}\right|}\,\mathrm{d}\bm{\alpha}
\label{eq:n_r,v to n_a,b}
\end{equation}
where $\mathrm{J}_{\bm{r}\rightarrow\bm{\beta}}$ is the Jacobian of the transformation from position vector $\bm{r}$ to the subset of the Keplerian elements $(\Omega,\omega,M)$, i.e.:
\begin{equation}
    \mathrm{J}_{\bm{r}\rightarrow\bm{\beta}}=
    \begin{bmatrix}
    \frac{\partial r_x}{\partial \Omega} & \frac{\partial r_x}{\partial \omega} & \frac{\partial r_x}{\partial M} \\
    \frac{\partial r_y}{\partial \Omega} & \frac{\partial r_y}{\partial \omega} & \frac{\partial r_y}{\partial M} \\
    \frac{\partial r_z}{\partial \Omega} & \frac{\partial r_z}{\partial \omega} & \frac{\partial r_z}{\partial M}
    \end{bmatrix}
\end{equation}
whose elements expression can be found in~\cite{Gonzalo2021}. The absolute value of the determinant of $\mathrm{J}_{\bm{r\rightarrow\beta}}$ takes the form:
\begin{equation}
    \left|\det{\mathrm{J}_{\bm{r}\rightarrow\bm{\beta}}^{(k)}}\right|=\frac{a^3e(1-e^2)^{3/2}\Bigl|\cos\left(\omega^{(k)}+f^{(k)}\right)\sin f^{(k)}\Bigr|\sin i}{(1+e\cos f^{(k)})^2}
\label{eq:jacobian}
\end{equation}
The summation in Eq.~(\ref{eq:n_r,v to n_a,b}) represents the four possible intersections between the target and fragments orbit, once the target position vector $\bm{r}_T$ is fixed and the subset of the Keplerian elements $\bm{\alpha}$ is given. The intersecting orbits are provided by the four combinations $(\bm{\alpha},\bm{\beta}^{(k)})$. In particular, the two solutions of the orbital radius equation:
\begin{equation}
    r_T=\frac{a(1-e^2)}{1+e\cos f}
\label{eq:orbital_radius}
\end{equation}
provide the two angular positions $f_1$ and $f_2$. The cotangent rule applied to the green spherical triangle of Figure~\ref{fig: intersection}, which is given in Eq.~(\ref{eq:Delta_Om}), allows the computation of the fragments orbital plane, through the solutions in right ascension of the ascending node $\Omega_1$ and $\Omega_2$. Finally, the application of the sine rule to the same spherical triangle, which states:
\begin{equation}
    \sin u_T\sin i_T=\sin u\sin i
\label{eq:orbit_orientation}
\end{equation}
fixes the fragments orbit orientation, provided by the solutions in argument of latitude $u_1$ and $u_2$. The resulting solutions $\bm{\beta}^{(k)}$ have the following characteristics:
\begin{equation}
\begin{split}
    &\bm{\beta}^{(1)}=(\Omega_1,\omega_1,M_1)\\
    &\bm{\beta}^{(2)}=(\Omega_1,\omega_2,M_2)\\
    &\bm{\beta}^{(3)}=(\Omega_2,\omega_3,M_1)\\
    &\bm{\beta}^{(4)}=(\Omega_2,\omega_4,M_2)
\end{split}
\label{eq:beta^k}
\end{equation}
and satisfy:
\begin{equation}
\begin{split}
    \omega_1+f_1(M_1)&=\omega_2+f_2(M_2)=u_1\\
    \omega_3+f_1(M_1)&=\omega_4+f_2(M_2)=u_2\\
    M_1&=-M_2
\end{split}
\end{equation}

Plugging Eq.~(\ref{eq:orbital_radius}) and Eq.~(\ref{eq:orbit_orientation}) into Eq.~(\ref{eq:jacobian}), the determinant of the Jacobian modifies as:
\begin{equation}
    \left|\det{\mathrm{J}_{\bm{r}\rightarrow\bm{\beta}}^{(k)}}\right|=r_T^2\,a\frac{e}{\sqrt{1-e^2}}\sqrt{1-g(a,e,r_T)^2}\sqrt{1-h(i,i_T,\omega_T,f_T)^2}\sin i=\frac{1}{\Psi(\bm{\alpha})}
\label{eq:detJ}
\end{equation}
where the functions $g$ and $h$ are the cosine of the true anomaly $f$ and the sine of the argument of latitude $u=\omega+f$, derived from Eq.~(\ref{eq:orbital_radius}) and Eq.~(\ref{eq:orbit_orientation}), and the function $\Psi$, which stands for the absolute value of the inverse of the determinant of $\mathrm{J}_{\bm{r}\rightarrow\bm{\beta}}$, is introduced to simplify the notation. As it can be noticed, the function $\Psi$ is independent of the value of the dependent Keplerian elements $\Omega, \omega, M$; thus, it can be taken outside of the summation of Eq.~(\ref{eq:n_r,v to n_a,b}).

Plugging Eq.~(\ref{eq:n_r,v to n_a,b}) into Eq.~(\ref{eq:eta_dot}) allows retrieving the impact rate directly from the the phase space density in Keplerian elements, $n_{\bm{\alpha}, \bm{\beta}}$. In addition, the use of a binning approach for the computation of the phase space density distribution, from the propagated bulk of characteristics, implies that the fragments density varies discretely over the phase space. As a result, the integration over the domain in the independent Keplerian elements $\bm{\alpha}$ can be split into a summation of integrals over the bins volume, over which the fragments density is constant. Therefore, the impact rate is approximated as follows.
\begin{equation}
    \dot{\eta}(\bm{\gamma}_T,M_T)=A_c\sum_{j=1}^{N_b^*}\left( \iiint_{V^{(j)}_{\bm{\alpha}}} \Psi(\bm{\alpha})\sum_{k=1}^4n_{\bm{\alpha},\bm{\beta}}^{(jk)}v_\mathrm{rel}^{(k)}(\bm{\alpha})\,\mathrm{d}\bm{\alpha}\right)
\label{eq:eta_dot_kep1}
\end{equation}
where $n_{\bm{\alpha},\bm{\beta}}^{(jk)}$ is the phase space density in the bin with center coordinates $\left[\bm{\alpha}^{(j)},\bm{\beta}^{(k)}(\bm{\alpha}^{(j)})\right]$, $v_\mathrm{rel}^{(k)}(\bm{\alpha})$ is the relative velocity between fragments and target for the k$^{\mathrm{th}}$ solution of intersection, given $\bm{\alpha}$, and $N_b^*$ is the number of bins, whose subset of Keplerian elements $(a,e,i)$ satisfies Eq.~(\ref{eq:orbital_radius}), Eq.~(\ref{eq:Delta_Om}) and Eq.~(\ref{eq:orbit_orientation}) for some combinations of $(\Omega,\omega,M)$. Therefore, by adopting a binning approach, the estimation of the impact rate reduces to the computation of the integral of Eq.~(\ref{eq:eta_dot_kep1}), at bin level, and to the summation of the contribution of each bin ensuring intersection with the target, and having a non-null density value. 

The integral of Eq.~(\ref{eq:eta_dot_kep1}) cannot be solved in closed form. Nevertheless, if the discretization in $(a,e,i)$ is sufficiently refined, the impact rate can be reduced to:
\begin{equation}
    \dot{\eta}(\bm{\gamma}_T,M_T)=A_c\sum_{j=1}^{N_b^*}\left(\sum_{k=1}^4\left(n_{\bm{\alpha},\bm{\beta}}^{(jk)}v_\mathrm{rel}^{(jk)}\right) \iiint_{V^{(j)}_{\bm{\alpha}}} \Psi(\bm{\alpha})\,\mathrm{d}\bm{\alpha}\right)
\label{eq:assumption}
\end{equation}
with:
\begin{equation}
    v_\mathrm{rel}^{(jk)}=\left|\left|\bm{v}\left(\bm{\alpha}^{(j)},\bm{\beta}^{(k)}(\bm{\alpha}^{(j)})\right)-\bm{v}_T\right|\right|
\end{equation}
which consists in approximating the relative velocity between the target and the fragments $v_\mathrm{rel}^{(k)}(\bm{\alpha})$ as the relative velocity in correspondence of the bins center $v_\mathrm{rel}^{(jk)}$, over each bin volume. Note that, the assumption of Eq.~(\ref{eq:assumption}) means that the spatial density function in correspondence of the target, $n_{\bm{r}}(\bm{r}_T)$, is computed without any approximation, net of the mathematical modelling of the phase space density $n_{\bm{\alpha},\bm{\beta}}$, while the relative velocity is measured discretely, in correspondence of the bins center. Indeed, the spatial density $n_{\bm{r}}(\bm{r}_T)$ can be obtained from Eq.~(\ref{eq:n_r,v to n_a,b}) through integration as follows.
\begin{equation}
\begin{split}
    n_{\bm{r}}(\bm{r}_T)&=\iiint_{\mathbb{R}^3}n_{\bm{r},\bm{v}}(\bm{r}_T,\bm{v})\,\mathrm{d}\bm{v}\\
    &=\sum_{j=1}^{N_b^*}\left(\sum_{k=1}^4\left(n_{\bm{\alpha},\bm{\beta}}^{(jk)}\right) \iiint_{V^{(j)}_{\bm{\alpha}}} \Psi(\bm{\alpha})\,\mathrm{d}\bm{\alpha}\right)
\end{split}
\end{equation}
In the following sections the assumption of Eq.~(\ref{eq:assumption}) is justified, estimating the error introduced by approximating the relative velocity between fragments and target object as bin-wise constant. Two analytical solutions of the integral of the function $\Psi(\bm{\alpha})$ are then proposed.

\subsubsection{Estimate of the model error}
\label{Estimate of the model error}

To justify the assumption of constant relative velocity over the bins volume, the Taylor expansion of $v_\mathrm{rel}$ around the bin center is considered:
\begin{equation}
    v_\mathrm{rel}^{(k)}(\bm{\alpha})=v_\mathrm{rel}^{(jk)}+\mathrm{J}_{v_\mathrm{rel}}^{(jk)}\left(\bm{\alpha}-\bm{\alpha}^{(j)}\right)+\frac{1}{2}\left(\bm{\alpha}-\bm{\alpha}^{(j)}\right)^T \mathrm{H}_{v_\mathrm{rel}}^{(jk)}\left(\bm{\alpha}-\bm{\alpha}^{(j)}\right)+\dots
\label{eq:vrel}
\end{equation}
where $\mathrm{J}_{v_\mathrm{rel}}^{(jk)}$ and $\mathrm{H}_{v_\mathrm{rel}}^{(jk)}$ are the Jacobian and the Hessian of the impact velocity function $v_\mathrm{rel}^{(k)}(\bm{\alpha})$, evaluated in the bins center. From now on the apexes, indicating the j$^{\mathrm{th}}$ bin and the k$^\mathrm{th}$ solution of intersection, are omitted for the sake of simplicity. The Jacobian $\mathrm{J}_{v_\mathrm{rel}}$ and Hessian $\mathrm{H}_{v_\mathrm{rel}}$ take the following form:
\begin{equation}
    \mathrm{J}_{v_\mathrm{rel}}=\frac{\bm{v}_\mathrm{rel}^T}{v_\mathrm{rel}}\,\mathrm{J}_{\bm{v}}
\label{eq:Jv_rel}
\end{equation}
\begin{equation}
    \mathrm{H}_{v_\mathrm{rel}}=\frac{\mathrm{J}_{\bm{v}}^T \mathrm{J}_{\bm{v}}}{v_\mathrm{rel}}-\frac{\mathrm{J}_{\bm{v}}^T\bm{v}_\mathrm{rel}\,\bm{v}_\mathrm{rel}^T\mathrm{J}_{\bm{v}}}{v_\mathrm{rel}^{\,3}}+\frac{\bm{v}_\mathrm{rel}^T}{v_\mathrm{rel}}\,\mathrm{H}_{\bm{v}}
\label{eq:Hv_rel}
\end{equation}
where $\mathrm{J}_{\bm{v}}$ and $\mathrm{H}_{\bm{v}}$ are Jacobian and Hessian of the fragments velocity vector function $\bm{v}(\bm{\alpha})$. Assuming that the step-sizes $\bm{\delta\alpha}$ are small enough, the velocity difference between the bin center and any point in the bin, $\bm{\delta v}(\bm{\alpha})$, can be written from the second order expansion of $\bm{v}(\bm{\alpha})$ around the bin center $\bm{\alpha}^{(j)}$, as follows.
\begin{equation}
    \bm{v}(\bm{\alpha})-\bm{v}=\bm{\delta v}(\bm{\alpha})\approx\mathrm{J}_{\bm{v}}\left(\bm{\alpha}-\bm{\alpha}^{(j)}\right)+\left(\bm{\alpha}-\bm{\alpha}^{(j)}\right)^T\mathrm{H}_{\bm{v}}\left(\bm{\alpha}-\bm{\alpha}^{(j)}\right)
\label{eq:v_exp}
\end{equation}
Introducing Eqs.~(\ref{eq:Jv_rel}) and~(\ref{eq:Hv_rel}) into Eq.~(\ref{eq:vrel}), it is possible to write:
\begin{equation}
\begin{split}
    v_{\mathrm{rel}}(\bm{\alpha}) \approx&\; v_{\mathrm{rel}} + \frac{\bm{v}_{\mathrm{rel}}^T}{v_{\mathrm{rel}}} \left[ \mathrm{J}_{\bm{v}}\left(\bm{\alpha}-\bm{\alpha}^{(j)}\right) + \frac{1}{2} \left(\bm{\alpha}-\bm{\alpha}^{(j)}\right)^T \mathrm{H}_{\bm{v}}\left(\bm{\alpha}-\bm{\alpha}^{(j)}\right)\right] + \\
    &\;\frac{1}{2} \left(\bm{\alpha}-\bm{\alpha}^{(j)}\right)^T \left[\frac{\mathrm{J}_{\bm{v}}^T \mathrm{J}_{\bm{v}}}{v_\mathrm{rel}} - \frac{\mathrm{J}_{\bm{v}}^T \bm{v}_\mathrm{rel} \bm{v}_\mathrm{rel}^T \mathrm{J}_{\bm{v}}}{v_\mathrm{rel}^3} \right] \left(\bm{\alpha}-\bm{\alpha}^{(j)}\right)
\end{split}
\end{equation}
The part in brackets of the first term is identified as $\bm{\delta v}=\bm{\delta v}(\bm{\bm{\alpha})}$ from Eq.~(\ref{eq:v_exp}). For the second term, because it is already quadratic in $(\bm{\alpha}-\bm{\alpha}^{(j)})$, the expansion in $\bm{v}(\bm{\bm{\alpha}})$ is truncated at the first order, i.e., the linear relation $\bm{\delta v} = \mathrm{J}_{\bm{v}}(\bm{\alpha}-\bm{\alpha}^{(j)})$ is used to reach:
\begin{equation}
    v_\mathrm{rel}(\bm{\alpha})\approx v_\mathrm{rel}+\frac{\bm{v}_\mathrm{rel}^T\,\bm{\delta v}}{v_\mathrm{rel}}+\frac{1}{2}\frac{\bm{\delta v}^T\bm{\delta v}}{v_\mathrm{rel}}-\frac{1}{2}\frac{\bm{\delta v}^T\bm{v}_\mathrm{rel}\,\bm{v}_\mathrm{rel}^T\bm{\delta v}}{v_\mathrm{rel}^{\,3}}
\label{eq:vrel1}
\end{equation}
Identified as $\rho=\rho(\bm{\alpha})$ the angle between the vectors $\bm{v}_\mathrm{rel}$ and $\bm{\delta v}(\bm{\alpha})$, Eq.~(\ref{eq:vrel1}) can be rewritten as:
\begin{equation}
    v_\mathrm{rel}(\bm{\alpha})\approx v_\mathrm{rel}+\delta v\cos\rho+\frac{1}{2}\frac{\delta v^2}{v_\mathrm{rel}}\sin^2\rho
\end{equation}
Thus, the normalized error introduced by the approximation of Eq.~(\ref{eq:assumption}) can be expressed as follows.
\begin{equation}
    \mathrm{Err.}(\bm{\alpha})=\frac{v_\mathrm{rel}(\bm{\alpha})-v_\mathrm{rel}}{v_\mathrm{rel}}\approx\xi\cos\rho+\frac{1}{2}\xi^2\sin^2\rho, \qquad \xi(\bm{\alpha})=\frac{\delta v(\bm{\alpha})}{v_\mathrm{rel}}
\label{eq:error}
\end{equation}
Since the step-sizes $\bm{\delta\alpha}$ are taken sufficiently small, the impact velocity $v_\mathrm{rel}$ is either greater or comparable to the difference in velocity between the bin center and any point belonging to the bin $\delta v(\bm{\alpha})$, for every bin in the domain, or, equivalently:
\begin{equation}
    \xi(\bm{\alpha})\lesssim 1
\end{equation}
However, it is worth recalling that the function $\xi(\bm{\alpha})$ may approach the unity only in those bins (if any) where the relative velocity between the target and the fragments is close to zero, i.e., when they move on very similar orbits in both shape and orientation. On the contrary, for any fragmentation scenario there exist many bins for which $\xi(\bm{\alpha}) \ll 1$, as the fragments spread out in a considerably vast domain. Therefore, considering that the impact rate is proportional to the impact velocity, the error expressed in Eq.~(\ref{eq:error}) is high only for those bins which provide a negligible contribution to the overall estimated impact rate. Finally, note that if $\xi(\bm{\alpha})$ is smaller than the unity, the following inequality for the error applies:
\begin{equation}
    \mathrm{Err.}(\bm{\alpha})<\xi(\bm{\alpha})
\end{equation}
This implies that the assumption of Eq.~(\ref{eq:assumption}) only slightly affects the accuracy of the method.

\subsubsection{Semi-analytical computation of the impact rate}
\label{Semi-analytical computation of the fragments flux}

The function $\Psi(\bm{\alpha})$, reported in Eq.~(\ref{eq:detJ}), can be written as the product among a constant, $r_T^3$, a function of semi-major axis and eccentricity, $\tilde{g}(a,e)$, and a function on inclination, $\tilde{h}(i)$, whose expressions are the following:
\begin{equation}
\begin{split}
    & \tilde{g}(a,e)=\frac{1}{a\sqrt{2ar_T-a^2(1-e^2)-r_T^2}}\\
    & \tilde{h}(i)=\frac{1}{\sqrt{\sin^2 i-\sin^2\phi_T}}
\end{split}
\end{equation}
where $\phi_T$ is the target latitude. As a result, the integrals in semi-major axis and eccentricity, $\mathcal{I}_{a/e}$, and inclination, $\mathcal{I}_i$, can be computed separately. The integration in inclination is firstly addressed.

The primitive $\mathcal{H}(i)$ of the function $\tilde{h}(i)$ reads as:
\begin{equation}
    \mathcal{H}(i)=\int\tilde{h}(i)\,\mathrm{d}i=\frac{1j}{\sin\phi_T}F(i,m), \qquad m=\frac{1}{\sin^2\phi_T}
\label{eq:primitive_g}
\end{equation}
where $1j$ is the imaginary unit, and $F(i,m)$ is the incomplete elliptic integral of the first kind with modulus $m\geq1\;\forall\phi_T$, whose value is in general complex, unless $m$ is the unity. By applying the reciprocal modulus transformation by Byrd and Friedman~\cite{Byrd1971}, the elliptic integral $F(i,m)$ can be written as sum of two elliptic integrals for which the solution is always real, as follows.
\begin{equation}
    F(i,m)=\sin\phi_T\left(K\left(\frac{1}{m}\right)-1j\,F\left(\beta(i),1-\frac{1}{m}\right)\right)
\label{eq:transformation}
\end{equation}
with:
\begin{equation}
    \beta(i)=\arcsin\left(\frac{\sqrt{m\sin^2i-1}}{\sin i\sqrt{m-1}}\right)
\end{equation}
where $K\left(\frac{1}{m}\right)$ is the complete elliptic integral of the first kind. The evaluation the primitive $\mathcal{H}(i)$ between the two extremes of integration, $i_1$ and $i_2$, provides the following expression for the integral in inclination $\mathcal{I}_i$:
\begin{equation}
    \mathcal{I}_i=F\left(\beta(i_2),1-\sin^2\phi_T\right)-F\left(\beta(i_1),1-\sin^2\phi_T\right)
\label{eq:Ii}
\end{equation}
As it can be observed, the imaginary part cancels out, as it multiplies the complete elliptic integral $K\left(\frac{1}{m}\right)$, which does not depend on inclination by definition. Note also that the primitive $\mathcal{H}(i)$ has a singularity when the following inequality applies:
\begin{equation}
    \sin i<|\sin\phi_T|
\label{eq:singularity_i}
\end{equation}
This condition is a physical singularity, as when Eq.~(\ref{eq:singularity_i}) is satisfied, the fragments orbit cannot intersect the target one, for any combination of $(\Omega,\omega,M)$. Indeed, the maximum latitude magnitude reachable by the fragments coincides with their orbital inclination. In general, the two extremes of integration in inclination, $i_1$ and $i_2$, can be obtained as follows.
\begin{equation}
\begin{cases}
    i_1=\arcsin\Bigl(\max(\sin i^-,|\sin\phi_T|)\Bigr)\\
    i_2=\pi-\arcsin\Bigl(\max(\sin i^+,|\sin\phi_T|)\Bigr)
\end{cases}
\end{equation}
where $i^-$ and $i^+$ indicate the lower and upper inclination boundaries for the considered bin.  

The integral in semi-major axis and eccentricity $\mathcal{I}_{a/e}$ is now considered. The primitive $\mathcal{G}(a,e)$ of the function $\tilde{g}(a,e)$ reads as:
\begin{equation}
    \mathcal{G}(a,e)=\iint \tilde{g}\,\mathrm{d}a\,\mathrm{d}e=\frac{1}{r_T}\left(e\arctan\Bigl(a(a-r_T)\tilde{g}(a,e)\Bigr)+\frac{a-r_T}{a} \ln\left(ae+\frac{1}{a\tilde{g}(a,e)}\right)\right)
\end{equation}
The integral in $\mathcal{I}_{a/e}$ is obtained evaluating the primitive $\mathcal{G}(a,e)$ at the two extremes of integration in semi-major, $a_1$ and $a_2$, and eccentricity, $e_1$ and $e_2$, as follows.
\begin{equation}
    \mathcal{I}_{a/e}=\mathcal{G}(a_1,e_1)-\mathcal{G}(a_1,e_2)-\mathcal{G}(a_2,e_1)+\mathcal{G}(a_2,e_2)
\end{equation}
Again, singular cases follow combinations of semi-major axis and eccentricity which cannot provide intersection with the target orbit. This condition verifies when either the fragments orbit perigee is larger than the target orbital radius $r_T$ or the apogee is smaller than it. The fragments orbits leading to the singularity satisfy the following inequality:
\begin{equation}
    e<\frac{|a-r_T|}{a}
\label{eq:singularity_ae}
\end{equation}
Unfortunately, contrary to the case of inclination, the constraint of intersection in semi-major axis and eccentricity divides all the bins crossed by the function of Eq.~(\ref{eq:singularity_ae}) into two non-rectangular shaped volumes. As a result, the primitive $\mathcal{G}(a,e)$ cannot be found analytically. To address this problem, two alternatives were identified:
\begin{itemize}
    \item[-] Analytical integration in semi-major axis and eccentricity, $\mathcal{I}_{a/e}$, for the bins not crossed by the function of Eq.~(\ref{eq:singularity_ae}), and numerical integration through sampling otherwise.
    \item[-] Analytical integration through change of variables to perigee $r_p$ and apogee $r_a$ radii, $\mathcal{I}_{r_p/r_a}$, over the entire domain.
\end{itemize}
Note that the integration through sampling is computationally heavy and less accurate. Therefore, since in principle the change of variable does not add complexity to the model, the first option must be chosen only if the primitive $\mathcal{G}(r_p, r_a)$ cannot be found. As demonstrated in the following, the primitive $\mathcal{G}(r_p, r_a)$ exists, even though it involves the evaluation of complex functions. This is the reason why the second option is preferred.

The change of variables $\tau(a,e)$, from semi-major axis $a$ and eccentricity $e$ to perigee $r_p$ and apogee $r_a$ radii, reads as:
\begin{equation}
    \tau(a,e):=\begin{cases}
        r_p=a(1-e)\\
        r_a=a(1+e)
    \end{cases}
\label{eq:a,e_to_rp,ra}
\end{equation}
To integrate in $r_p$ and $r_a$, the propagated characteristics need to be interpolated in a grid defined in the new variables, to preserve the rectangular shape of the bins when crossed by the function of Eq.~(\ref{eq:singularity_ae}). Note that the change of variables does not modify the integrand; indeed, one should consider that the density transformation from Cartesian to Keplerian elements of Eq.~(\ref{eq:n_r,v to n_a,b}) applies when $\bm{\alpha}$ refers to both the subset $(a,e,i)$ and $(r_p,r_a,i)$. This is the reason why the determinant of the Jacobian of the transformation $\tau(a,e)$ must not be added to the integration. The new integrand is simply obtained applying the change of variables of Eq.~(\ref{eq:a,e_to_rp,ra}) to the function $\tilde{g}$, which modifies as follows.
\begin{equation}
    \tilde{g}\left(\tau^{-1}(a,e)\right)=\frac{2}{(r_p+r_a)\sqrt{(r_p+r_a)r_T-r_pr_a-r_T^2}}
    \label{eq:integrand_rp_ra}
\end{equation}

The analytical integration of Eq.~(\ref{eq:integrand_rp_ra}) is not straightforward. Furthermore, the integral becomes improper when either the upper limit of $r_p$ or the lower limit of $r_a$ coincide with $r_T$, complicating even its numerical integration. Both issues can be addressed with an additional change of variables, reducing the integrand to the inverse of a hyperbolic paraboloid. Different sets of integration variables can be proposed, corresponding to different geometrical representations of the paraboloid; while some preserve the shape of the integration domain, others provide simpler expressions for the integrand. The domain-preserving case is considered first, with the change of variables:
\begin{equation}
    \tau_X^{-1}(r_p,r_a):=\begin{cases}
        r_p = r_T - 2 r_T X_p^2\\
        r_a = r_T + 2 r_T X_a^2
    \end{cases}
\label{eq:paraboloid_change1}
\end{equation}
Each one of the new variables $(X_p,X_a)$ depends only on one of the original variables $(r_p,r_a)$, preserving the rectangular shape of the integration domain and the independence of the integration limits. From the physical bounds $r_p \in ]0,r_T]$ and $r_a \in [r_T,\infty[$, it follows that $X_p \in [X_p(r_{p_2}),X_p(r_{p_1})] \in [0,1/\sqrt{2}[$ and $X_a \in [X_a(r_{a_1}),X_a(r_{a_2})] \in [0,\infty[$. Introducing this change of variables and the Jacobian of the transformation into Eq.~(\ref{eq:integrand_rp_ra}), integrand $\tilde{g}$ takes the form:
\begin{equation}
    \tilde{g}\left(X_p,X_a\right)=\frac{-8}{1+X_a^2-X_p^2}
    \label{eq:integrand_paraboloid1}
\end{equation}
The integrable singularity for $r_p,r_a=r_T$ has vanished. There is still a singularity for $1+X_a^2-X_p^2=0$, but it falls out of the physical domain of the problem. After some manipulations, an analytical primitive for $\tilde{g}\left(X_p,X_a\right)$ is obtained:
\begin{equation}
    \mathcal{G}(X_p,X_a) = 4[ \mathrm{Li}_2\left(X_c\right) + \mathrm{Li}_2\left(\overline{X_c}\right) - \mathrm{Li}_2\left(-X_c\right) - \mathrm{Li}_2\left(-\overline{X_c}\right) ] = 8 \Re \left[\mathrm{Li}_2\left(X_c\right) - \mathrm{Li}_2\left(-X_c\right) \right]
    \label{eq:primitive_paraboloid1}
\end{equation}
\begin{equation*}
    X_c = - \left( X_a - \sqrt{1+X_a^2}\right) \left( X_p + 1j\sqrt{1-X_p^2} \right)
\end{equation*}
where $\overline{z}$ denotes the complex conjugate of $z$, $\Re [z]$ is the real part of $z$, and $\mathrm{Li}_2\left(z\right)$ is the dilogarithm, or Spence's function \cite{loxton1984special}:
\begin{equation}
    \mathrm{Li}_2\left(z\right) = - \int_0^z \frac{\ln\left(1-u\right)}{u} \mathrm{d}u = \sum_{k=1}^\infty \frac{z^k}{k^2}
\end{equation}
for complex $z$, where the series is convergent only for $|z|<1$. Note that the number of dilogarithm evaluations in the last expression of Eq.~(\ref{eq:primitive_paraboloid1}) has been halved using the relation $\mathrm{Li}_2\left(\overline{z}\right)=\overline{\mathrm{Li}_2\left(z\right)}$. The integral $\mathcal{I}_{r_p/r_a}$ is finally obtained evaluating the primitive at the two extremes of integration in $(X_p,X_a)$, function of the ones in $(r_p,r_a)$:
\begin{equation}
    \mathcal{I}_{r_p/r_a}=\mathcal{G}(X_{p_1},X_{a_1})-\mathcal{G}(X_{p_1},X_{a_2})-\mathcal{G}(X_{p_2},X_{a_1})+\mathcal{G}(X_{p_2},X_{a_2})
    \label{eq:I_rp_ra_paraboloid1}
\end{equation}
which involves the computation of 8 dilogarithms of complex argument.

\begin{figure}[!ht]
     \centering
     \includegraphics[width=0.6\textwidth]{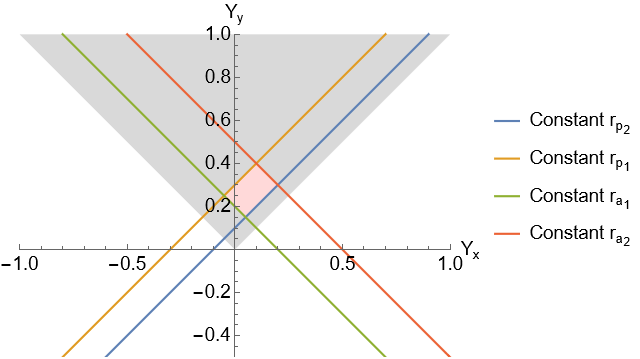}
     \caption{Integration domain in variables $(Y_x,Y_y)$.}
     \label{fig:I_rp_ra__SecondChange_IntegrationDomain}
\end{figure}

The presence of complex arguments in Eq.~(\ref{eq:primitive_paraboloid1}) is related to the negative Gaussian curvature of the hyperbolic paraboloid, and cannot be avoided. However, the numerical evaluation of dilogarithms is significantly more costly for complex arguments than for real ones, so it is convenient to reduce their presence. This is achieved with a new change of variables that leverages the fact that the hyperbolic paraboloid in Eq.~(\ref{eq:integrand_paraboloid1}) is a rectangular one:
\begin{equation}
    \tau_Y^{-1}(X_p,X_a):=\begin{cases}
        X_p = Y_y - Y_x\\
        X_a = Y_y + Y_x
    \end{cases}
\label{eq:paraboloid_change2}
\end{equation}
leading to a simpler integrand:
\begin{equation}
    \tilde{g}\left(Y_x,Y_y\right)=\frac{16}{1+4 Y_x Y_y}
    \label{eq:integrand_paraboloid2}
\end{equation}
This change of variables introduces a functional dependency between the original variables, so the integration limits are no longer independent. The new integration domain $\mathcal{B}$ is a parallelogram, bounded by the lines for constant $r_p$, $Y_y^{p_{1,2}}$, and the lines for constant $r_a$, $Y_y^{a_{1,2}}$:
\begin{equation}
\begin{split}
    Y_y^{p_{1,2}} &= X_{p_{1,2}} + X_x\\
    Y_y^{a_{1,2}} &= X_{a_{1,2}} - X_x
\end{split}
    \label{eq:bound_paraboloid2}
\end{equation}
Figure~\ref{fig:I_rp_ra__SecondChange_IntegrationDomain} shows a schematic of the domain. While the particular values of $r_{p_{1,2}}$ and $r_{a_{1,2}}$ will change for each bin, the relative position of the lines is preserved. Moreover, $\mathcal{B}$ is always contained in the semi-infinite plane $Y_y > |Y_x|$.

Green's Theorem is used to reduce the area integral over $\mathcal{B}$ to a line integral along its boundary $\partial \mathcal{B}$:
\begin{equation}
    \mathcal{I}_{r_p/r_a}=\iint_{\mathcal{B}}\frac{16}{1+4 Y_x Y_y} \mathrm{d}Y_x \mathrm{d}Y_y = - 4 \oint_{\partial \mathcal{B}} \frac{\ln (1+4 Y_x Y_y) }{Y_x} \mathrm{d}Y_x
    \label{eq:paraboloid2_lineIntegral}
\end{equation}
The line integral has to be evaluated over the 4 boundary segments in counter-clockwise direction, substituting $Y_y$ with the corresponding one from Eq.~(\ref{eq:bound_paraboloid2}). This reduces Eq.~(\ref{eq:paraboloid2_lineIntegral}) to the integral of the logarithm of a second degree polynomial of $Y_x$, divided by $Y_x$. Integration by parts allows to reduce it again to dilogarithms, involving the roots of the polynomial. For the constant $r_p$ boundaries, the polynomial roots are complex conjugates and the primitive given as function of $(X_p,X_a)$ is:
\begin{equation}
    \mathcal{G}^p(X_{p_{1,2}},X_a) = 8 \Re \left[ \mathrm{Li}_2\left( (X_{p_{1,2}}-X_a) (X_{p_{1,2}}+1j\sqrt{1-X^2_{p_{1,2}}})  \right) \right]
\end{equation}
while for the constant $r_a$ boundaries the polynomial roots are real and the primitive is:
\begin{equation}
    \mathcal{G}^a(X_p,X_{a_{1,2}}) = 4 \left[ \mathrm{Li}_2\left( (X_p-X_{a_{1,2}}) (X_{a_{1,2}}+\sqrt{1+X^2_{a_{1,2}}})\right) + \mathrm{Li}_2\left( (X_p-X_{a_{1,2}}) (X_{a_{1,2}}-\sqrt{1+X^2_{a_{1,2}}})\right) \right]
\end{equation}
Consequently, the new change of variables allows to limit the dilogarithms of complex arguments to half of the primitive evaluations. The integral $\mathcal{I}_{r_p/r_a}$ is obtained evaluating over $\partial \mathcal{B}$ in counter-clockwise direction:
\begin{equation}
\begin{split}
    \mathcal{I}_{r_p/r_a} &= \left[ \mathcal{G}^p(X_{p_2},X_{a_2}) - \mathcal{G}^p(X_{p_2},X_{a_1}) \right] + \left[ \mathcal{G}^a(X_{p_1},X_{a_2}) - \mathcal{G}^a(X_{p_2},X_{a_2}) \right] \\
    & + \left[ \mathcal{G}^p(X_{p_1},X_{a_1}) - \mathcal{G}^p(X_{p_1},X_{a_2}) \right] + \left[ \mathcal{G}^a(X_{p_2},X_{a_1}) - \mathcal{G}^a(X_{p_1},X_{a_1}) \right] 
\end{split}
\end{equation}
involving 4 dilogarithms of complex argument and 8 of real argument, compared to the 8 dilogarithms of complex argument for Eq.~(\ref{eq:I_rp_ra_paraboloid1}).

\section{Evaluation of the effects of real breakup events}
\label{Evaluation of the effects of real breakup events}

This section is devoted to the application of the model presented in Section~\ref{Fragmentation modeling and propagation} and Section~\ref{Debris density-based collision risk assessment} to the evaluation of the hazard caused by two real fragmentation events. The first is the breakup of the US payload (P/L) NOAA-16 in Sun-Synchronous Orbit (SSO); instead, the second is the fragmentation of the Russian Rocket Body (R/B) AMC 14 BRIZ-M on a high-elliptical orbit. The effect of the fragmentation clouds is monitored in terms of impact rate and collision probability with the rocket body SL-6 in SSO, which appears in the list of 50 statistically-most-concerning derelict objects in LEO proposed by McKnight et al.~\cite{McKnight2021}. The considered target slow-varying Keplerian elements are reported in Table~\ref{tab:kep SL-6}.

\begin{table}[hbt!]
\caption{\label{tab:kep SL-6} SL-16 slow-varying Keplerian elements.}
\centering
\begin{tabular}{ccccc}
\hline
$a$ [km] & $e$ [-] & $i$ [deg] & $\Omega$ [deg] & $\omega$ [deg]\\
\hline
7186 & 0.00090 & 98.31 & 315.59 & 256.72\\
\hline
\end{tabular}
\end{table}

For both the fragmentation events, the density distribution at breakup epoch is firstly depicted and commented. The evolution of the fragments cloud is analyzed, showing the distributions at some time epochs, and the main effects of the orbital perturbations are discussed. The collision probability with the target object is computed under different assumptions on the cloud and target dynamics, eventually considering the complete description of the cloud in the phase space of slow-varying orbital elements and the target orbit evolution. 

\subsection{Effects of the NOAA-16 fragmentation in Sun-synchronous orbit}
\label{Effects of the NOAA-16 fragmentation in Sun-synchronous orbit}
This event was the second known breakup of a NOAA-series spacecraft. The payload was launched on 21$^\mathrm{st}$ September 2000, as part of the Polar Operational Environmental Satellite series of U.S. weather satellites, and operated until 2005. The fragmentation occurred at 09:50 GMT on 25$^\mathrm{th}$ November 2015, most likely caused by a battery explosion~\cite{Tan2017}. The spacecraft breakup generated a considerable number of fragments, 136 of which were sufficiently large to be tracked and catalogued by the Joint Space Operations Center~\cite{FragHistory}. The satellite had a mass of 1475 kg and was orbiting a Sun-synchronous orbit. The fragmentation coordinates are reported in Table~\ref{tab:kep NOAA-16}.

\begin{table}[hbt!]
\caption{\label{tab:kep NOAA-16} NOAA-16 P/L Keplerian elements at fragmentation epoch.}
\centering
\begin{tabular}{cccccc}
\hline
$a$ [km] & $e$ [-] & $i$ [deg] & $\Omega$ [deg] & $\omega$ [deg] & $f$ [deg]\\
\hline
7226 & 0.00113 & 98.93 & 35.00 & 133.56 & 24.88\\
\hline
\end{tabular}
\end{table}

\subsubsection{NOAA-16 debris cloud evolution}
\label{NOAA-16 debris cloud evolution}

The simulation here proposed considers fragments in the range 1 cm - 1 m. The number of generated fragments predicted by the NASA SBM, in case of an explosion, depends on the parameter $S$, as defined in~\cite{Krisko2011}. Its value is set according to the expression reported in~\cite{Letizia2018}, where the parameter $S$ is related to the object mass $M_P$ as follows.
\begin{equation}
    S=\begin{cases}
        k\frac{M_P\; [\textrm{kg}]}{10000\; [\textrm{kg}]} \quad \textrm{if}\;kM_P< 10000 \; \textrm{kg}\\
        1 \qquad \quad \quad \;\;\, \textrm{if}\; kM_P\geq10000 \; \textrm{kg}
    \end{cases}
\end{equation}
with $k=1$ for payloads and $k=9$ for rocket bodies. Thus, for P/L NOAA-16, the parameter $S$ is set to 0.1475, which results in 1401 ejected fragments.

\begin{figure}[!ht]
     \centering
     \includegraphics[width=0.6\textwidth]{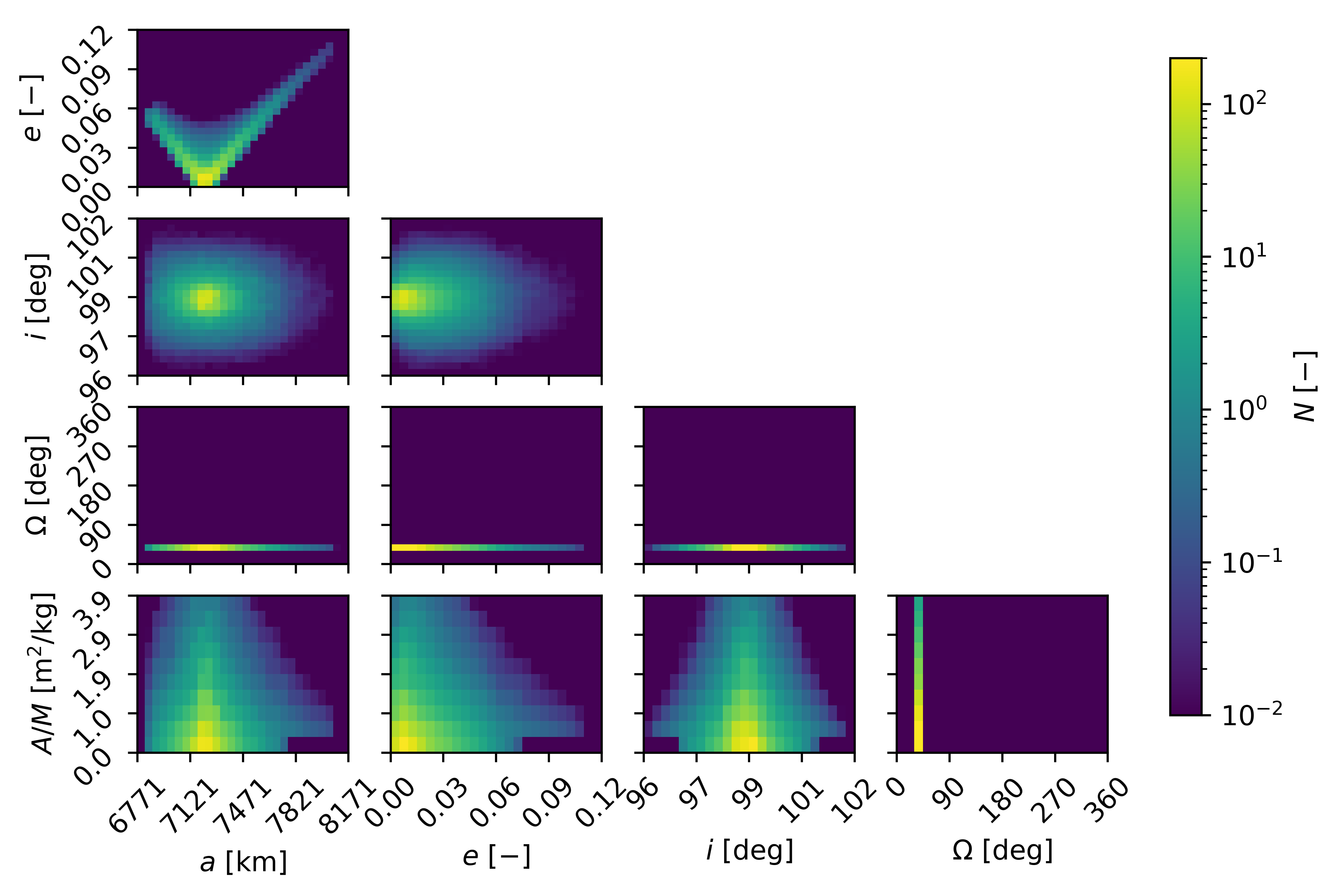}
     \caption{NOAA-16 P/L fragmentation - Density distribution in \texorpdfstring{$\bm{(a,e,i,\Omega,A/M})$}{vars} at fragmentation epoch.}
     \label{fig:NOAA-16_dist0}
\end{figure}

Figure~\ref{fig:NOAA-16_dist0} shows the initial density distribution in the subset of Keplerian elements $(a,e,i,\Omega)$ and area-to-mass ratio $A/M$. Instead, the cloud is assumed to be randomized over argument of periapsis $\omega$ and mean anomaly $M$. The randomization over $\omega$ is considered because of the small eccentricity of the parent orbit, which causes the fragments to spread almost uniformly over a range of 360 degrees in $\omega$. On the other hand, the fast angular variable $M$ is not accounted as interpolation variable, as the objective of this analysis is the estimation of the long-term behavior of the debris cloud. Indeed, as motivated in~\cite{McKnight1990}, the difference in the fragments orbital period induces the formation of a toroid around the Earth after few orbital revolutions. Note that the colorbar depicts the total number of fragments $N$ in a bin in the two given variables, for any values of the other elements.

As it can be observed, the cloud assumes the typical V-shape distribution of a LEO fragmentation in the semi-major axis-eccentricity domain. This peculiar shape is caused by the small eccentricity of the parent orbit, which bounds the fragments distribution above the curve:
\begin{equation}
    e=\frac{|a-r_P|}{a}
\label{eq:e_frag}
\end{equation}
with $r_P$ orbital radius of the parent object. The curve of Eq.~(\ref{eq:e_frag}) constrains the fragments perigee and apogee to be smaller and larger than $r_P$, respectively. The ejected fragments distribute over a range of approximately 6 degrees in inclination and 2.5 degrees in right ascension of the ascending node. The amplitude of the cloud in $i$ and $\Omega$, for the case of a circular parent orbit, only depends on the latitude of the fragmentation point, with the two elements behaving in opposite ways. In particular, the closer the fragmentation occurs to the equatorial plane, the more widely the fragments spread over inclination and the narrower is the cloud domain in right ascension of the ascending node. The opposite effect results from a fragmentation near the poles. Note that in the limit cases of a fragmentation on the equatorial plane and over the poles, all the ejected fragments would share the same right ascension of the ascending node and inclination, respectively.

\begin{figure*}[!ht]
     \centering
     \begin{subfigure}[b]{0.45\textwidth}
         \centering
         \includegraphics[width=\textwidth]{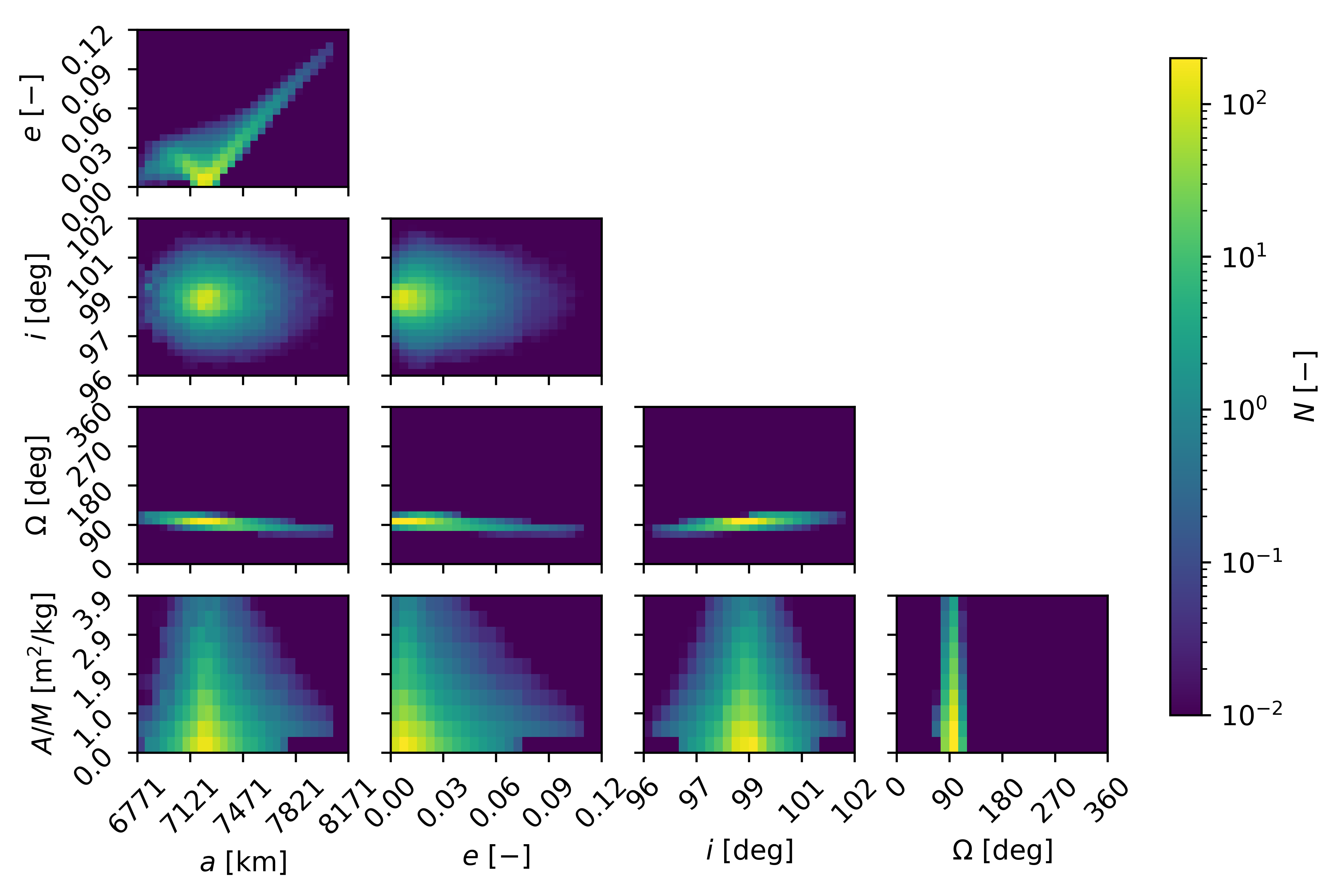}
         \caption{Epoch: 2 months after fragmentation}
         \label{fig:NOAA-16_dist1}
     \end{subfigure}
     \begin{subfigure}[b]{0.45\textwidth}
         \centering
         \includegraphics[width=\textwidth]{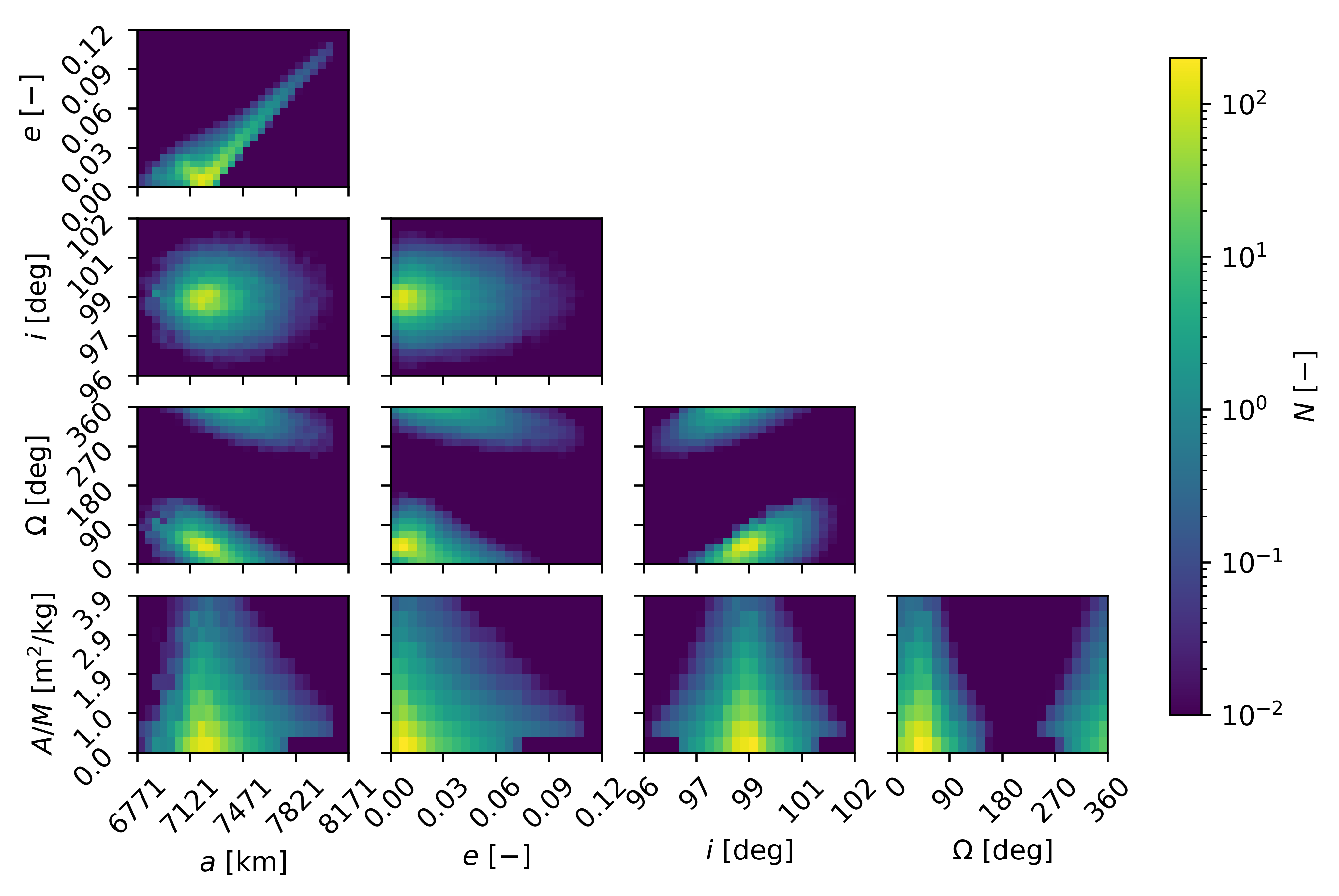}
         \caption{Epoch: 1 year after fragmentation}
         \label{fig:NOAA-16_dist2}
     \end{subfigure}\\
     \begin{subfigure}[b]{0.45\textwidth}
         \centering
         \includegraphics[width=\textwidth]{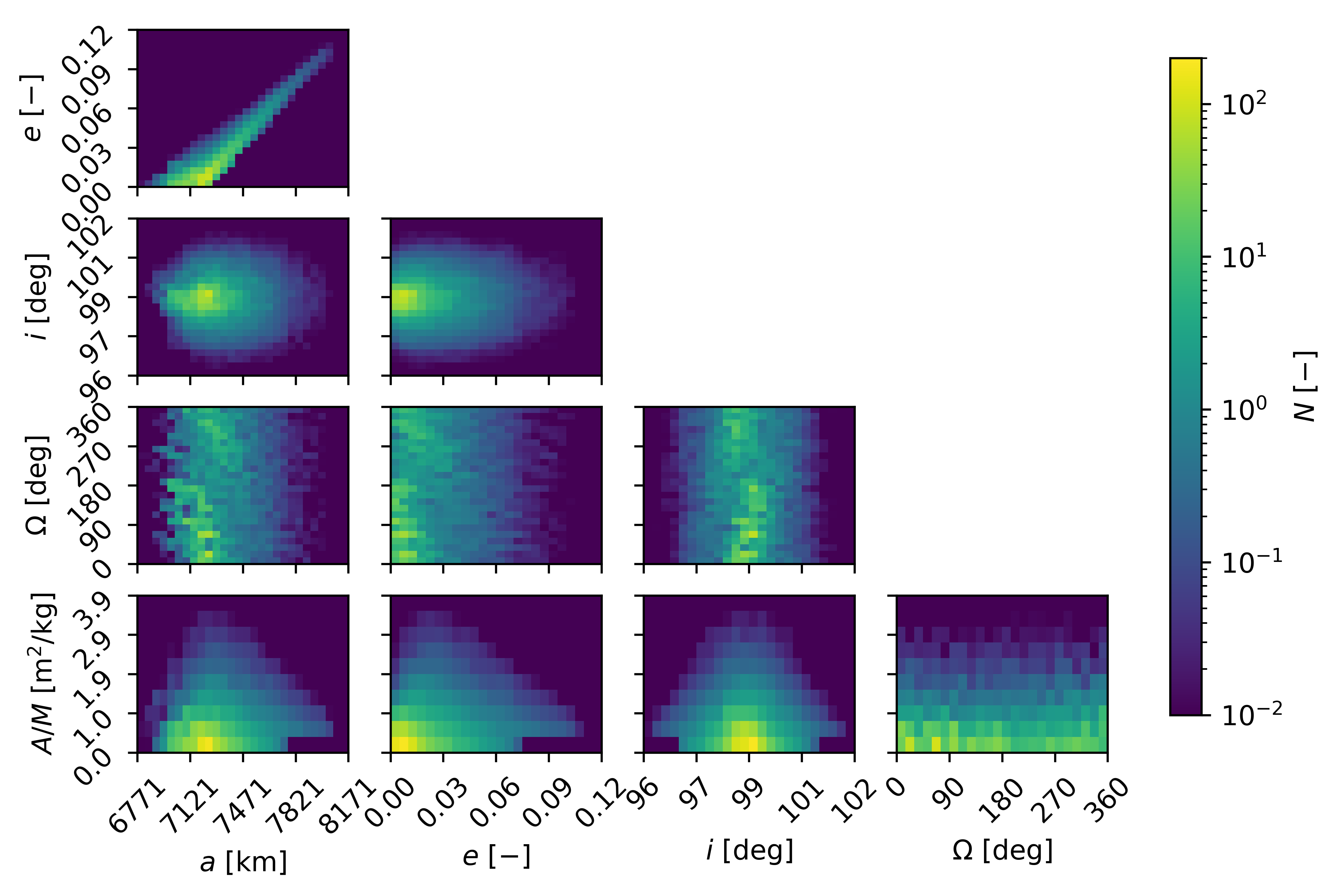}
         \caption{Epoch: 5 years after fragmentation}
         \label{fig:NOAA-16_dist3}
     \end{subfigure}
     \begin{subfigure}[b]{0.45\textwidth}
         \centering
         \includegraphics[width=\textwidth]{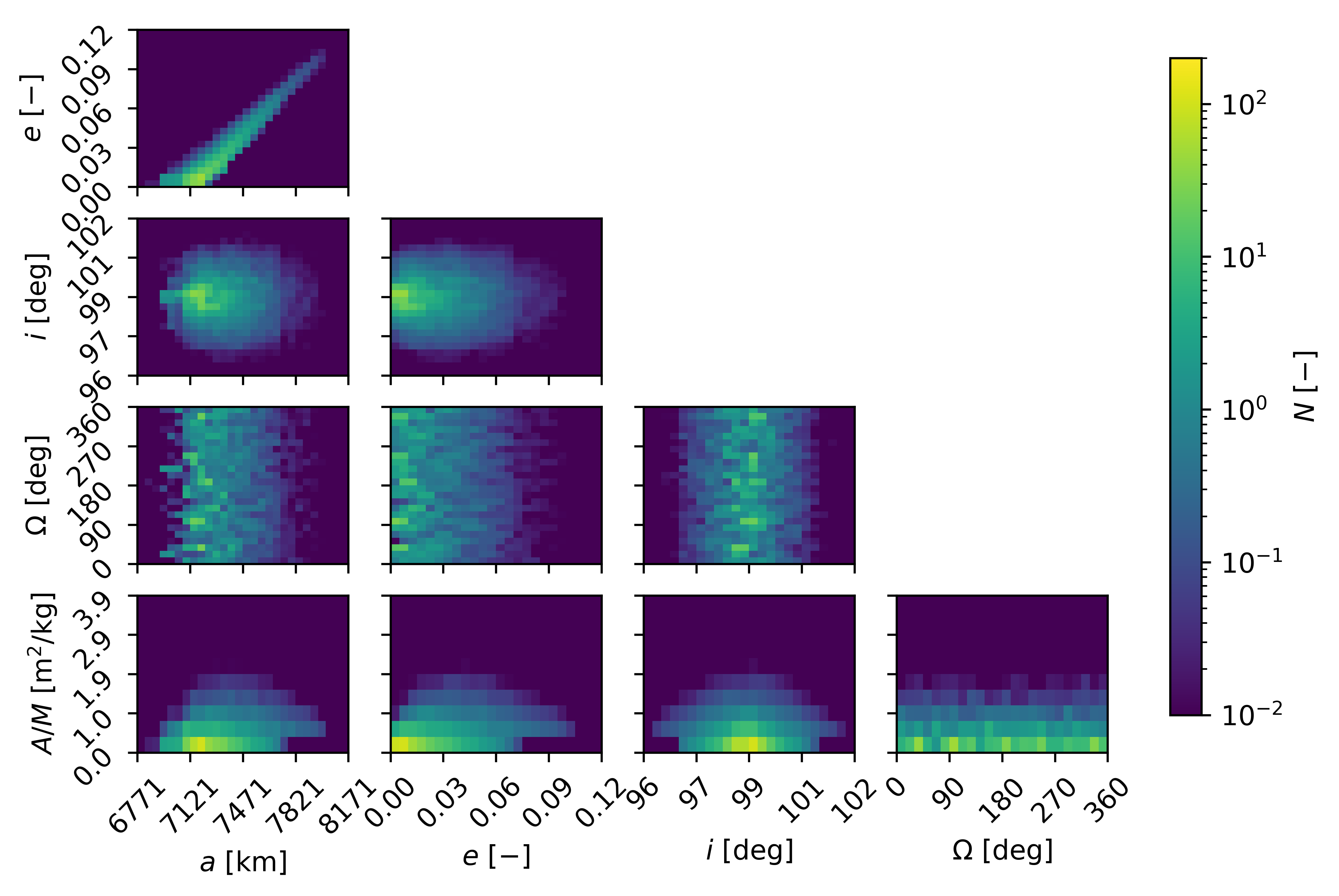}
         \caption{Epoch: 15 years after fragmentation}
         \label{NOAA-16_dist4}
     \end{subfigure}
     \caption{NOAA-16 P/L fragmentation - Density distribution in \texorpdfstring{$\bm{(a,e,i,\Omega,A/M})$}{vars} over time.}
     \label{fig:NOAA-16_dist_prop}
\end{figure*}

The debris density is propagated along 20000 characteristics curves, whose initial conditions are uniformly extracted from the initial distribution of Figure~\ref{fig:NOAA-16_dist0}. The considered force model accounts for atmospheric drag, J$_2$ perturbation, solar radiation pressure and luni-solar perturbation. The 5D density distributions are retrieved according to a 1-month time discretization, allowing to monitor both the short- and long-term dynamical behaviour of the cloud. In Figure~\ref{fig:NOAA-16_dist_prop} the fragments distribution 2 months, 1 year, 5 and 15 years after fragmentation is depicted.

By looking at Figure~\ref{fig:NOAA-16_dist_prop} the following considerations can be done:
\begin{itemize}
    \item[-] The fragments residing in the left leg of the V-shape distribution quickly reenter the atmosphere, under the effect of a higher atmospheric density.
    \item[-] As expected, the high area-to-mass ratio fragments are the most affected by atmospheric drag. As a result, the upper part of the distribution in $A/M$ vanishes after 15 years.
    \item[-] A complete randomization over right ascension of the ascending node is not achieved even after 15 years of propagation, as some high-density regions stand out over the distribution in $\Omega$. It is worth further noticing that after 1 year the fragments with the highest rate of change in $\Omega$ have not yet reached the slowest ones. Thus, it can be already inferred the error one may commit in assuming the distribution to be randomized in $\Omega$ throughout the whole simulation time.
    \item[-] The fragments inclination is barely affected by solar radiation pressure and luni-solar perturbation, because of the negligible force exerted by the two disturbances below 2000 km altitude.
\end{itemize}

\subsubsection{Collision risk from a 1D debris cloud in orbital radius r}
\label{Collision risk from a 1D debris cloud in orbital radius r}

\begin{figure*}[!ht]
     \centering
     \begin{subfigure}[b]{0.33\textwidth}
         \centering
         \includegraphics[width=\textwidth]{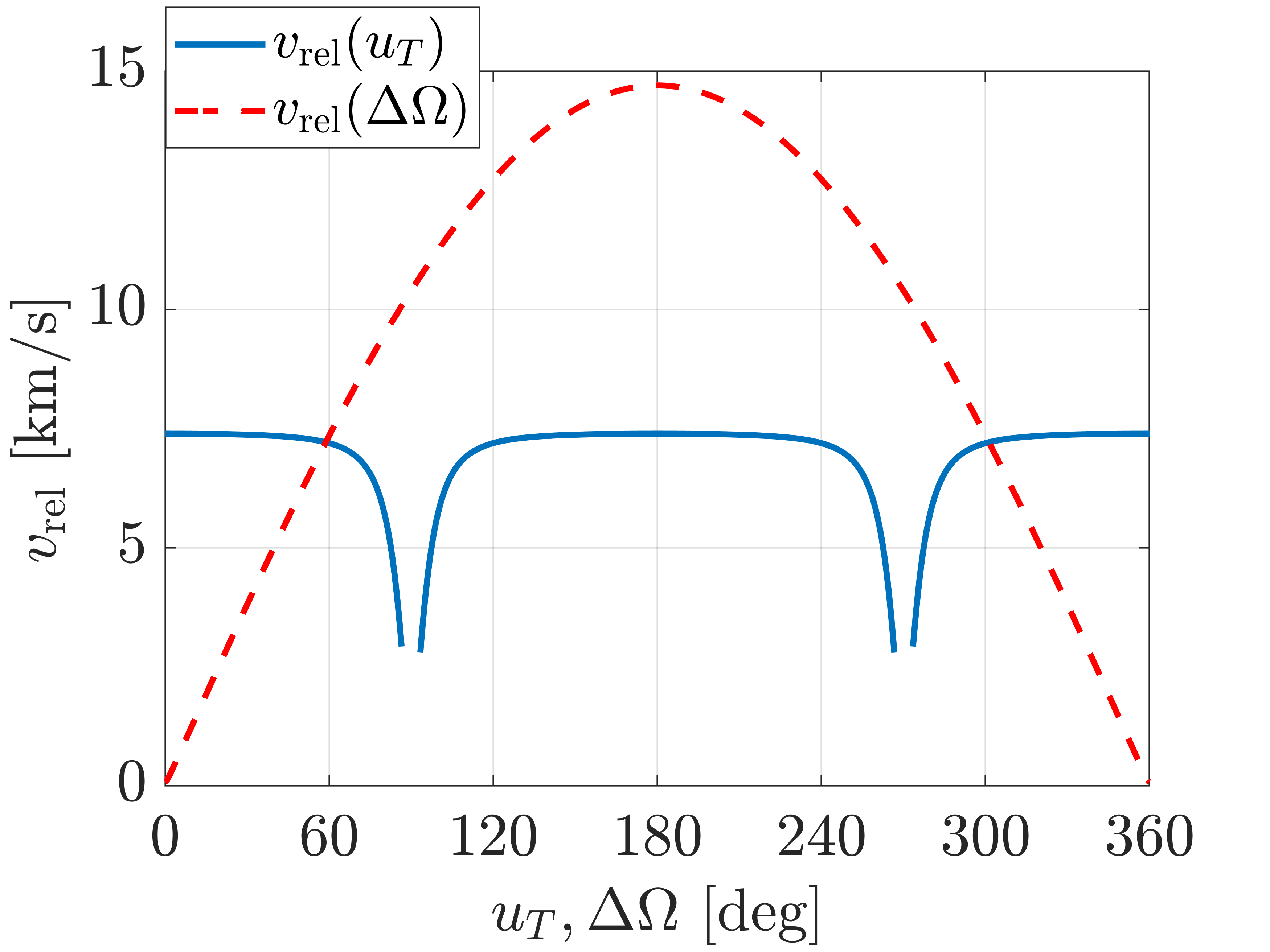}
         \caption{Impact velocity \texorpdfstring{$\bm{v_\mathrm{rel}{}}$}{vrel}}
         \label{fig:NOAA-16_vrel}
     \end{subfigure}
     \begin{subfigure}[b]{0.33\textwidth}
         \centering
         \includegraphics[width=\textwidth]{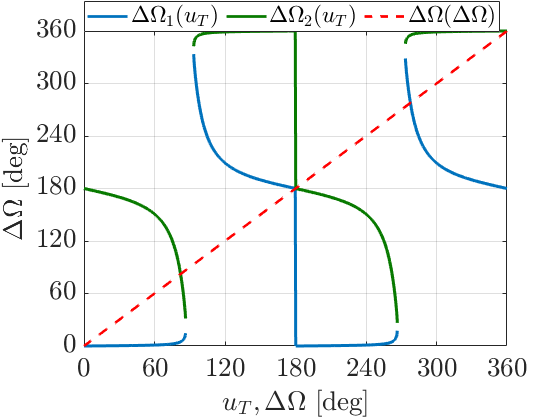}
         \caption{Difference in right ascension \texorpdfstring{$\bm{\Delta\Omega}$}{dOm}}
         \label{fig:NOAA-16_dOm}
     \end{subfigure}
     \begin{subfigure}[b]{0.33\textwidth}
         \centering
         \includegraphics[width=\textwidth]{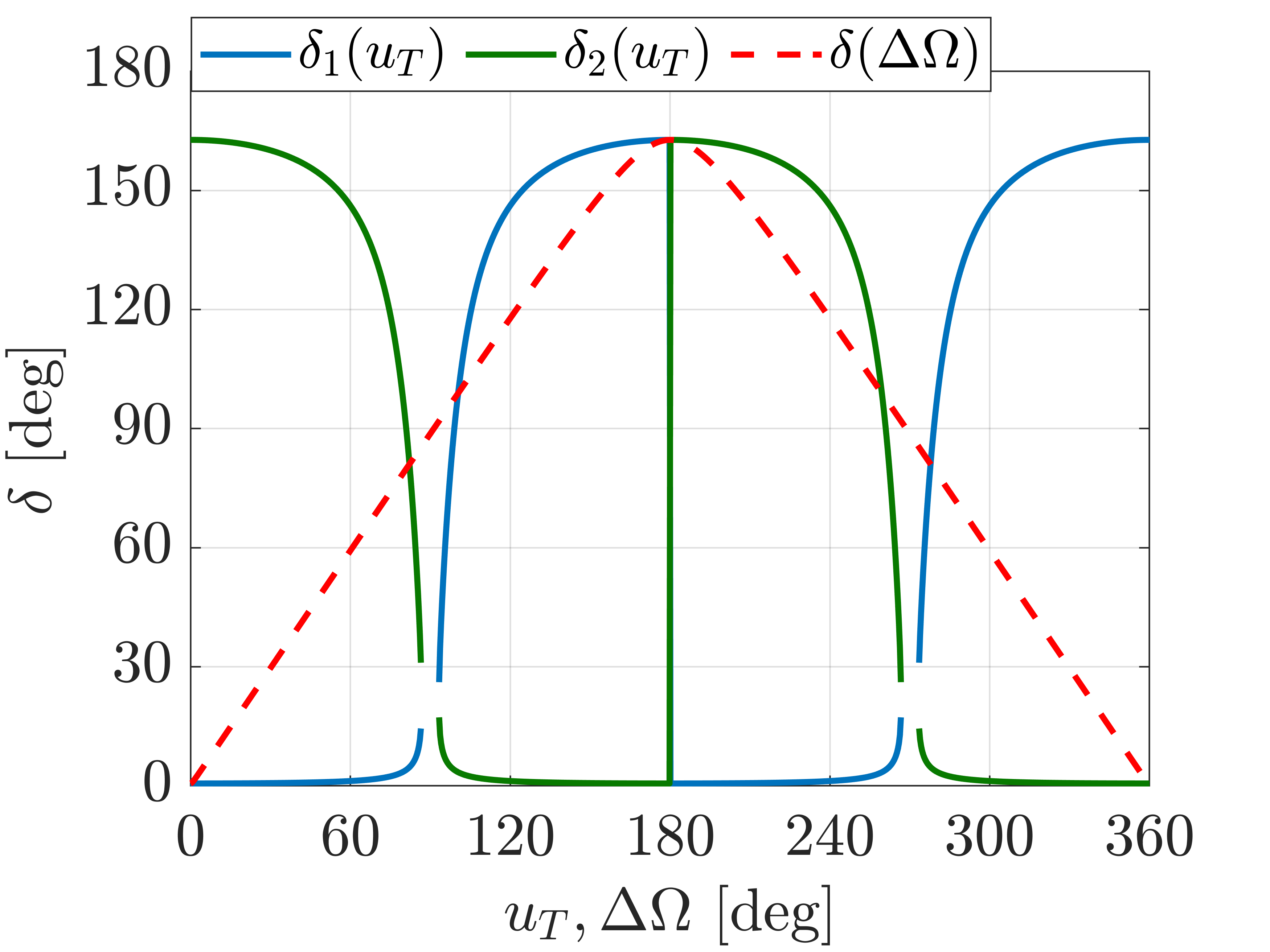}
         \caption{Angle between the velocity vectors \texorpdfstring{$\bm{\delta}$}{\textdelta}}
         \label{fig:NOAA-16_delta}
     \end{subfigure}
     \caption{NOAA-16 P/L fragmentation - Impact velocity, difference in right ascension of the ascending node and angle between fragments and target velocity vector as function of the target argument of latitude.}
     \label{fig:NOAA-16_vrel,dOm,delta}
\end{figure*}

The impact rate is computed averaging the fragments flux against the target cross section, over the target mean anomaly, according to Eq.~(\ref{eq:eta_dot_c1}). The 1D spatial density function $n_{\bm{r}}$ is retrieved from the 4D distribution in Keplerian elements through the Keplerian to Cartesian coordinates transformation, and it varies discretely with orbital radius $r$, as a result of a 1D interpolation through binning. Following the approach proposed in~\cite{Letizia2015b}, to compute the average impact velocity of Eq.~(\ref{eq:Delta_v_av}), the fragments orbits are assumed to have the same inclination as the parent one. As a result, the average impact velocity $v_\mathrm{rel}$, as well as its product with the latitude-dependent function $\beta$, remain constant over time, as they do not depend on the evolution of the fragments distribution. Therefore, the average impact velocity is only responsible for the impact rate magnitude, while its trend over time $t$ is only dependent on the evolution of the spatial density function, evaluated at the target orbital radius $r_T$, $n_{\bm{r}}(t, r_T)$. Figure~\ref{fig:NOAA-16_vrel} shows the impact velocity as function of the target argument of latitude $u_T$, and Figure~\ref{fig:NOAA-16_dOm} and Figure~\ref{fig:NOAA-16_delta} the associated values of difference in right ascension of the ascending node $\Delta\Omega$ and angle between fragments and target velocity vectors $\delta$, computed through Eq.~(\ref{eq:Delta_Om}) and Eq.~(\ref{eq:cos_delta}). The subscripts $(\cdot)_1$ and $(\cdot)_2$ identify the two possible solutions of intersection. Note that the incorrect solutions, which results from considering $\Delta\Omega$ and $M_T$ as linearly related~\cite{Letizia2015b}, corresponding to Eq.~(\ref{eq:vrel_Letizia}), are also displayed with the red dashed line.

\begin{figure*}[!ht]
     \centering
     \begin{subfigure}[b]{0.45\textwidth}
         \centering
         \includegraphics[width=\textwidth]{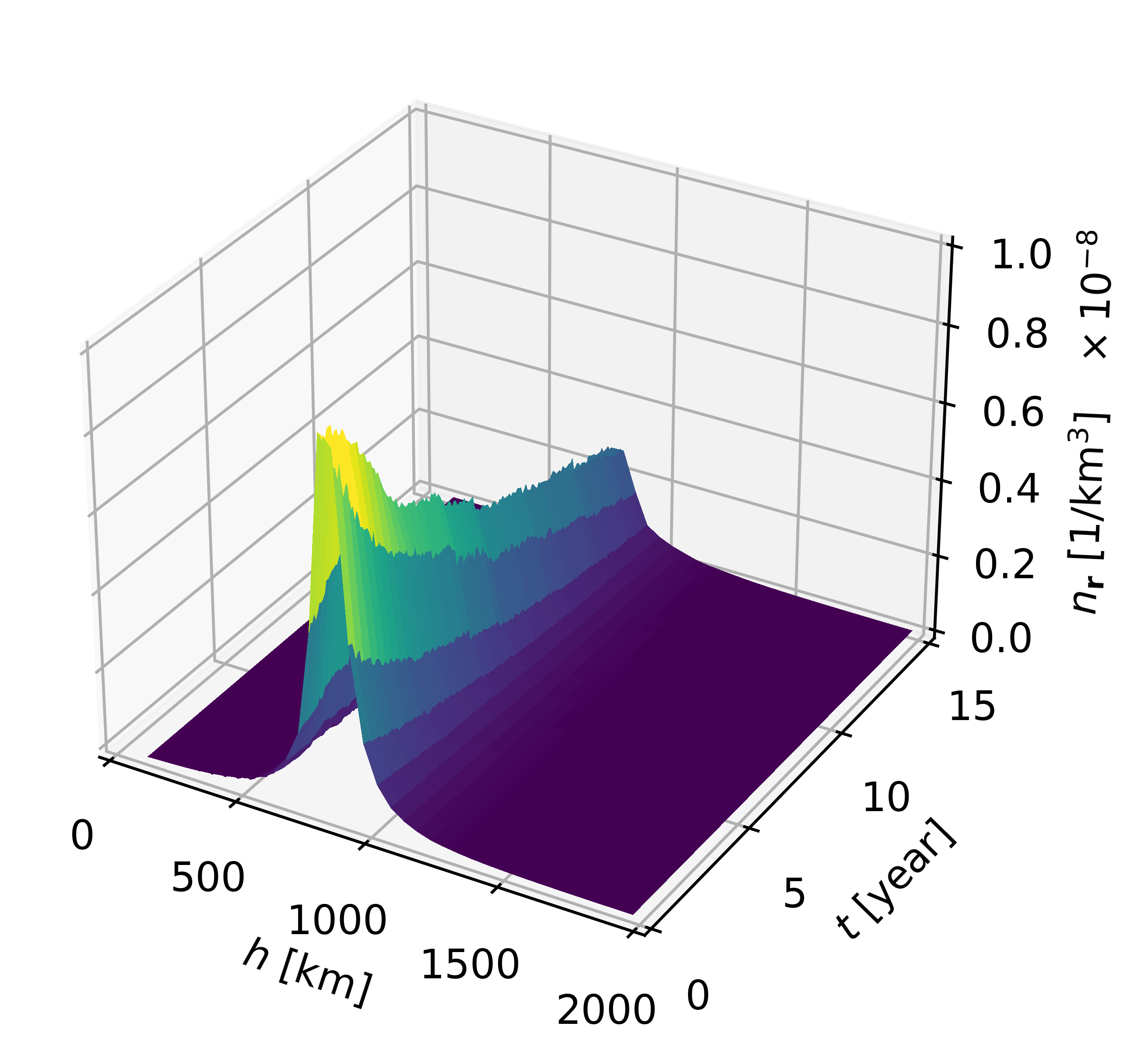}
         \caption{Dependency on altitude}
         \label{fig:NOAA-16_spatial_R}
     \end{subfigure}
     \begin{subfigure}[b]{0.45\textwidth}
         \centering
         \includegraphics[width=\textwidth]{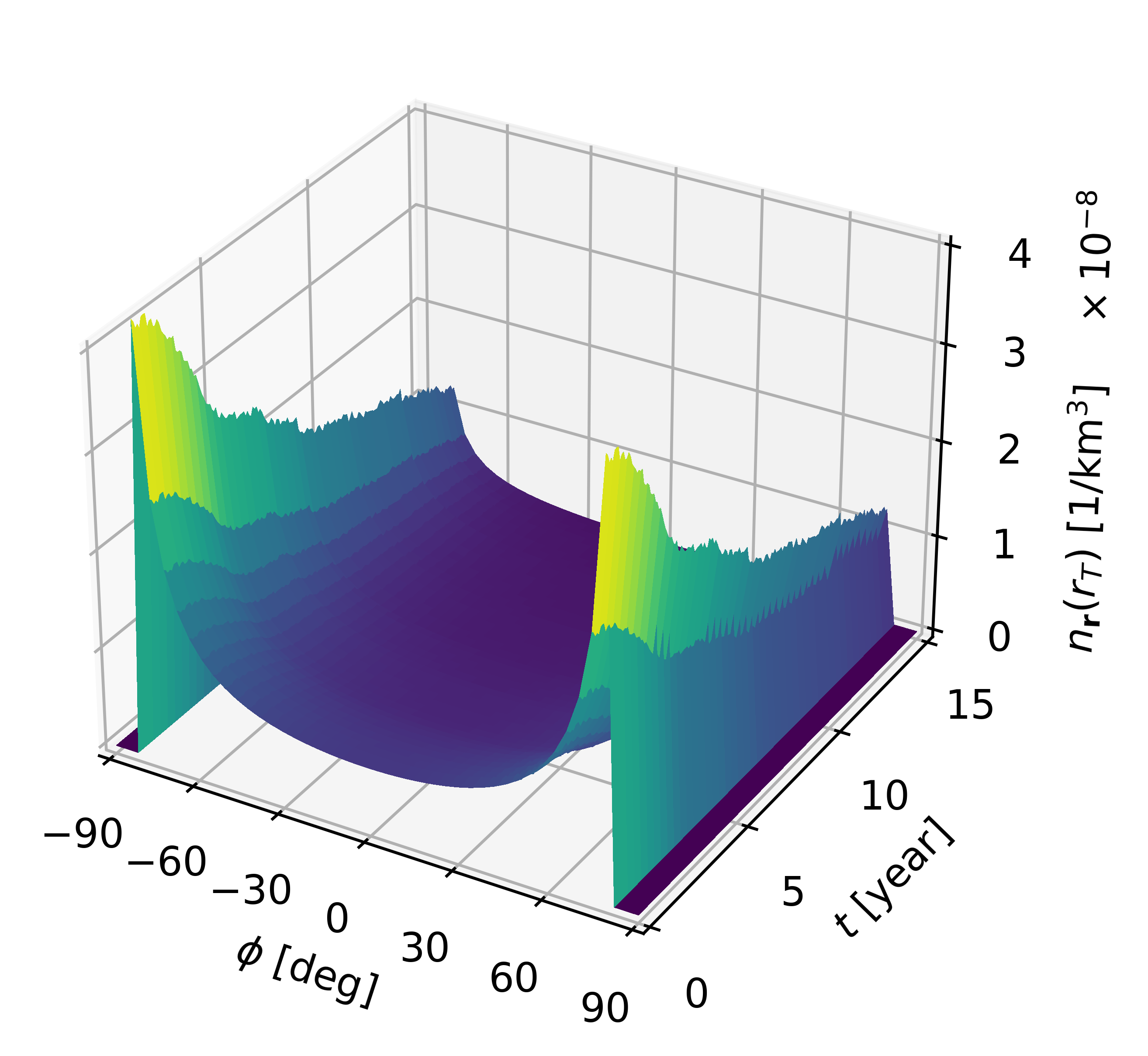}
         \caption{Dependency on latitude}
         \label{fig:NOAA-16_spatial_phi}
     \end{subfigure}
     \caption{NOAA-16 P/L fragmentation - Spatial density as function of altitude, latitude, and time.}
     \label{fig:NOAA-16_spatial_R_phi}
\end{figure*}

As it can be inferred, the different approach between the model in~\cite{Letizia2015b} and the newly proposed method dramatically changes the resulting profile of the impact velocity. It is worth noticing that, for a narrow range of values of the target argument of latitude $u_T$, no solution is found. Indeed, since the target covers a wider range in latitude, because of its lower inclination with respect to the parent orbit, there exist values of $u_T$ for which the target latitude is greater than the fragments one, for any value of the fragments argument of latitude $u$. As a result, no intersection is geometrically possible. Figure~\ref{fig:NOAA-16_spatial_R_phi} depicts the evolution of the spatial density function over time, as function of altitude $h$ and latitude $\phi$.

\begin{figure}[!ht]
     \centering
     \includegraphics[width=0.5\textwidth]{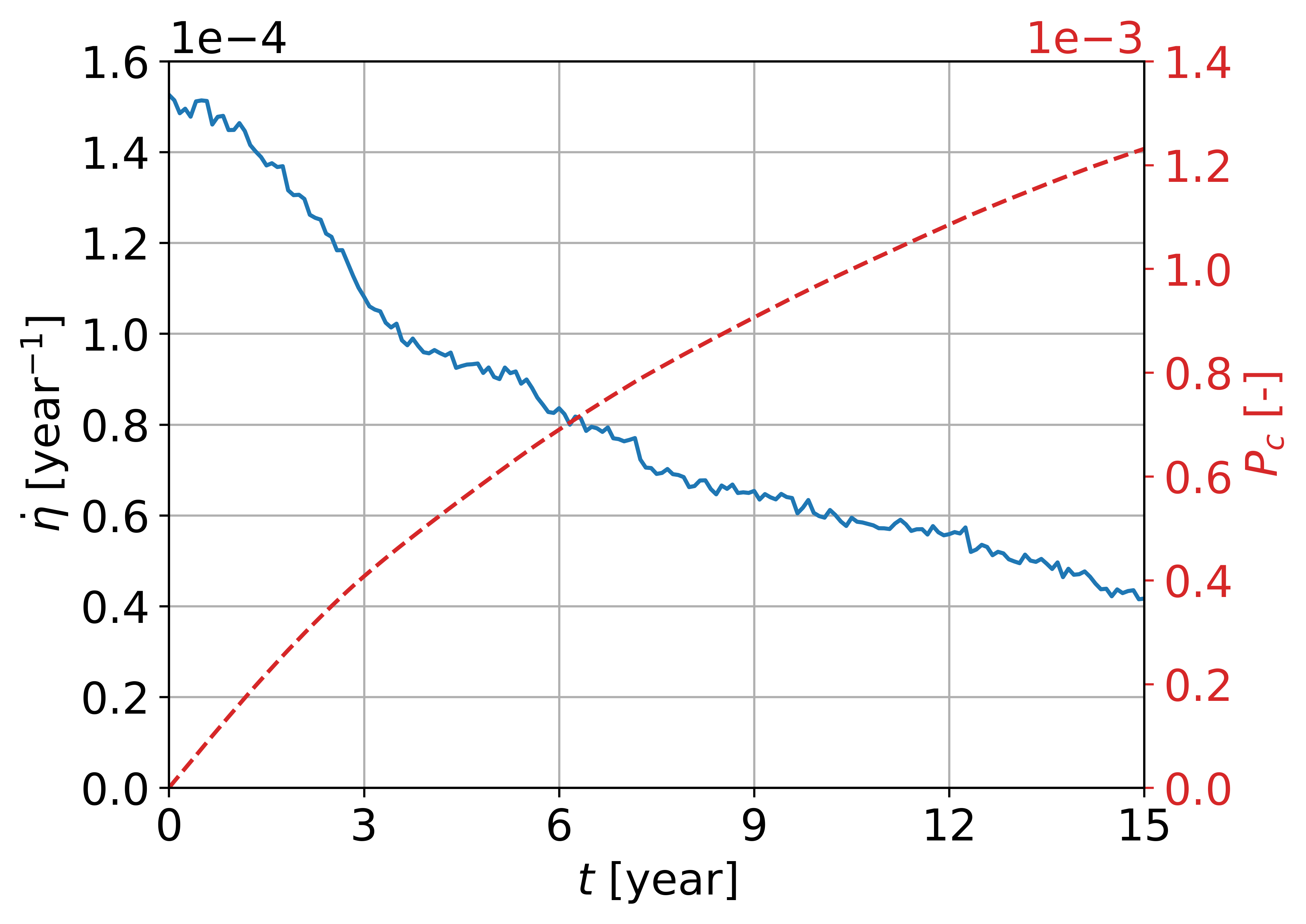}
     \caption{NOAA-16 P/L fragmentation - Impact rate and collision probability with SL-6 R/B over time from the 1D spatial density function.}
     \label{fig:NOAA-16_Pc1D}
\end{figure}

Combing the results of Figure~\ref{fig:NOAA-16_vrel} and Figure~\ref{fig:NOAA-16_spatial_R_phi}, the profile of the impact rate can be obtained and, as a consequence, the estimated collision probability, according to the Poisson distribution of Eq.~(\ref{eq:Pc}). The two profiles as function of time are shown in Figure~\ref{fig:NOAA-16_Pc1D}.

As it can be noted, the decrease of the impact rate over time comes as a consequence of the lowering of the spatial density function at $r=r_T$, caused by the effect of atmospheric drag.

\subsubsection{Collision risk from a 3D debris cloud in Keplerian elements a, e, i}
\label{Collision risk from a 3D debris cloud in Keplerian elements a, e, i}

\begin{figure*}[!ht]
     \centering
     \begin{subfigure}[b]{0.45\textwidth}
         \centering
         \includegraphics[width=\textwidth]{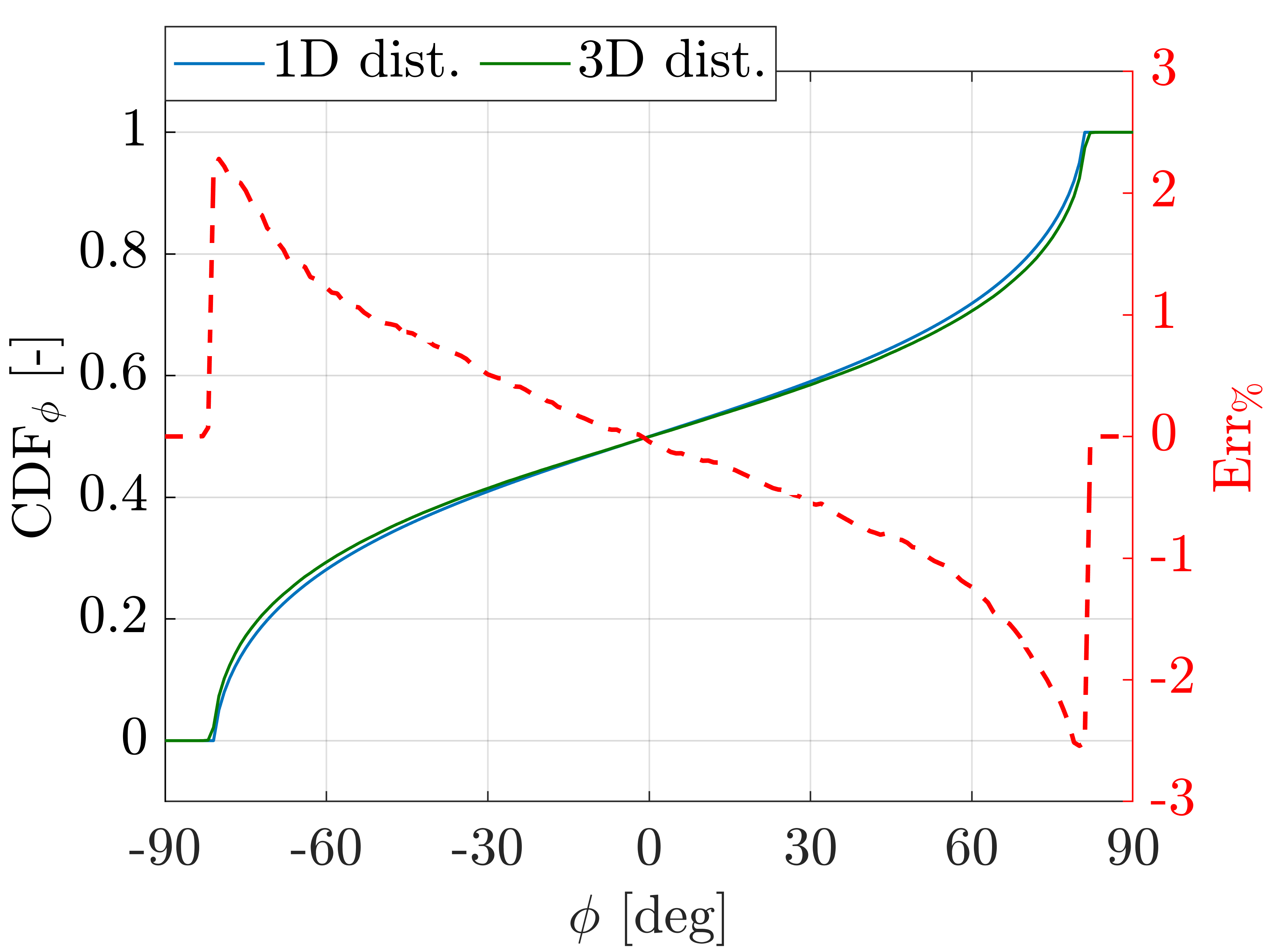}
         \caption{Normalized cumulative number of fragments as function of latitude}
         \label{fig:NOAA-16_3Dvs1D_CDFphi}
     \end{subfigure}
     \begin{subfigure}[b]{0.45\textwidth}
         \centering
         \includegraphics[width=\textwidth]{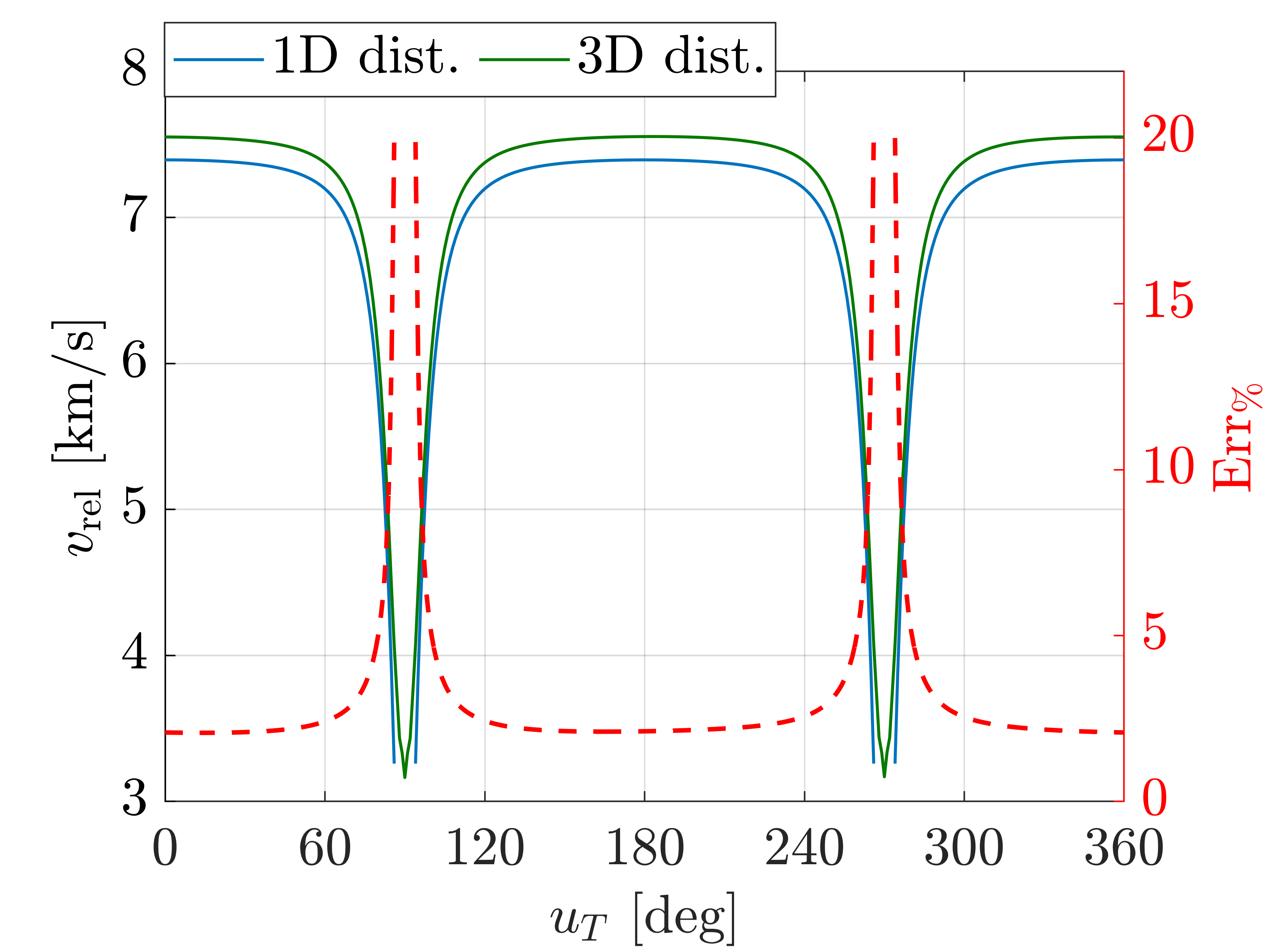}
         \caption{Impact velocity as function of target argument of latitude}
         \label{fig:NOAA-16_3Dvs1D_dv}
     \end{subfigure}
     \caption{NOAA-16 P/L fragmentation - Comparison between 1D and 3D formulations.}
     \label{fig:NOAA-16_3Dvs1D}
\end{figure*}

The 4D density distribution in Keplerian elements of Section~\ref{NOAA-16 debris cloud evolution} is here randomized over right ascension of the ascending node. Therefore, the associated spatial density function is constant over longitude $\lambda$. Nevertheless, two main improvements are added with respect to the solution proposed in Section~\ref{Collision risk from a 1D debris cloud in orbital radius r}, namely:
\begin{itemize}
    \item[-] The characterization of the fragments in inclination allows the accurate estimation of the spatial distribution over latitude $\phi$.
    \item[-] From the distribution of the fragments in the independent Keplerian elements $(a,e,i)$, the impact velocity can be computed as a discrete function of both phase space and time. As demonstrated in Section~\ref{Estimate of the model error}, if the grid is fine enough, the accuracy of the model is guaranteed.
\end{itemize}
Figure~\ref{fig:NOAA-16_3Dvs1D} shows the effect of the additional complexity. In particular, Figure~\ref{fig:NOAA-16_3Dvs1D_CDFphi} displays the normalized cumulative distribution of fragments at fragmentation epoch as function of the latitude $\phi$, $\mathrm{CDF}_\phi$, resulting from the different modelling of the debris cloud. It is computed as follows.
\begin{equation}
    \mathrm{CDF}_\phi=\frac{1}{N}\int_{-\pi/2}^\phi\int_0^{2\pi}\int_{R_E}^\infty n_{\bm{r}}\,\mathrm{d}r\mathrm{d}\lambda\mathrm{d}\phi
\end{equation}
with $N$ total number of fragments and $R_E$ Earth radius. Note that the dashed red line is the percentage error of the 1D formulation with respect to the 3D model, computed according to the following relation:
\begin{equation}
    \mathrm{Err}^{\mathrm{CDF}_\phi}_\%=100\cdot(\mathrm{CDF}_\phi^{\mathrm{3D}}-\mathrm{CDF}_\phi^{\mathrm{1D}})
\end{equation}

As it can be observed, the 1D formulation underestimates the number of fragments at high latitudes; this result is expected, as all the fragments are assumed to share the same inclination as the parent one, which constrains them to distribute over the following range in latitude:
\begin{equation}
\begin{cases}
    \Delta\phi=[-i,i]\qquad\mathrm{if}\;i\leq\frac{\pi}{2}\\
    \Delta\phi=[i-\pi,\pi-i]\qquad\mathrm{if}\;i>\frac{\pi}{2}
\end{cases}
\end{equation}
On the contrary, as part of the fragments are actually injected on orbits with lower inclination with respect to the parent one (Figure~\ref{fig:NOAA-16_dist0}), the 3D distribution covers a wider range in latitude. As a result, since the total integral of the density is preserved between the two formulations, the 1D model overestimates the number of fragments at low latitudes. 

Instead, Figure~\ref{fig:NOAA-16_3Dvs1D_dv} shows a comparison in terms of estimated average impact velocity as function of the target argument of latitude $u_T$. As it can be noted, the characterization of the fragments over inclination allows the debris cloud to potentially impact the target, for any value of $u_T$. It is also worth observing that the 3D model estimates an average impact velocity higher than the 1D model, for any target position along its orbit. The error on the relative velocity, represented by the red dashed line on the plot, is monitored through the following relation:
\begin{equation}
    \mathrm{Err}^{v_{\mathrm{rel}}}_{\%}=100\cdot\frac{v^{\mathrm{3D}}_{\mathrm{rel}}-v^{\mathrm{1D}}_{\mathrm{rel}}}{\max_{u_T}\left(v^{\mathrm{3D}}_{\mathrm{rel}}\right)}
\end{equation}

\begin{figure}[!ht]
     \centering
     \includegraphics[width=0.5\textwidth]{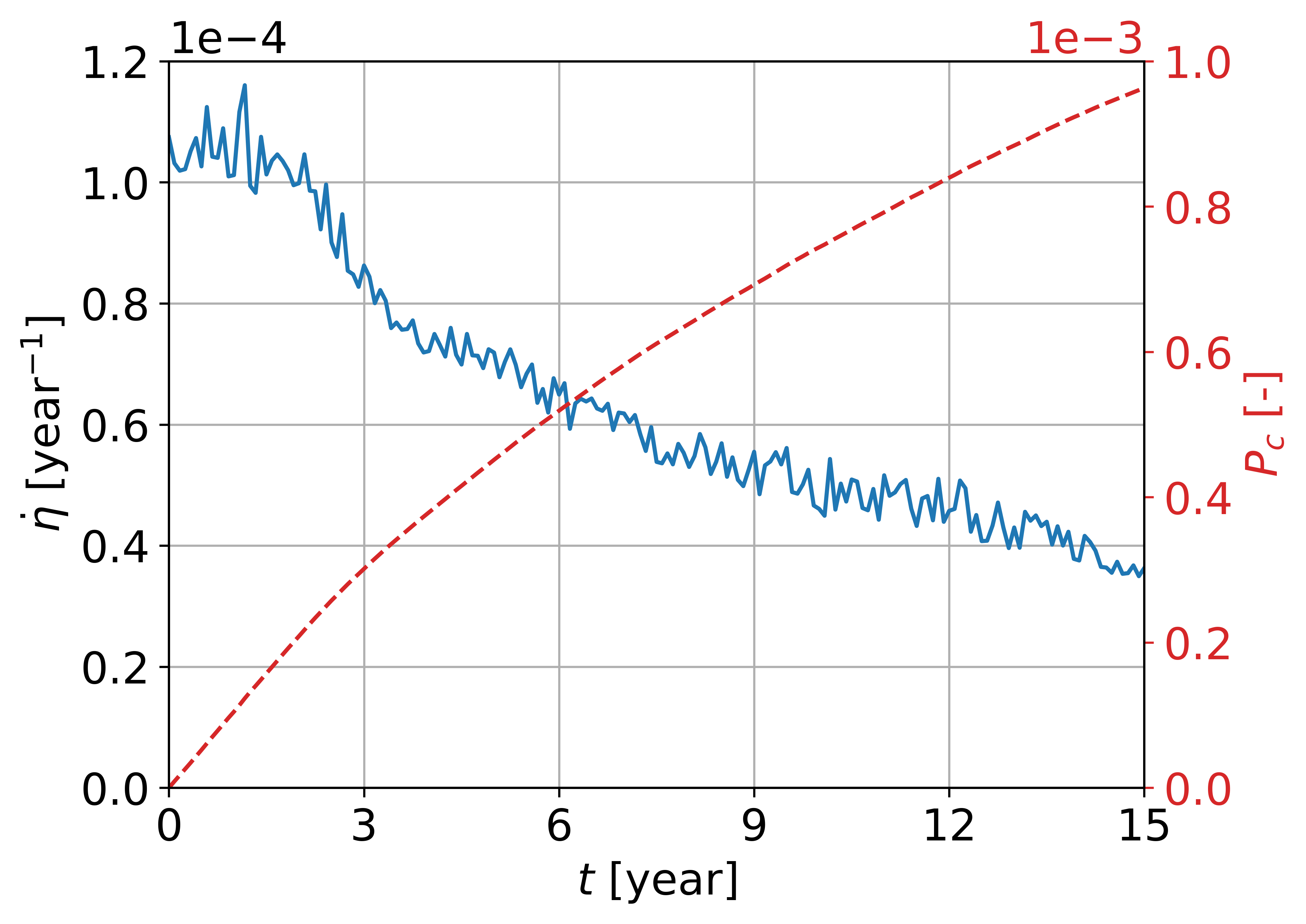}
     \caption{NOAA-16 P/L fragmentation - Impact rate and collision probability with SL-6 R/B over time from the 3D phase space density function.}
     \label{fig:NOAA-16_Pc3D}
\end{figure}

The impact rate is here computed directly from the density in Keplerian elements, according to Eq.~(\ref{eq:assumption}). Note that, as the distribution is randomized over all the Euler angles, the density value is independent of the dependent Keplerian elements $\bm{\beta}$; hence, the phase space density $n_{\bm{\alpha},\bm{\beta}}$ can be taken outside of the inner summation. Figure~\ref{fig:NOAA-16_Pc3D} shows the impact rate and the probability of collision as function of time.

As it can be observed, the predicted cumulative collision probability after 15 years is lower than the one predicted by the 1D model reported in Figure~\ref{fig:NOAA-16_Pc1D}. Again, the result was expected; indeed, the 3D formulation estimates a higher number of fragments at high latitudes, where the average impact velocity is low (the minima of the impact velocity $v_{\mathrm{rel}}$ are monitored at $u_T$ equal to either 90 deg or 270 deg). On the contrary, the high impact velocity region is occupied by a lower concentration of fragments with respect to the 1D model. As the impact rate comes as a combined effect of impact velocity and debris density, an overall lower probability of collision is estimated.

\subsubsection{Collision risk from a 4D debris cloud in Keplerian elements a, e, i, \texorpdfstring{$\Omega$}{\textOmega}}
\label{Collision risk from a 4D debris cloud in Keplerian elements a, e, i, Om}

\begin{figure}[!ht]
     \centering
     \includegraphics[width=0.5\textwidth]{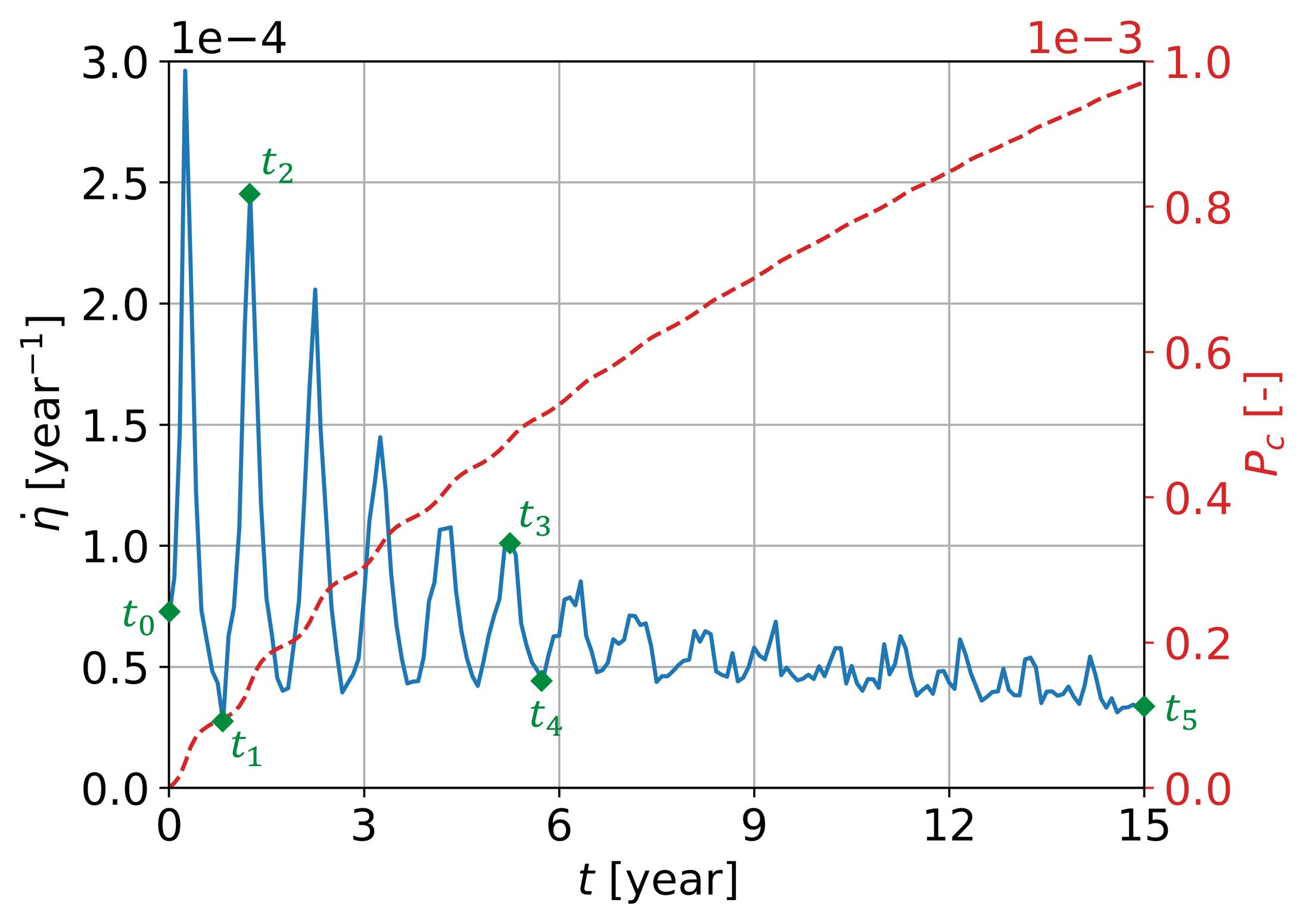}
     \caption{NOAA-16 P/L fragmentation - Impact rate and collision probability with SL-6 R/B over time from the 4D phase space density function.}
     \label{fig:NOAA-16_Pc4D}
\end{figure}

The collision probability is finally computed accounting for the debris distribution over right ascension of the ascending node. This means that for a given subset of Keplerian elements $\bm{\alpha}$ and fixed target position $\bm{r}_T$, there exist two possible density values $n_{\bm{\alpha},\bm{\beta}}$ associated to the orbits with right ascension $\Omega_1$, $\Omega_2$ of Eq.~(\ref{eq:beta^k}), guaranteeing intersection. As a result, the cloud evolution in $\Omega$ is expected to massively impact the profile of the impact rate. Note that the characterization of the debris cloud over right ascension of the ascending node implies that the fragments are no longer uniformly spread in longitude $\lambda$. Figure~\ref{fig:NOAA-16_Pc4D} depicts the impact rate and the probability of collision with rocket body SL-6 as function of time. 

\begin{figure*}[!ht]
     \centering
     \begin{subfigure}[b]{0.33\textwidth}
         \centering
         \includegraphics[width=\textwidth]{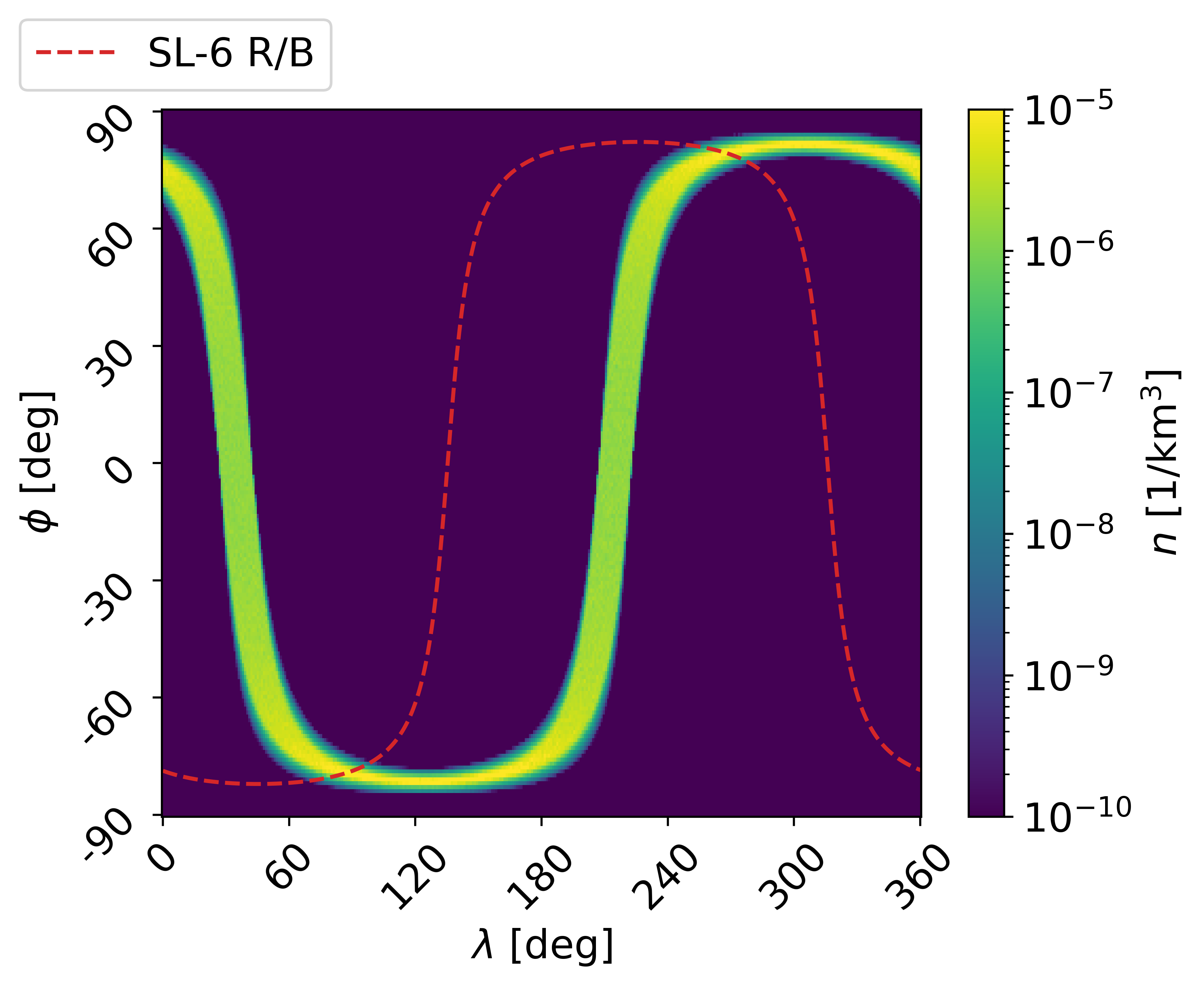}
         \caption{Epoch: $\bm{t_0}$}
         \label{fig:NOAA-16_spatial_t0}
     \end{subfigure}
     \begin{subfigure}[b]{0.33\textwidth}
         \centering
         \includegraphics[width=\textwidth]{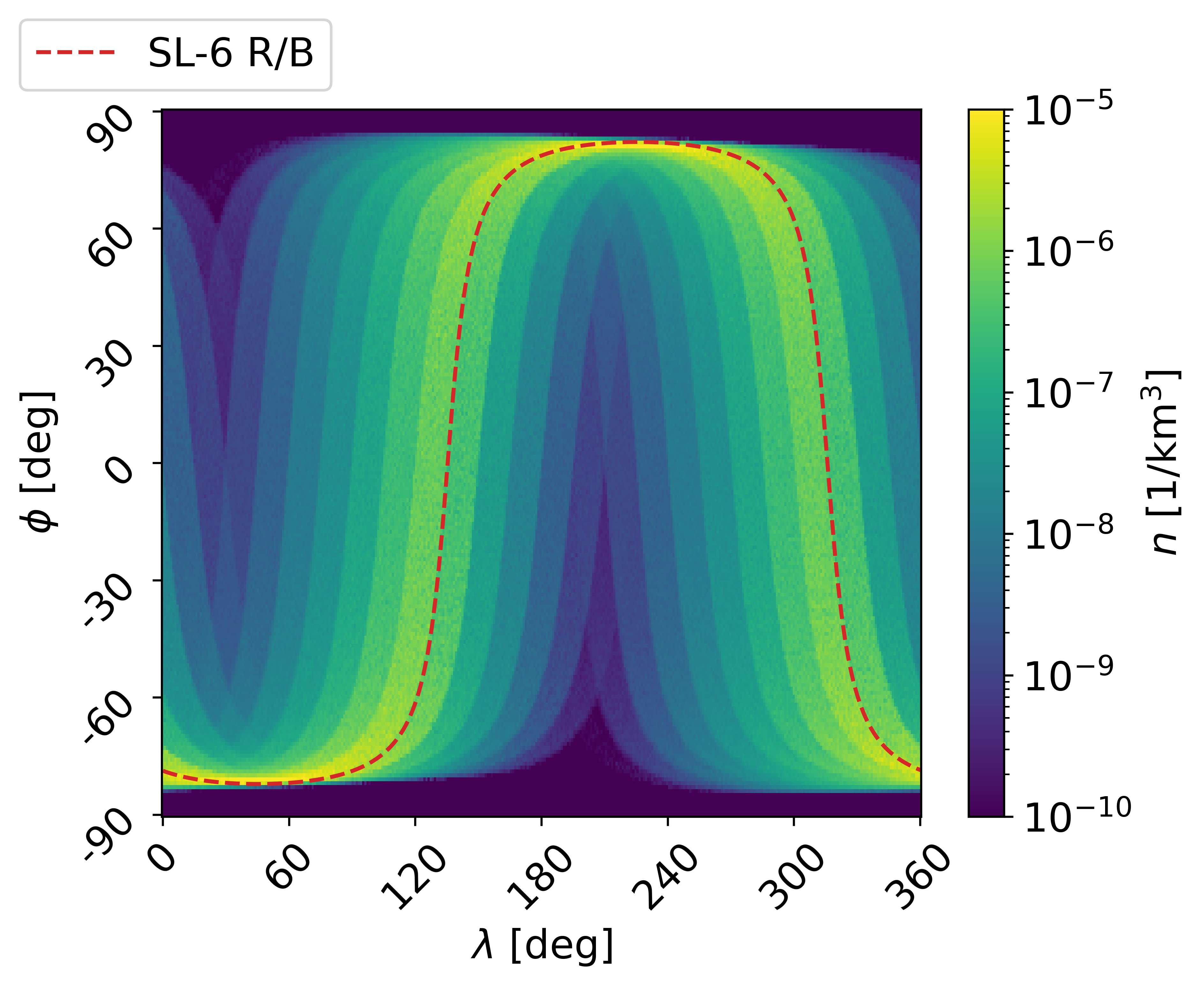}
         \caption{Epoch: $\bm{t_1}$}
         \label{fig:NOAA-16_spatial_t1}
     \end{subfigure}
     \begin{subfigure}[b]{0.33\textwidth}
         \centering
         \includegraphics[width=\textwidth]{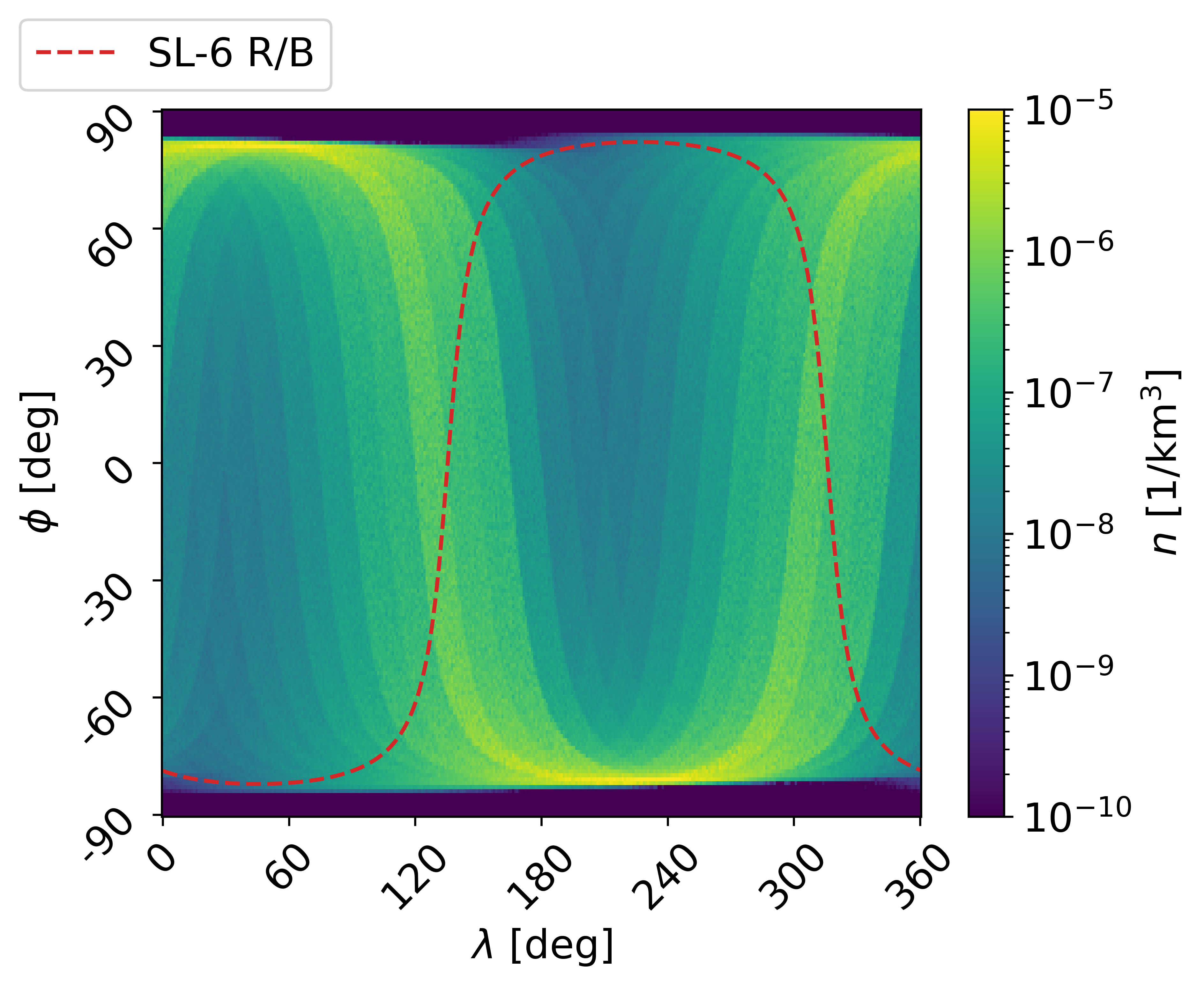}
         \caption{Epoch: $\bm{t_2}$}
         \label{fig:NOAA-16_spatial_t2}
     \end{subfigure}
     \begin{subfigure}[b]{0.33\textwidth}
         \centering
         \includegraphics[width=\textwidth]{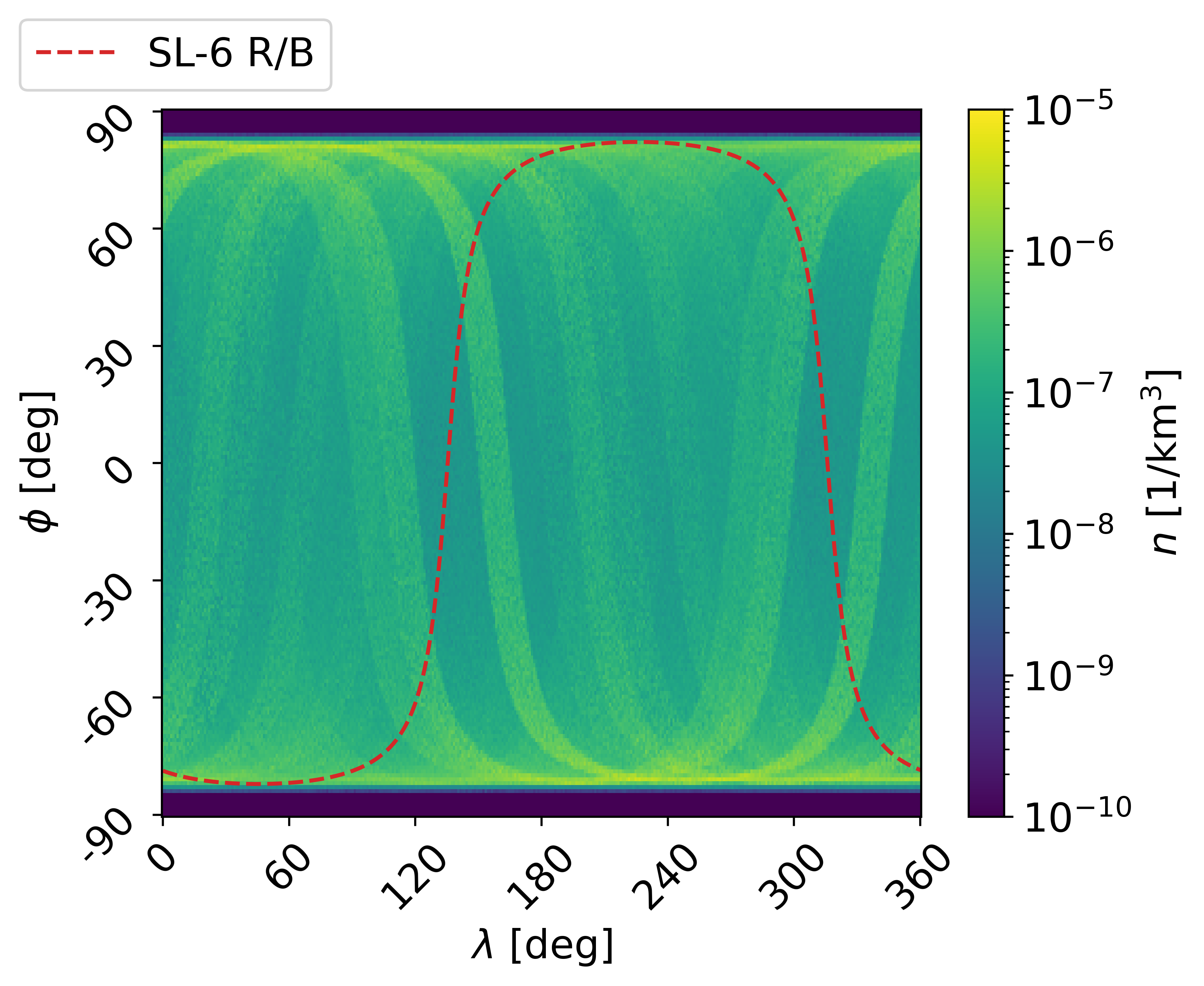}
         \caption{Epoch: $\bm{t_3}$}
         \label{fig:NOAA-16_spatial_t3}
     \end{subfigure}
     \begin{subfigure}[b]{0.33\textwidth}
         \centering
         \includegraphics[width=\textwidth]{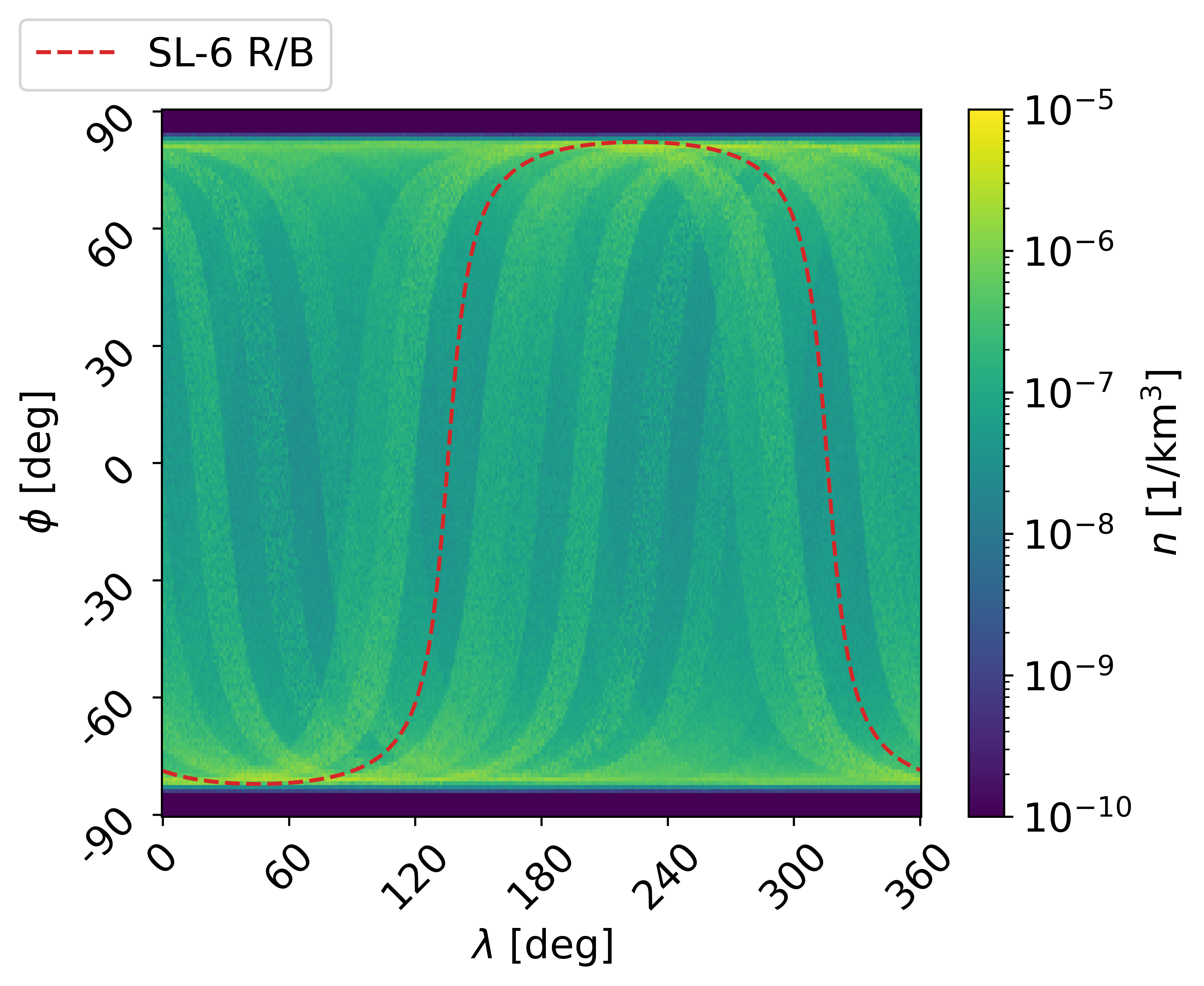}
         \caption{Epoch: $\bm{t_4}$}
         \label{fig:NOAA-16_spatial_t4}
     \end{subfigure}
     \begin{subfigure}[b]{0.33\textwidth}
         \centering
         \includegraphics[width=\textwidth]{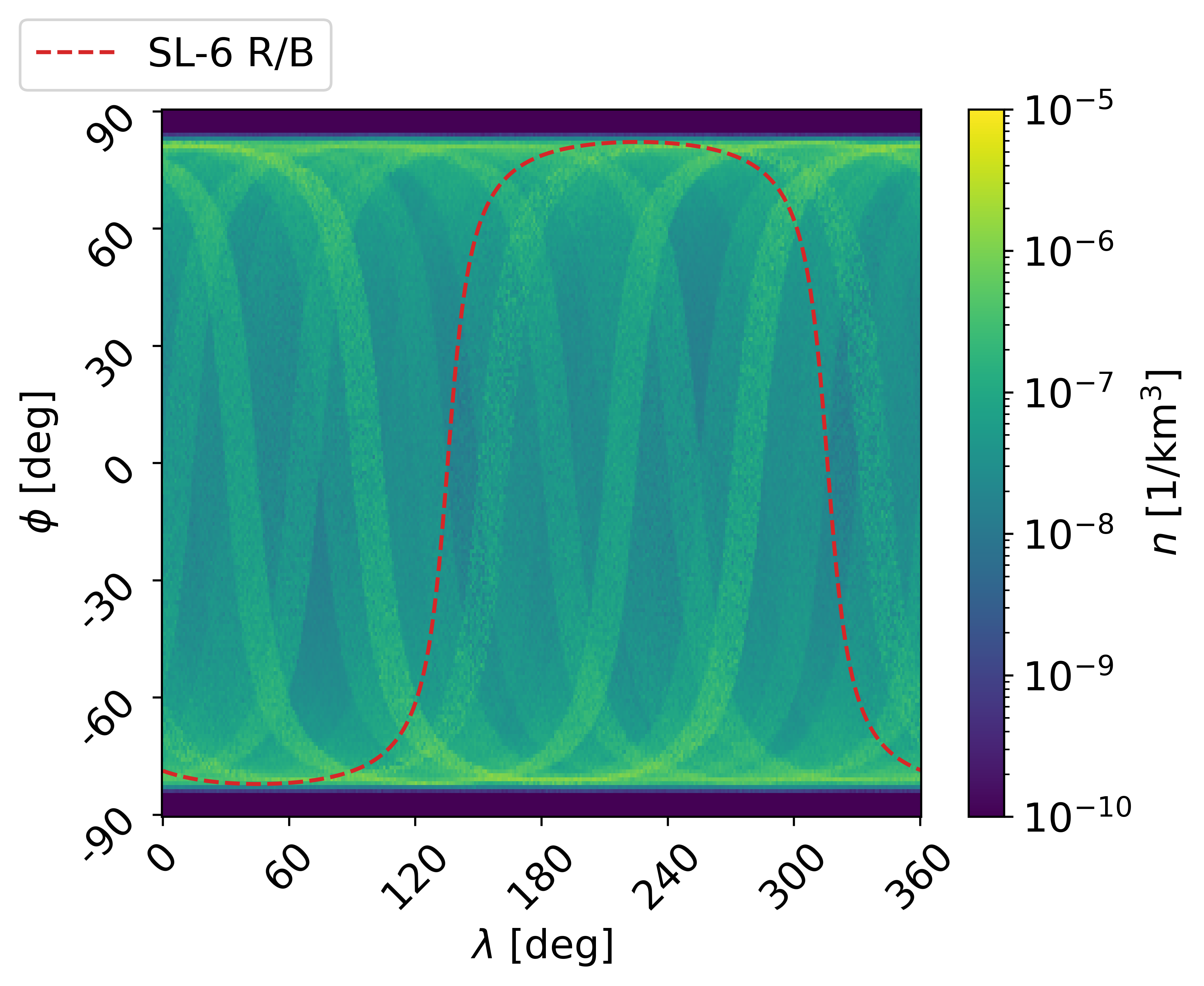}
         \caption{Epoch: $\bm{t_5}$}
         \label{fig:NOAA-16_spatial_t5}
     \end{subfigure}
     \caption{NOAA-16 P/L fragmentation - Spatial density as function of longitude $\bm{\lambda}$ and latitude $\phi$ over time.}
     \label{fig:NOAA-16_spatial_long_lat}
\end{figure*}

As it can be observed, the profile of the impact rate has dramatically changed. It is now characterized by an oscillatory behaviour over time with a characteristic period of roughly 1 year, whose amplitude decreases over time. The reason of these oscillations can be understood by looking at the evolution of the spatial density as function of longitude $\lambda$ and latitude $\phi$ over time, relative to the fixed target orbit, reported in Figure~\ref{fig:NOAA-16_spatial_long_lat}. Note that the distributions refer to the epochs $t_0,\dots,t_5$ highlighted in Figure~\ref{fig:NOAA-16_Pc4D}.

The randomization process over right ascension of the ascending node is clearly observable. Note that at epoch $t_2$ a unique high-density region still stands out from the fragments distribution. After the 15 years propagation, even though the distribution is more uniform over longitude, some high-density orbits can be still distinguished. The maxima in the impact rate profile (epochs $t_2$ and $t_3$) verify when the most crowded orbits have a shift of 180 deg in right ascension of the ascending node $\Omega$ with respect to the target. Instead, when they share the same $\Omega$ as the target (epochs $t_1$ and $t_4$) a minimum is found in the impact rate evolution. Note that the maxima (and minima) occur every 1 year because, since the fragments move on a quasi-SSO orbit, the rate of change in $\Omega$ is roughly 360 deg/year. The decrease in the amplitude of oscillation is a consequence of the randomization over right ascension, which causes the target to move over an approximately uniformly distributed debris cloud in longitude.

\subsection{Effects of the AMC 14 BRIZ-M fragmentation in high-elliptical orbit}
\label{Effects of the AMC 14 BRIZ-M fragmentation in high-elliptical orbit}

The fragmentation event occurred at 05:53 GMT on 13$^\mathrm{th}$ October 2010, 31 months after launch, when the rocket body was orbiting on a highly elliptical orbit~\cite{FragHistory}. The cause was identified in a malfunction of the propulsion system. The Space Surveillance Network (SSN) catalogued 106 large fragments, even though they were supposed to be many more~\cite{FragHistory}. Indeed, the parent orbit perigee was located in the southern hemisphere, out of the SSN coverage, which made the tracking of the ejected fragments difficult. The Keplerian elements of the rocket body at fragmentation epoch are reported in Table~\ref{tab:kep BRIZ-M}. The object had a mass of 2510 kg~\cite{FragHistory}.

\begin{table}[hbt!]
\caption{\label{tab:kep BRIZ-M} AMC 14 BRIZ-M R/B Keplerian elements at fragmentation epoch.}
\centering
\begin{tabular}{cccccc}
\hline
$a$ [km] & $e$ [-] & $i$ [deg] & $\Omega$ [deg] & $\omega$ [deg] & $f$ [deg]\\
\hline
19981 & 0.64859 & 48.94 & 195.24 & 287.15 & 31.97\\
\hline
\end{tabular}
\end{table}

\subsubsection{AMC 14 BRIZ-M debris cloud evolution}
\label{AMC 14 BRIZ-M debris cloud evolution}

\begin{figure}[!ht]
     \centering
     \includegraphics[width=0.6\textwidth]{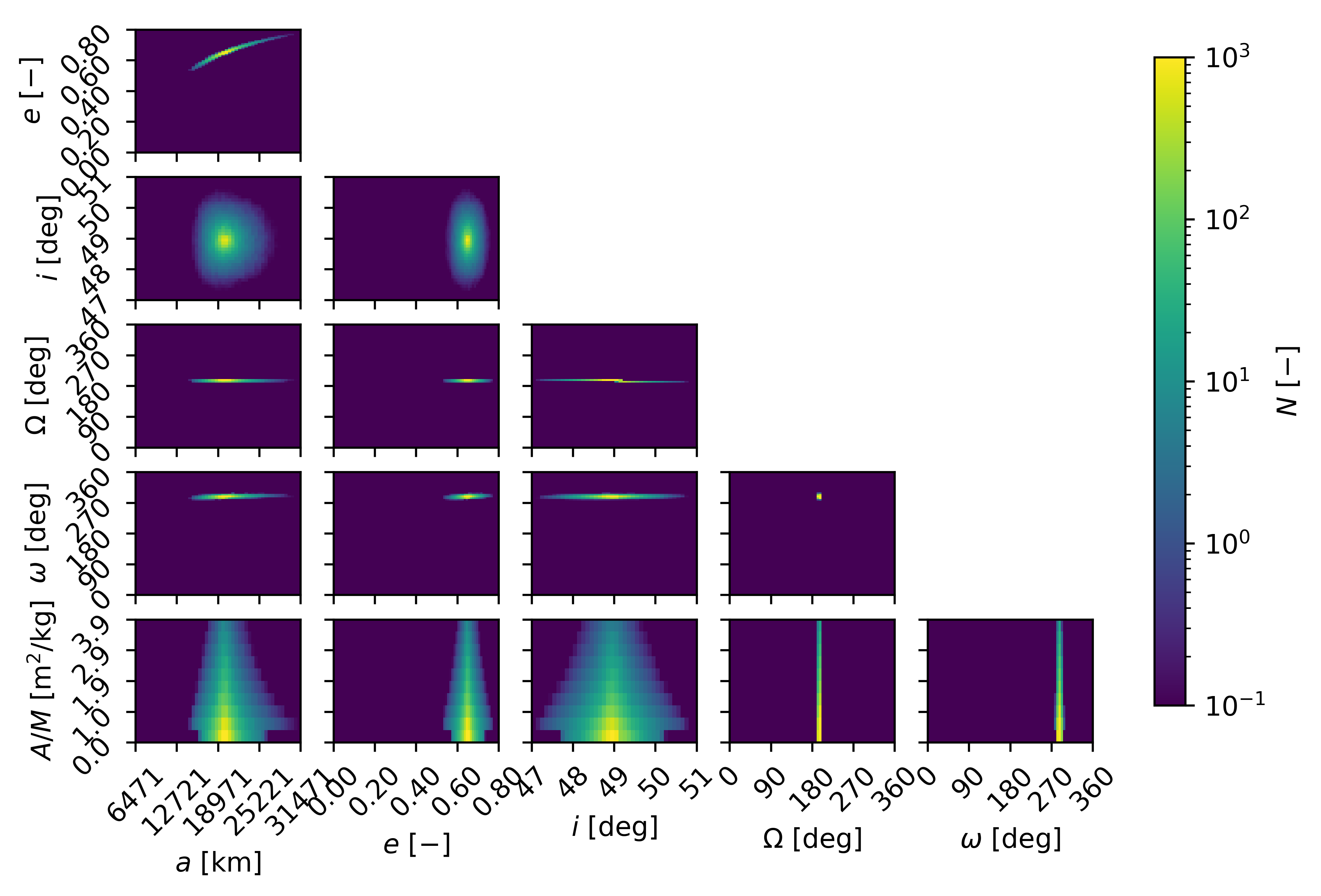}
     \caption{AMC 14 BRIZ-M R/B fragmentation - Density distribution in \texorpdfstring{$\bm{(a,e,i,\Omega,\omega,A/M})$}{vars} at fragmentation epoch.}
     \label{fig:BRIZ-M_dist0}
\end{figure}

As for the previous fragmentation scenario, fragments in the range 1 cm - 1 m are considered. According to the object mass and type, the explosion results in 9503 ejected fragments~\cite{Krisko2011}. Figure~\ref{fig:BRIZ-M_dist0} shows the initial density distribution in the phase space of slow-varying orbital elements $(a,e,i,\Omega,\omega)$ and area-to-mass ratio $A/M$. As always, the debris cloud is assumed to be randomized over mean anomaly. 

\begin{figure*}[!ht]
     \centering
     \begin{subfigure}[b]{0.45\textwidth}
         \centering
         \includegraphics[width=\textwidth]{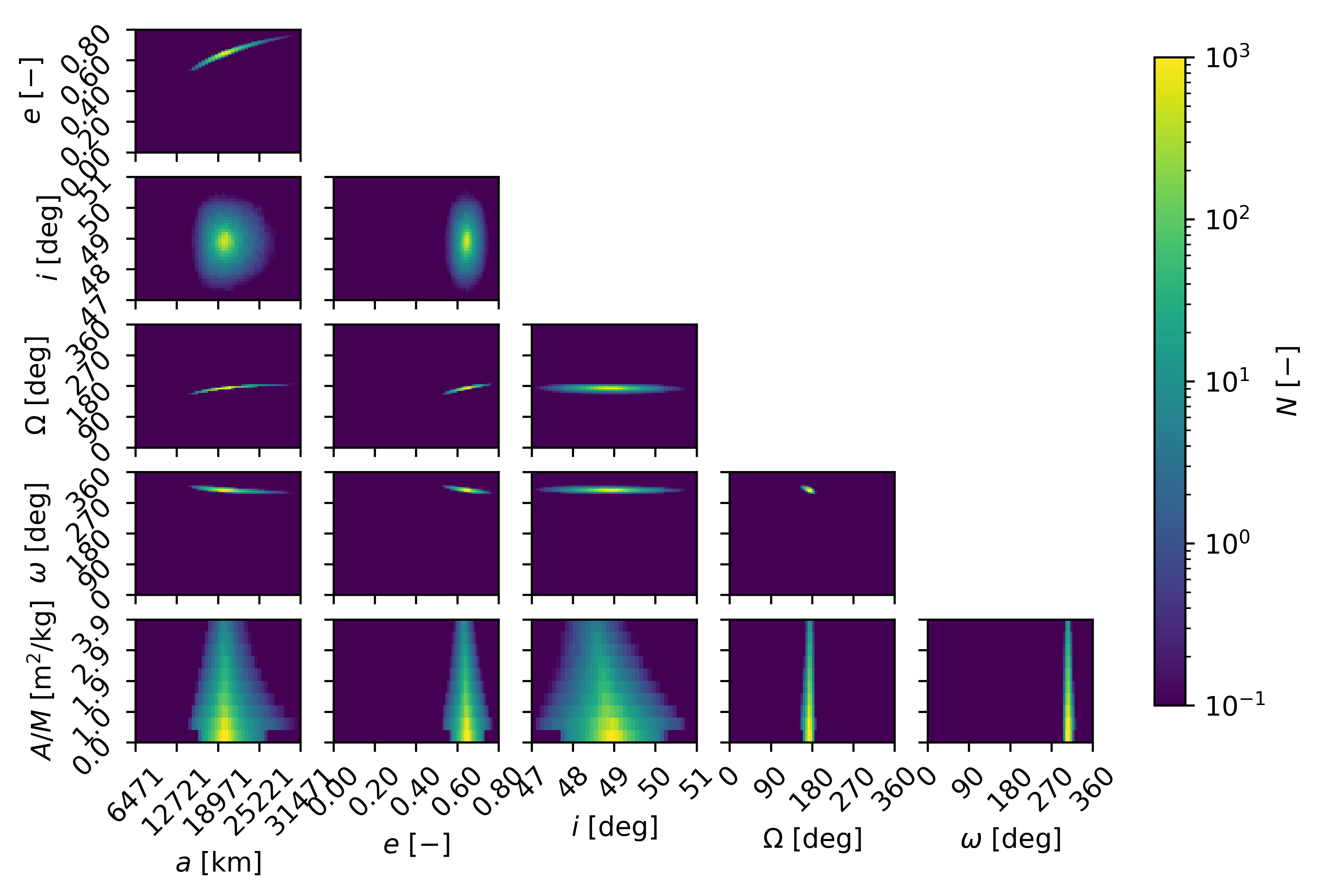}
         \caption{Epoch: 2 months after fragmentation}
         \label{fig:BRIZ-M_dist1}
     \end{subfigure}
     \begin{subfigure}[b]{0.45\textwidth}
         \centering
         \includegraphics[width=\textwidth]{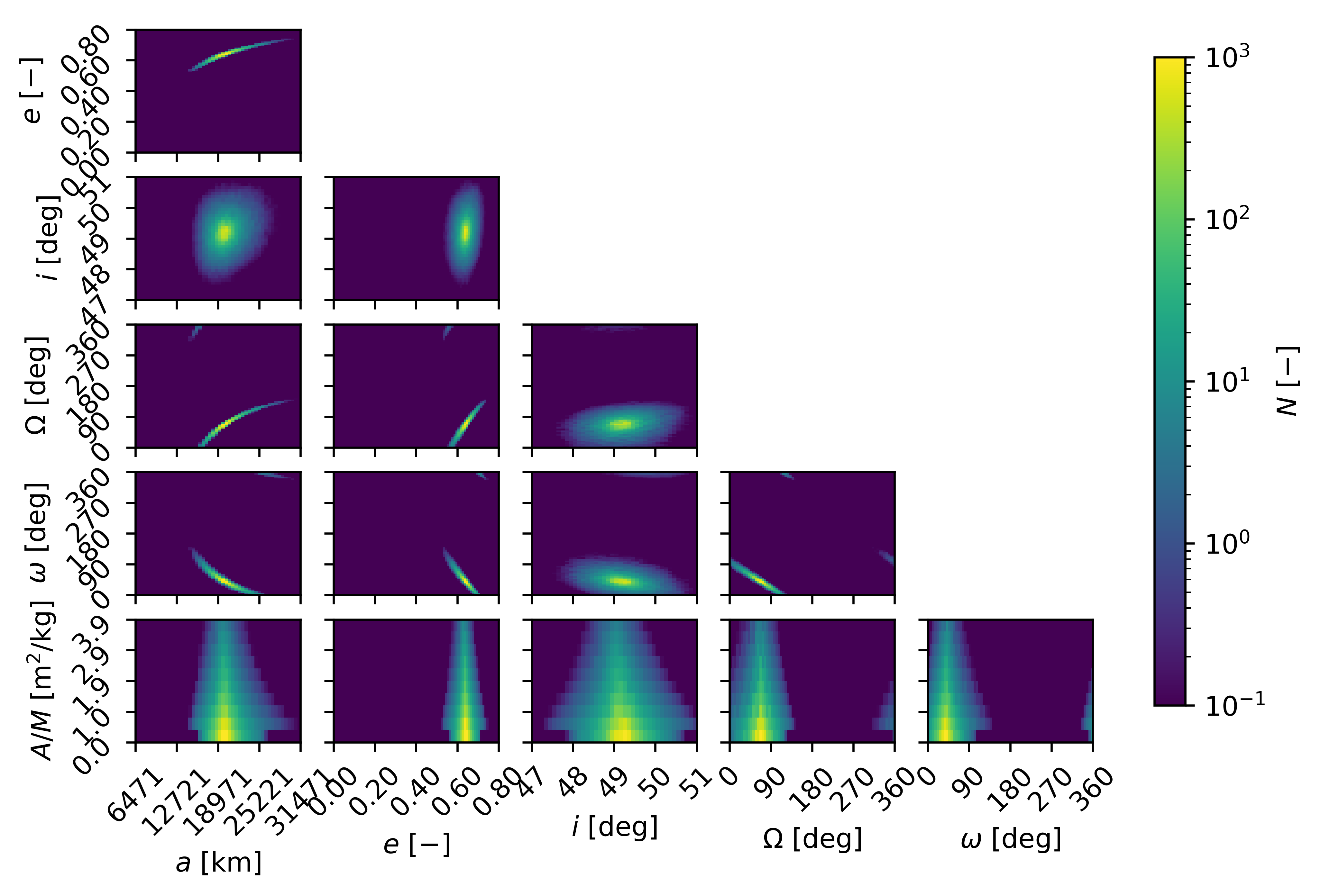}
         \caption{Epoch: 1 year after fragmentation}
         \label{fig:BRIZ-M_dist2}
     \end{subfigure}\\
     \begin{subfigure}[b]{0.45\textwidth}
         \centering
         \includegraphics[width=\textwidth]{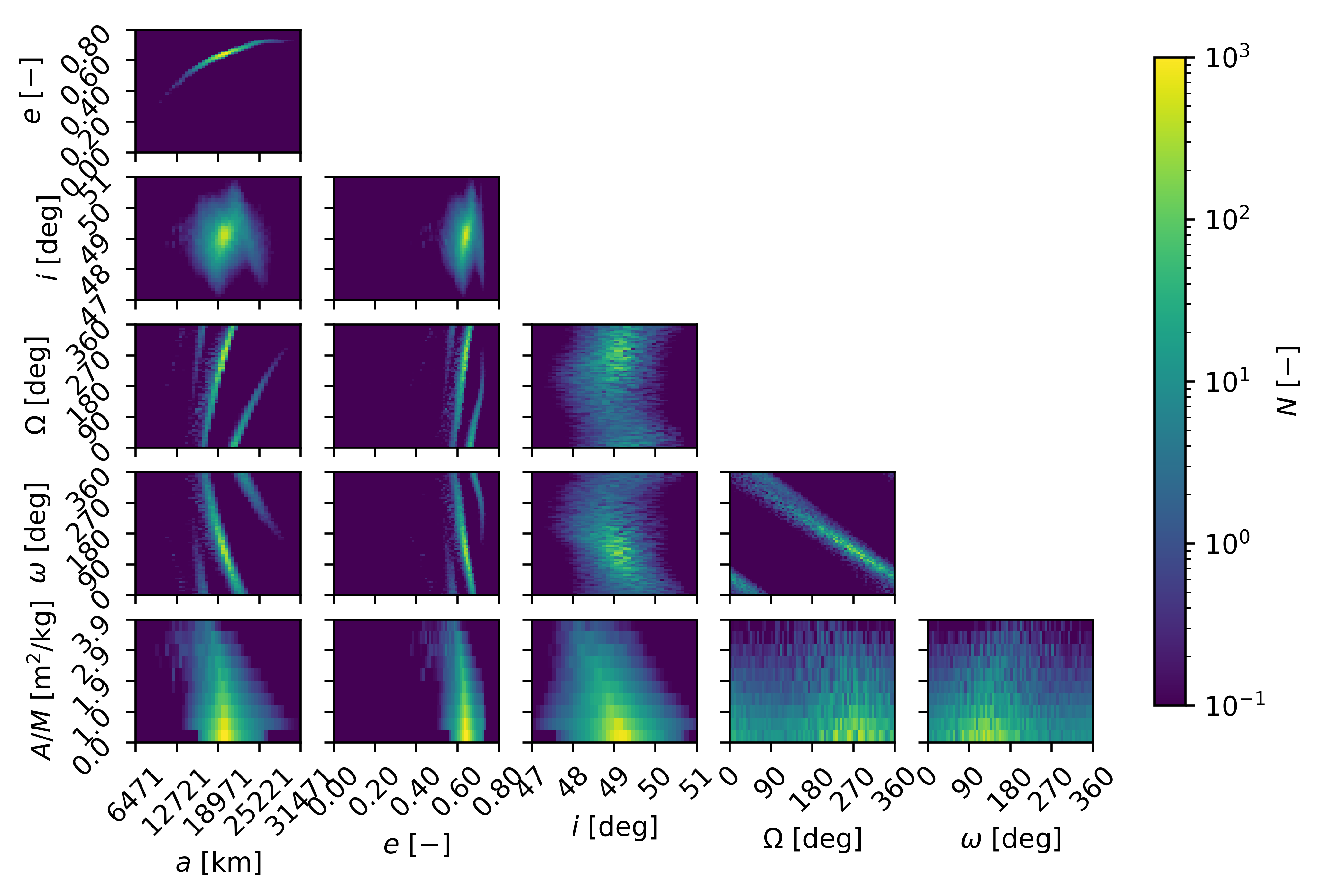}
         \caption{Epoch: 5 years after fragmentation}
         \label{fig:BRIZ-M_dist3}
     \end{subfigure}
     \begin{subfigure}[b]{0.45\textwidth}
         \centering
         \includegraphics[width=\textwidth]{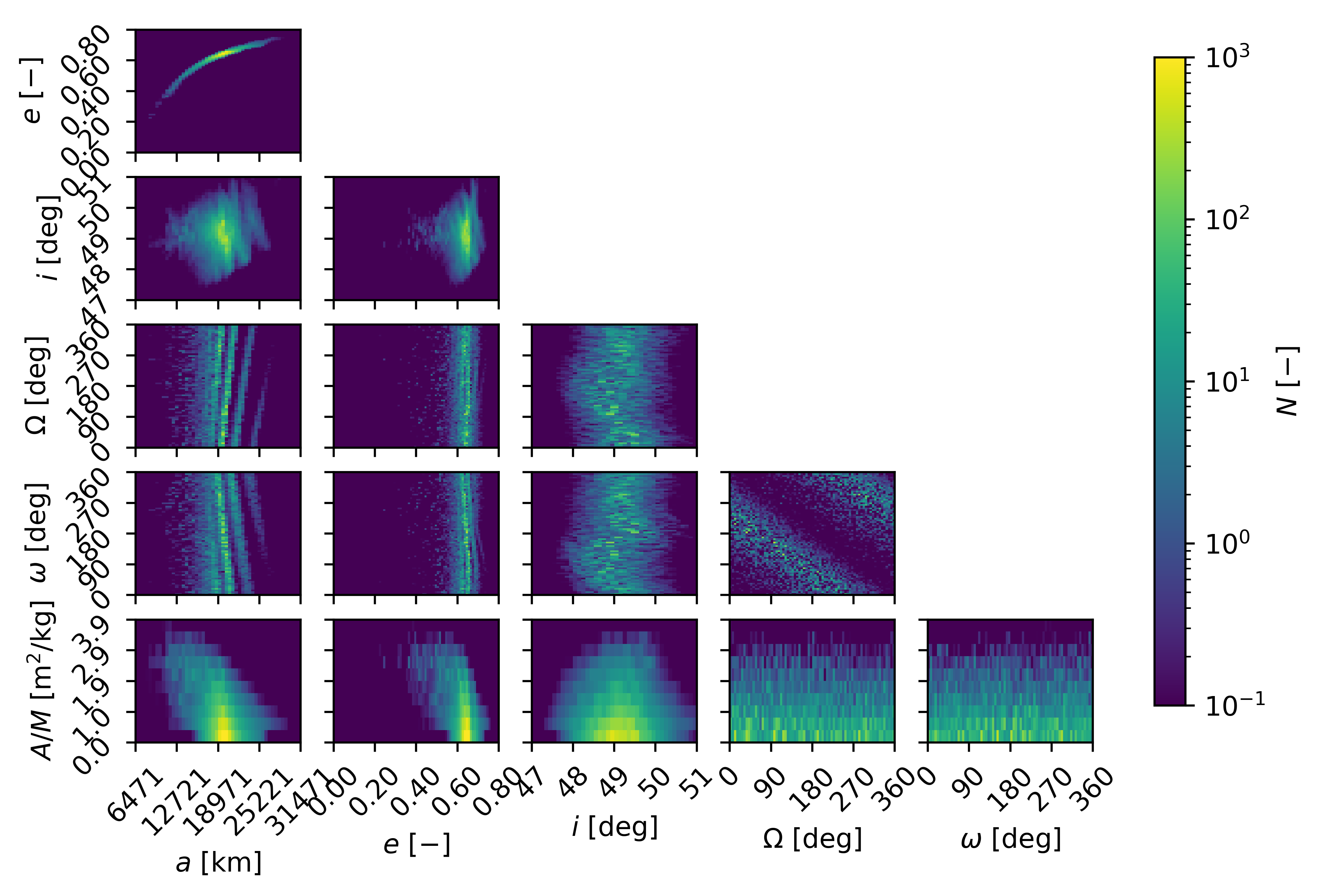}
         \caption{Epoch: 15 years after fragmentation}
         \label{BRIZ-M_dist4}
     \end{subfigure}
     \caption{AMC 14 BRIZ-M R/B fragmentation - Density distribution in \texorpdfstring{$\bm{(a,e,i,\Omega,\omega,A/M})$}{vars} over time.}
     \label{fig:BRIZ-M_dist_prop}
\end{figure*}

As it can be observed, the cloud covers a much wider range in semi-major axis and eccentricity; indeed, the higher is the parent specific orbital energy, the easier it is to modify the orbit shape. Instead, the fragments occupy a tiny domain over both right ascension of the ascending node and argument of periapsis. Figure~\ref{fig:BRIZ-M_dist_prop} depicts the fragments distribution 2 months, 1 year, 5 and 15 years after fragmentation, resulting from the propagation of the debris density along 20000 characteristic lines. The same force model as in Section~\ref{NOAA-16 debris cloud evolution} is considered.

It is worth underlining the following differences with respect to the fragmentation scenario described in Section~\ref{Effects of the NOAA-16 fragmentation in Sun-synchronous orbit}:
\begin{itemize}
    \item[-] The debris cloud is far from being randomized over the Euler angles, even after 15 years of propagation. On the contrary, in the $(\Omega,\omega)$ phase space, the fragments distribute over a diagonal line, which gets thicker and thicker as time passes. As demonstrated in~\cite{Giudici2023}, the angular coefficient of this line-like distribution can be approximated as the ratio between the long-term rate of change $\dot{\Omega}$ and $\dot{\omega}$, caused by the $J_2$ perturbation.
    \item[-] Third-body perturbation and solar radiation pressure notably affect the cloud evolution in inclination.
    \item[-] Despite the low-altitude of the parent orbit perigee, only a small fraction of the debris cloud is considerably affected by atmospheric drag. As a result, only the fragments with the highest area-to-mass ratio reenter the atmosphere over the 15 years propagation time.
\end{itemize}

\subsubsection{Collision risk from a 5D debris cloud in Keplerian elements a, e, i, \texorpdfstring{$\Omega$}{\textOmega}, \texorpdfstring{$\omega$}{\textomega}}
\label{Collision risk from a 5D debris cloud in Keplerian elements a, e, i, Om, om}

\begin{figure}[!ht]
     \centering
     \includegraphics[width=0.5\textwidth]{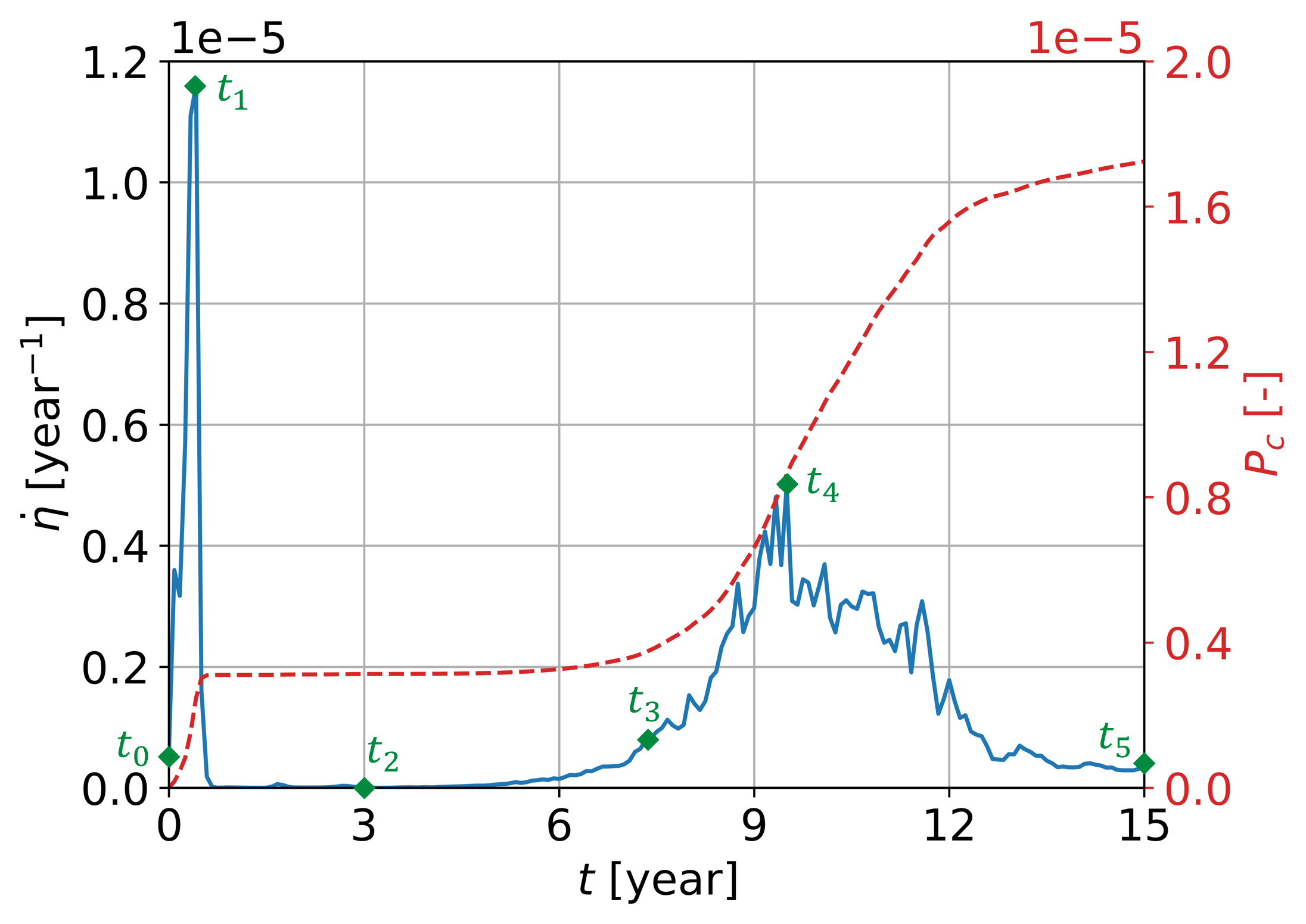}
     \caption{AMC 14 BRIZ-M R/B fragmentation - Impact rate and collision probability with SL-6 R/B over time from the 5D phase space density function.}
     \label{fig:BRIZ-M_Pc5D}
\end{figure}

\begin{figure*}[!ht]
     \centering
     \begin{subfigure}[b]{0.45\textwidth}
         \centering
         \includegraphics[width=\textwidth]{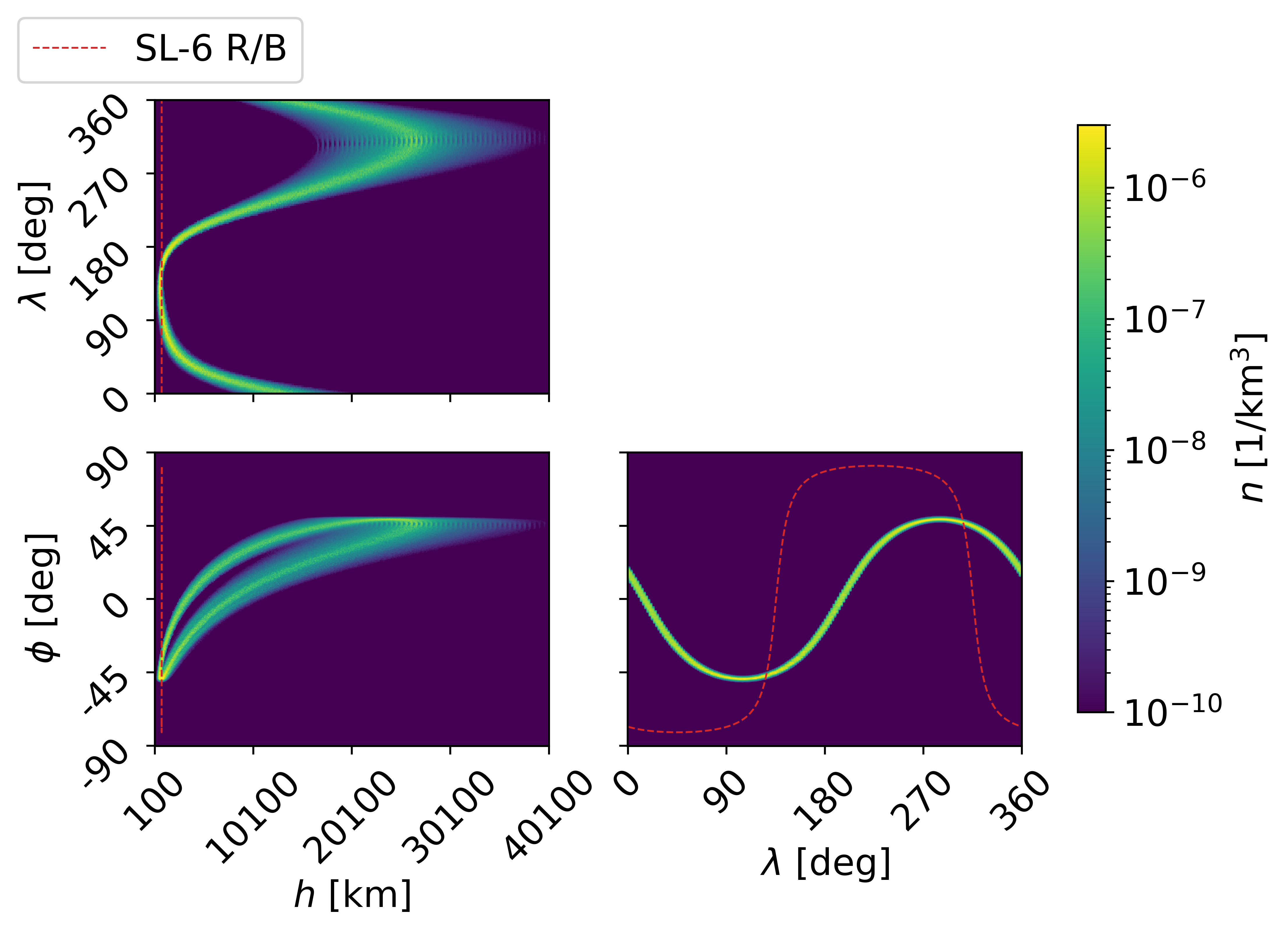}
         \caption{Epoch: $\bm{t_0}$}
         \label{fig:BRIZ-M_spatial_t0}
     \end{subfigure}
     \begin{subfigure}[b]{0.45\textwidth}
         \centering
         \includegraphics[width=\textwidth]{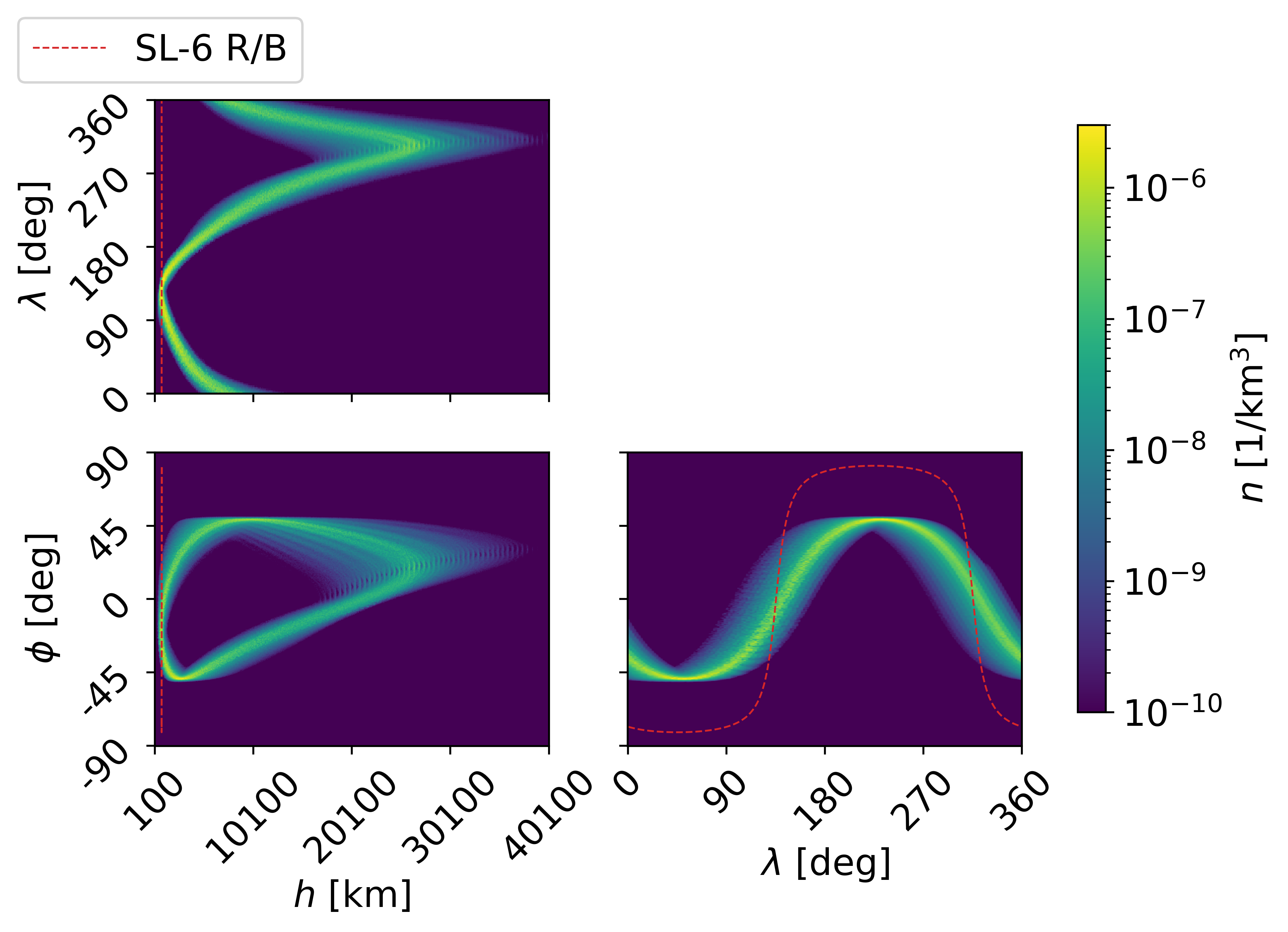}
         \caption{Epoch: $\bm{t_1}$}
         \label{fig:BRIZ-M_spatial_t1}
     \end{subfigure}
     \begin{subfigure}[b]{0.45\textwidth}
         \centering
         \includegraphics[width=\textwidth]{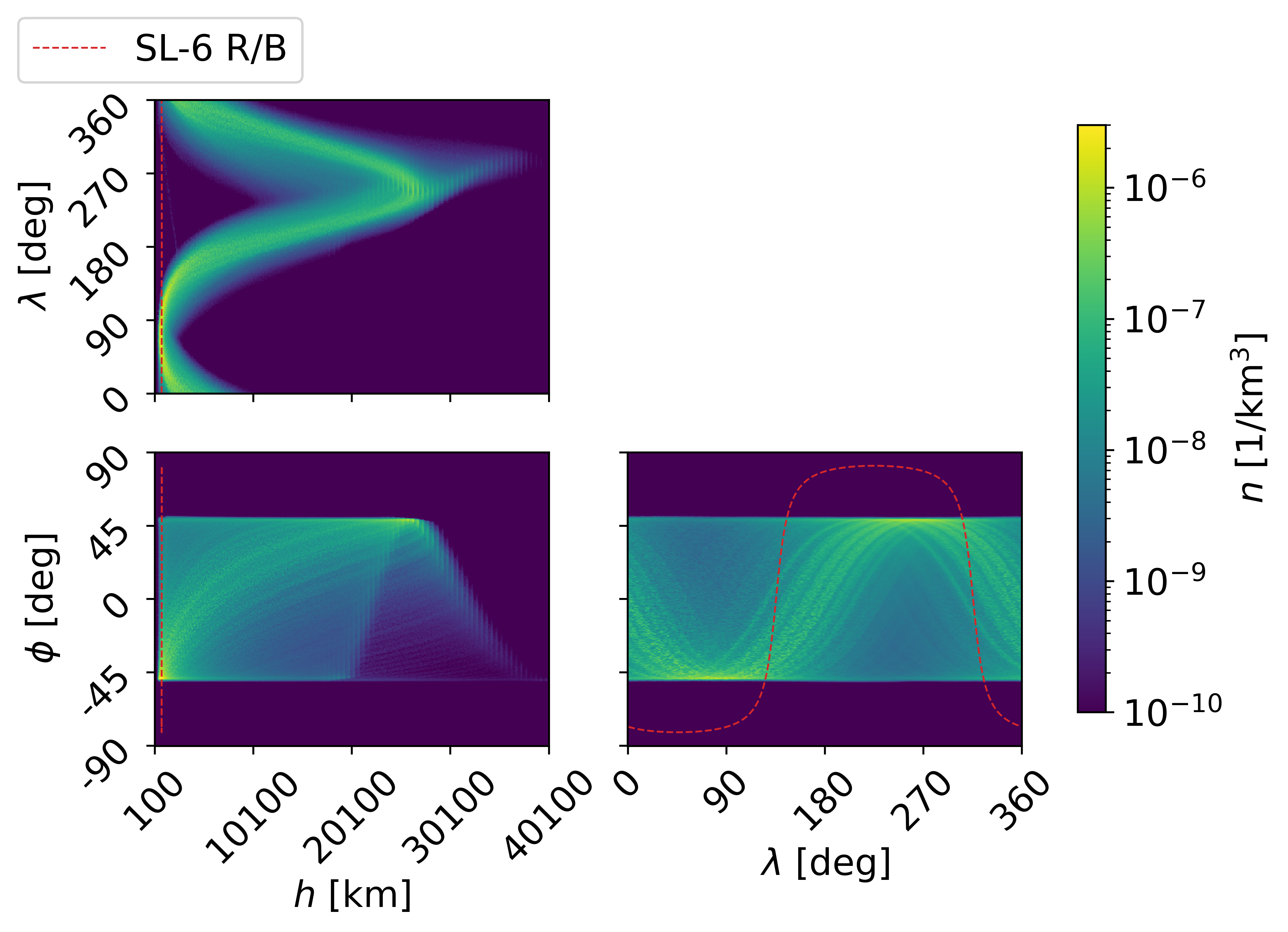}
         \caption{Epoch: $\bm{t_2}$}
         \label{fig:BRIZ-M_spatial_t2}
     \end{subfigure}
     \begin{subfigure}[b]{0.45\textwidth}
         \centering
         \includegraphics[width=\textwidth]{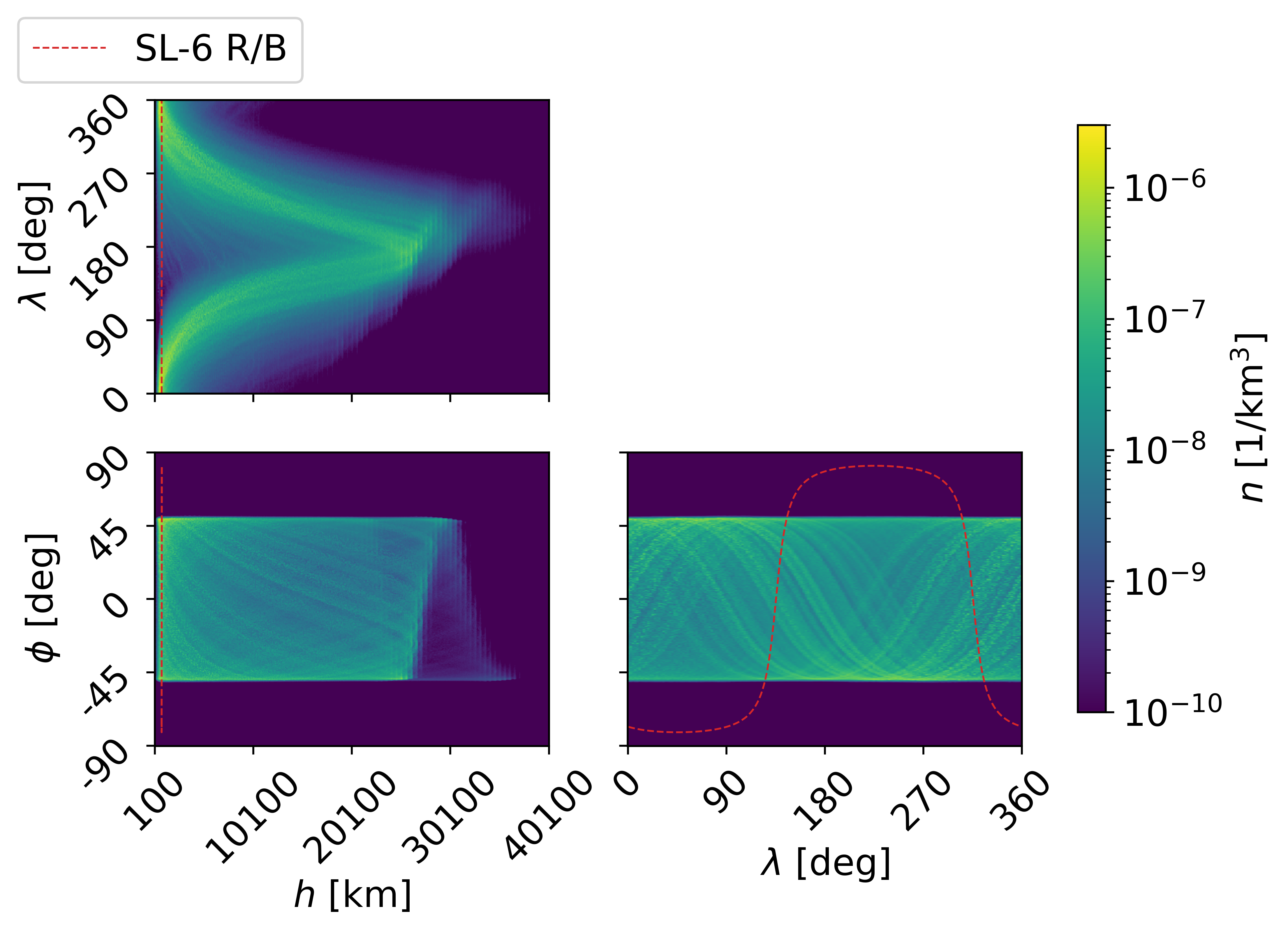}
         \caption{Epoch: $\bm{t_3}$}
         \label{fig:BRIZ-M_spatial_t3}
     \end{subfigure}
     \begin{subfigure}[b]{0.45\textwidth}
         \centering
         \includegraphics[width=\textwidth]{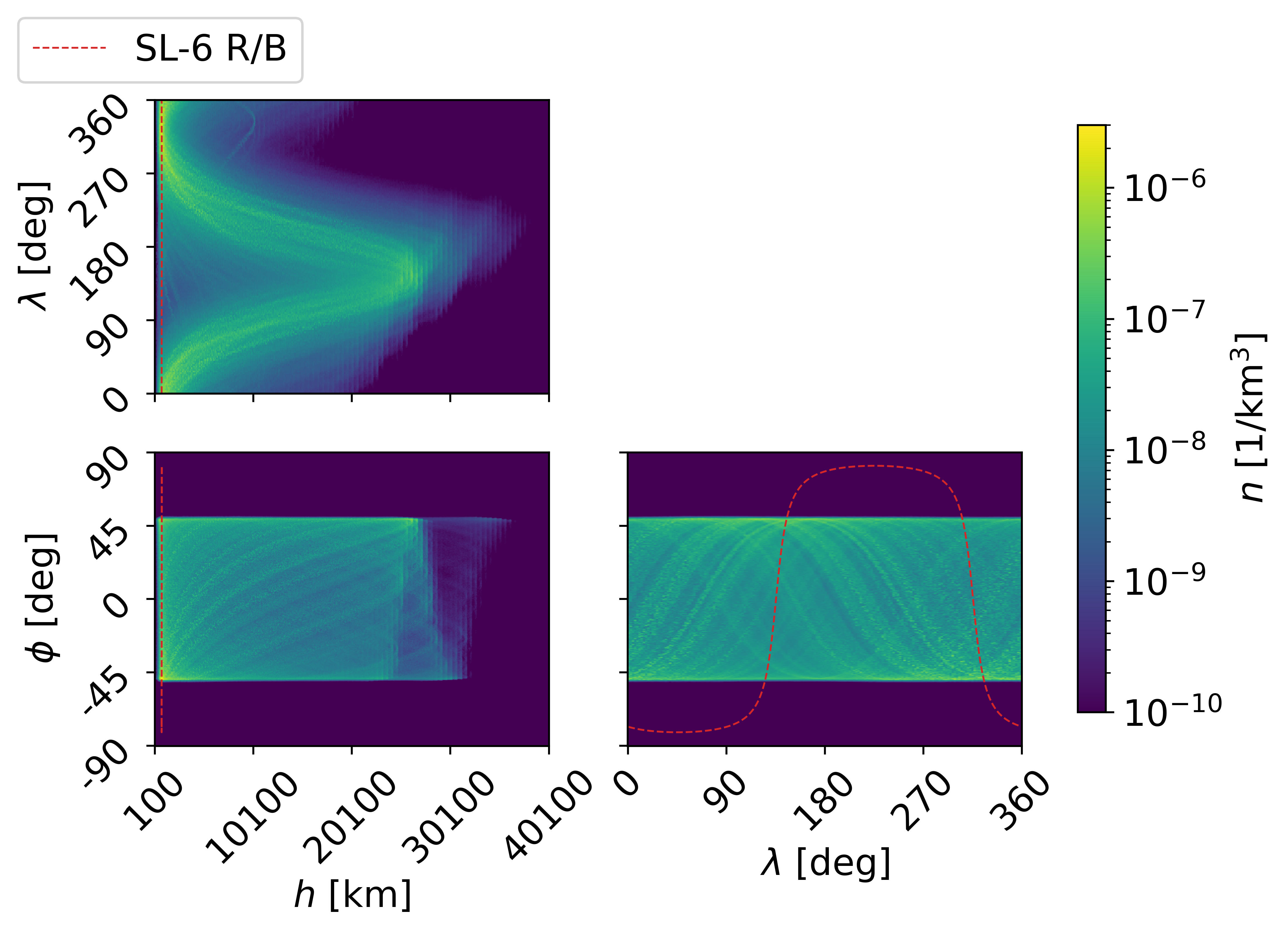}
         \caption{Epoch: $\bm{t_4}$}
         \label{fig:BRIZ-M_spatial_t4}
     \end{subfigure}
     \begin{subfigure}[b]{0.45\textwidth}
         \centering
         \includegraphics[width=\textwidth]{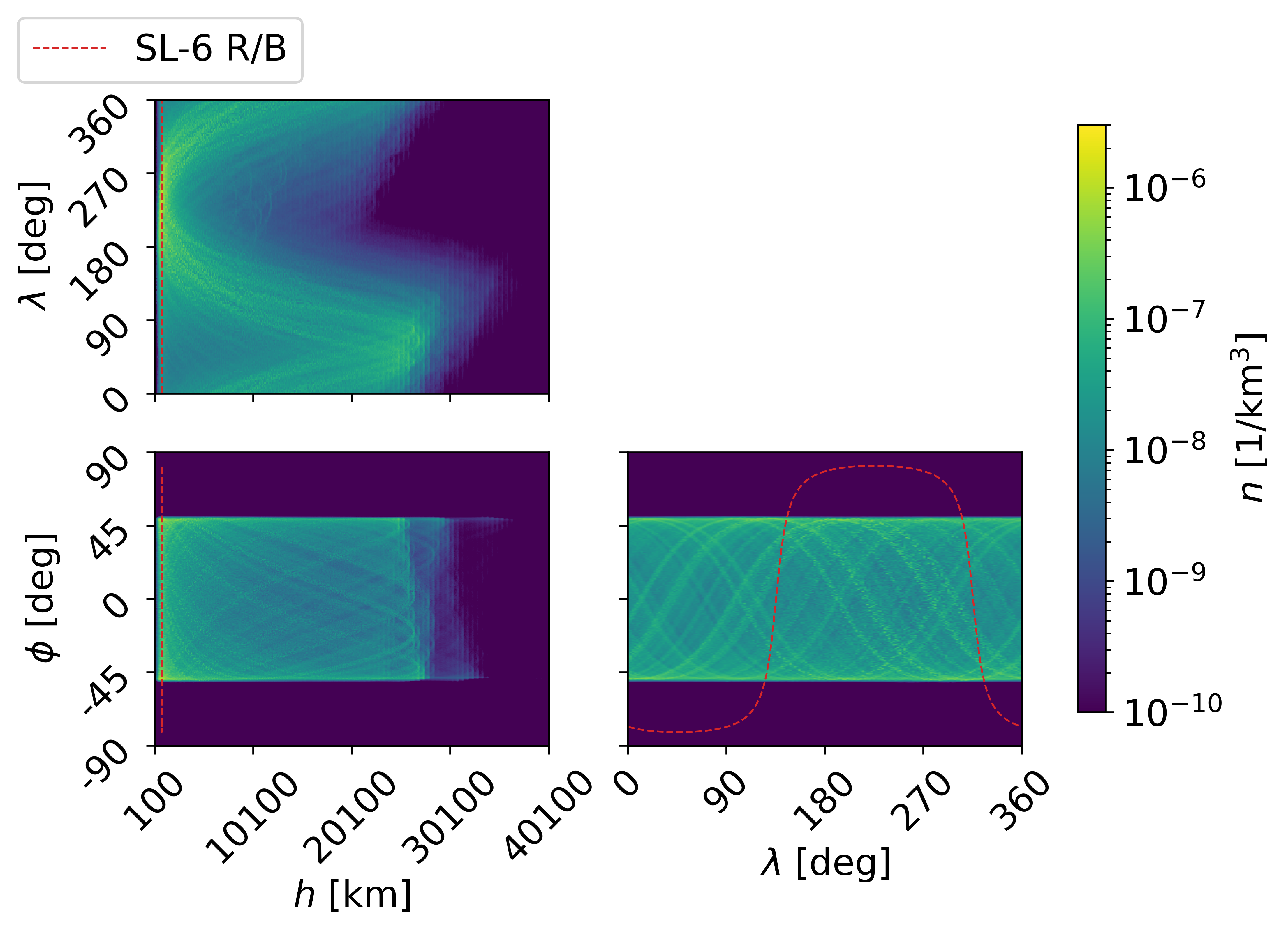}
         \caption{Epoch: $\bm{t_5}$}
         \label{fig:BRIZ-M_spatial_t5}
     \end{subfigure}
     \caption{AMC 14 BRIZ-M R/B fragmentation - Spatial density as function of altitude $\bm{h}$ longitude $\bm{\lambda}$ and latitude $\phi$ over time.}
     \label{fig:BRIZ-M_spatial3D}
\end{figure*}

The collision probability is here computed considering the debris distribution over the slow-varying Keplerian elements $(a,e,i,\Omega,\omega)$. As a result, the summation of Eq.~(\ref{eq:assumption}) is carried out over the four possible intersecting orbits, according to the two values of right ascension of the ascending node $\Omega_1$ and $\Omega_2$, and to the four values of argument of periapsis $\omega_1,\dots,\omega_4$. The profiles of the impact rate and collision probability over time are reported in Figure~\ref{fig:BRIZ-M_Pc5D}. Again, for a better comprehension of the results, the spatial density function at the epochs highlighted in Figure~\ref{fig:BRIZ-M_Pc5D} is also shown in Figure~\ref{fig:BRIZ-M_spatial3D}. 

As it can be noticed, the impact rate, after a narrow peak within the first year of cloud evolution, is almost null for a period of approximately 4 years. Indeed, note that the fragments inclination guarantees a considerably smaller coverage over latitude. Thus, the intersection with the target is possible for a limited range of longitude values $\lambda\in\mathcal{D}_\lambda$. In addition, the fragments move on highly eccentric orbits and the cloud is not randomized over the Euler angles, which means that only a small part of the distribution is in the altitude range covered by the target. In order for the fragments to possibly impact the target, the following condition must verify:
\begin{equation}
    \exists\lambda\in\mathcal{D}_\lambda : n_{\bm{r}}(r_T,\lambda) \neq 0
\end{equation}
At time $t_2$, the debris cloud is at the target altitude for $\lambda\in[30, 110]$ deg. However, for such values of longitude, the target latitude is outside the range covered by the cloud. As a result, no intersection is geometrically possible. In other words, the zero-impact rate period of Figure~\ref{fig:BRIZ-M_Pc5D} coincides with the time needed for the main bulk of fragments to reacquire an orbital plane and orientation capable of providing intersection with the target orbit. Note that, as time passes, even tough the high-density orbits can be still distinguished in the spatial distribution over altitude $h$ and longitude $\lambda$, the cloud covers a wider and wider range of longitude values, at the target altitude, eventually spreading over the entire domain of 360 deg. When this condition verifies, a non-null impact rate is always obtained. Note that, because of the very slow randomization process, if the propagation time were increased, the profile of the impact rate would be characterized by notable oscillations, which would be attenuated in a much longer period of time compared to the case of Figure~\ref{fig:NOAA-16_Pc4D}.

\subsubsection{Collision risk with a naturally evolving target}
\label{Collision risk with a naturally evolving target}

As a last analysis, the natural evolution of the target orbit is also taken into account in the assessment of the collision risk with the debris cloud. Note that this additional feature is of crucial importance; indeed, one should consider that the objects causing the highest hazard for the proliferation of space debris are uncontrolled derelicts as rocket bodies or mission related objects~\cite{McKnight2021}, which naturally evolve under the effect of orbital perturbations.

\begin{figure}[!ht]
     \centering
     \includegraphics[width=0.5\textwidth]{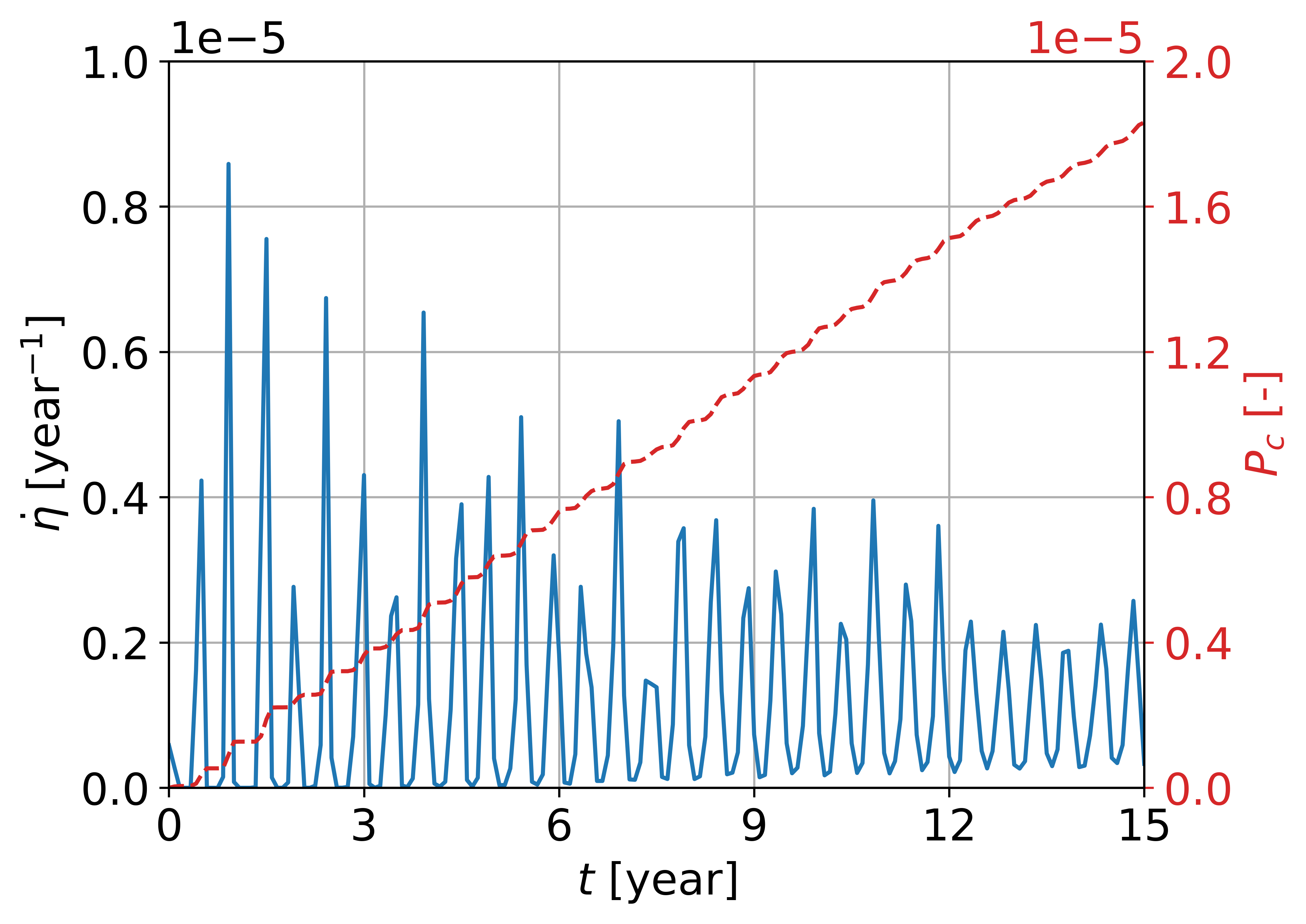}
     \caption{AMC 14 BRIZ/M R/B fragmentation - Impact rate and collision probability with SL-6 R/B over time from the 5D phase space density function. Target orbit evolution included.}
     \label{fig:BRIZ-M_Pc5D_m}
\end{figure}

The Keplerian elements of the target object SL-6 R/B are propagated under the same dynamical model adopted for the debris cloud. The average impact rates and collision probabilities, computed every 1 month, are evaluated considering the target Keplerian elements propagated at the same time epoch. Figure~\ref{fig:BRIZ-M_Pc5D_m} shows the resulting profiles of impact rate and collision probability with SL-6 R/B.

Even though the profile of the impact rate is now the result of the relative evolution of debris cloud and target orbit, it can be immediately inferred that the precession of the target orbit node is driving the profile; indeed, the peaks in the impact rate are found with an approximately 6-months repetition, which coincides with half revolution of the node for a Sun-synchronous orbit. Note that the profile is characterized by several zero-impact rate periods, which are found until the cloud covers the 360 deg range in longitude, at the target altitude, as explained in Section~\ref{Collision risk from a 5D debris cloud in Keplerian elements a, e, i, Om, om}. As the fragments get more spread over right ascension of the ascending node and argument of periapsis, the oscillations reduce in amplitude.

\section{Conclusions}
\label{Conclusions}

The evaluation of the collision risk posed by fragmentation clouds, evolving under any complex dynamics, is a delicate task. Historically, this objective was achieved via semi-deterministic approaches, which suffer of a high computational cost when centimetre- or millimetre-sized particles are modeled. On the other hand, the more efficient probabilistic methods demanded the introduction of simplifying assumptions on the orbital dynamics and impact geometry between fragments and target satellite. This paper proposed a new density-based formulation for an efficient and accurate estimation of the collision hazard caused by a debris cloud, described through a multi-dimensional phase space density function in Keplerian elements. The probabilistic debris cloud propagation model, derived in a previous work by the authors, was adopted to compute the evolution of the fragments density over time. The resulting phase space density function, which discretely varies over both space and time, was provided as input to the collision risk model. This part was the core of the paper. A novel approach for the estimation of the impact rate with a target satellite was here presented. It analytically transforms the six-dimensional phase space density function into the three-dimensional spatial density function, needed for the evaluation of the flux of fragments over the target area. The impact velocity was approximated as bin-wise constant, as it was mathematically demonstrated to only slightly affect the accuracy of the method. The resulting analytical formulation guarantees flexibility to the modeling of any impact geometry in any arbitrarily complex orbital regime, as well as greater efficiency with respect to semi-deterministic approaches. The model was applied to the evaluation of the collision risk posed by real fragmentation events on an uncontrolled rocket body. The additional dynamical features, which the model is able to characterize with respect to previous probabilistic formulation, were consecutively included and discussed.

\section{Acknowledgements}

This project has received funding from the European Research Council (ERC) under the European Union's Horizon 2020 research and innovation programme (grant agreement No 679086 - COMPASS) and from the European Space Agency contract 4000133981/21/D/KS.

\bibliography{References}

\end{document}